\documentclass[reprint,onecolumn, 
 aps,
 pra,
 showkeys,
 nofootinbib
]{revtex4-1}
\usepackage{amsfonts}
\usepackage{amssymb}
\usepackage{amsmath}
\usepackage{graphicx}
\usepackage{hyperref}
\usepackage{subfigure}
\usepackage{dcolumn}
\usepackage{bm}
\usepackage{setspace}
\usepackage[hang]{footmisc}
\usepackage{multirow}
\usepackage[english]{babel}

\def\equationautorefname~#1\null{(#1)\null}

\graphicspath{{./figures/},{./figures.new.davide.temp/}}

\begin{document}
\title{Rough contact mechanics for graded bulk rheology: The role of small-scale wavelengths on rubber friction}

\author{Michele Scaraggi$^{1}$}
\thanks{\mbox{Corresponding author}}
\email[]{michele.scaraggi@unisalento.it}
\author{Davide Comingio$^{1}$}
\affiliation{$^{1}$DII,Universit\'a del Salento, 73100 Monteroni-Lecce, Italy}

\date{\today}


\begin{abstract}
We present a numerical model for the prediction of the rough contact mechanics
of a viscoelastic block, with graded rheology, in steady sliding contact with a randomly rough rigid surface.
In particular, we derive the effective surface response of a stepwise or
continuously-graded block in the Fourier domain, which is then embedded in a Fourier-based residuals molecular dynamic formulation
of the contact mechanics.
Finally we discuss on the role of small-scale wavelengths on rubber friction and contact area, and we demonstrate that
the rough contact mechanics exhibits effective interface properties which converge to asymptotes upon increase of the small-scale roughness content,
when a realistic rheology of the confinement is taken into account.
\end{abstract}
\maketitle
%\makenomenclature
%
%\pacs{PACS number}
%\volumeyear{year}
%\volumenumber{number}
%\issuenumber{number}
%\eid{identifier}
%\date[Date text]{date}
%\received[Received text]{date}

%\revised[Revised text]{date}

%\accepted[Accepted text]{date}

%\published[Published text]{date}

%\startpage{101}
%\endpage{102}

%\tableofcontents
%
%
\section{Introduction}
\label{introduction}

The rough contact mechanics of solids exhibiting graded rheology at confinement is
of major interest for both technological (e.g. seals, rubber friction, tribology of protective coatings)
and biological (tissue engineering, bio-lubrication in the-bone cartilage contact) applications, to cite few.
On the theoretical side, the graded functionalization of surfaces has been long believed to provide
the short distance cutoff, in the range of available contact roughness wavelengths (usually extending to the atomic distance,
given the fractal nature of the real surfaces), as the physical threshold over which smaller asperities do not
contribute to the effective properties of the contact. As an example, in a generic dry rubber contact,
surface contamination\cite{Greenwood300,SCHWEITZ1986289,Persson20013840} as well as a {\it thin skin}\cite{Persson20013840,schipper}
on the rubber surface, e.g. generated as a consequence of a steady-state wear,
can strongly alter the effective surface properties of the confinement. Typically, this contamination layer
will avoid the smallest roughness asperities to contribute to the energy dissipation, thus
reducing the friction force with respect to the ideal value. 
Nevertheless, literature provides no strong theoretical support to this aspect,
mainly due to the lack of (both analytical and numerical) mean field modelling of rough contact mechanics
in presence of graded rheology at confinement, apart few investigations\cite{Persson2012,Paggi2011696}.
In particular, in Ref. \cite{Paggi2011696}, a GW-like (many-asperities) formulation of the rough contact mechanics
under some simplified graded-response assumptions is presented, whereas in Ref. \cite{Persson2012}
the Persson's multiscale contact theory\cite{Persson20013840} has been extended
and the contact area calculated in the case of an elastic coating
bonded onto an elastic half space. In particular, in \cite{Persson2012}
the $M_{zz}(\omega,\mathbf{q})$ function, which provides in the Fourier domain the (out-of-plane) surface displacement response
as a function of the contact pressure field, is derived for the case of a coating bonded
onto a half space, under the assumption of frequency-independent Poisson coefficient.

In this work we will make use of the field decomposition suggested in \cite{Persson20013840}
to determine the effective $M_{zz}(\omega,\mathbf{q})$ (in the Fourier domain) function for the more general case of a stepwise or
continuously-graded block. The Poisson coefficient for the generic layer is assumed frequency-dependent, which makes the theory of more general applicability. This effective surface response will be then implemented in a Fourier-based
residuals molecular dynamic (RMD) formulation of the contact
mechanics\cite{Persson2014,scaraggi.in.prep,scaraggi.in.prep.true}, however, the same function might be easily embedded in
the Persson's mean field analytical contact model as well.
The RMD model will then be adopted to a focussed investigation of the role of small-scale roughness wavelengths on the rubber friction and contact area.
We numerically show that the rough contact exhibits effective interface properties which converge to asymptotes upon increase of the small-scale roughness content,
when a realistic rheology of the confinement, which includes a graded rheology, is taken into account. For the rubber contact case a graded rheology
has been recently experimentally shown in Ref. \cite{schipper}, where the authors clearly demonstrate the existence of a modified surface layer with strongly different
properties than the bulk, inspiring indeed our theoretical investigation.

The manuscript is organized as follows. In Sec. \ref{BEM} we summarize the
BEM (boundary element method) numerical scheme adopted for the investigation of a steady-sliding rough
interaction characterized by arbitrary rheological properties, whereas in
Sec. \ref{navier.section} we more specifically focus on the calculation of
the surface displacement Green's function, in the Fourier domain $%
M_{zz}\left( \mathbf{q},\omega \right) $, for a generic block with stepwise
graded (isotropic) viscoelastic rheology. In Sec. \ref{results} we apply the
numerical model to the investigation of the role of the (wear) modified
surface layer (MSL) on the rubber friction and contact area for the
simplest case of a rubber bulk covered by an elastic coating, in steady
sliding contact onto a randomly rough rigid surface. Finally, in Sec. \ref%
{discussion} we more generally discuss on the role of the graded bulk
rheology in rough contact mechanics, whereas in Sec. \ref{conclusions} the
conclusions follow. In Appendix \ref{appendix.1} we solve the Navier's
equation for a homogeneously-viscoelastic infinitely-wide slab (of finite
thickness) in the quasi-static deformation dynamics. Analytical relations
are derived for the surface response of coated bulks in the most general
case of non-constant (in the frequency domain) Poisson's ratio. We show
that, even for the simplest case of coating on bulk, the adoption of constant
Poisson ratio (i.e. independent from frequency) lead to qualitatively different results
with respect to the more general case of frequency dependent Poisson ratio, in
a range of roughness wavelengths. In Appendix \ref{appendix.2}
the surface response for the general case of continuously-graded
viscoelastic rheology is formulated in term of a set of non-linear
differential equations, which is solved in the two representative cases of
linear and sinusoidal variation of the confined elastic properties of a
bulk. The results are then compared with the predictions of the
stepwise-graded theory (Sec. \ref{navier.section}) as applied to a discretized version of the confinement
(at different numbers of divisions in sub-layers)
showing, for the sinusoidal variation, that the number of layers needed for
the stepwise-graded surface response to converge to the continuously-graded
predictions can be (relatively) very large.

\section{Summary of the numerical scheme for a steady-sliding rough
interaction}

\label{BEM}

We consider the case of a rigid, \textit{periodically-rough} surface (of $%
L_{0}$ periodic length, in both $x$- and $y$-direction, with small
wavelength cut-off frequency $q_{0}=2\pi /L_{0}$) in steady sliding
adhesionless contact with a graded body characterized by linear rheology,
under isothermal and frictionless conditions. We assume the small
deformation regime to apply, as well as a small square slope roughness $%
h\left( \mathbf{x}\right) $, with $m_{2}=\left\langle \nabla h(\mathbf{x}%
)^{2}\right\rangle \ll 1$ ($\left\langle h\right\rangle =0$). 
\begin{figure}[tbh]
\centering
\subfigure[]{
\includegraphics[
                        width=0.407\textwidth,
                        angle=0
                ]{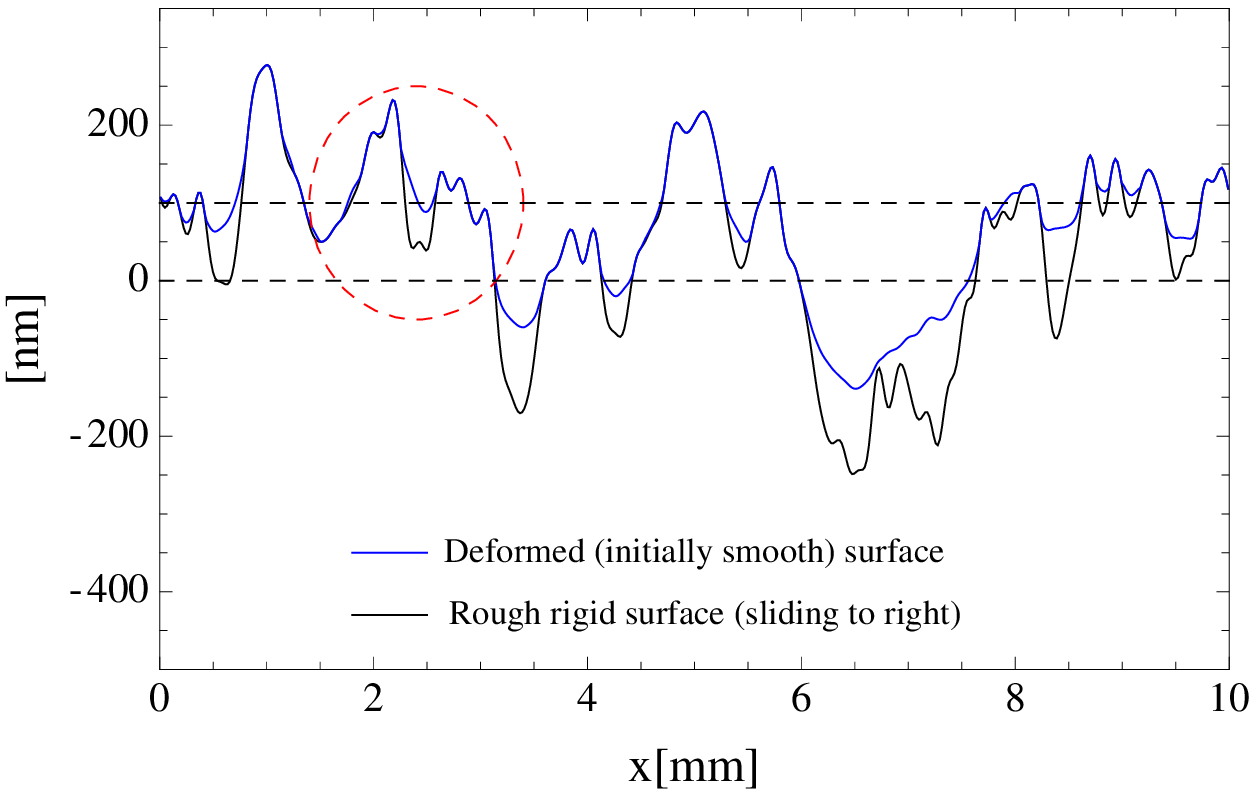} \label{geometry.eps}
                } \qquad 
\subfigure[]{
\includegraphics[
                        width=0.4\textwidth,
                        angle=0
                ]{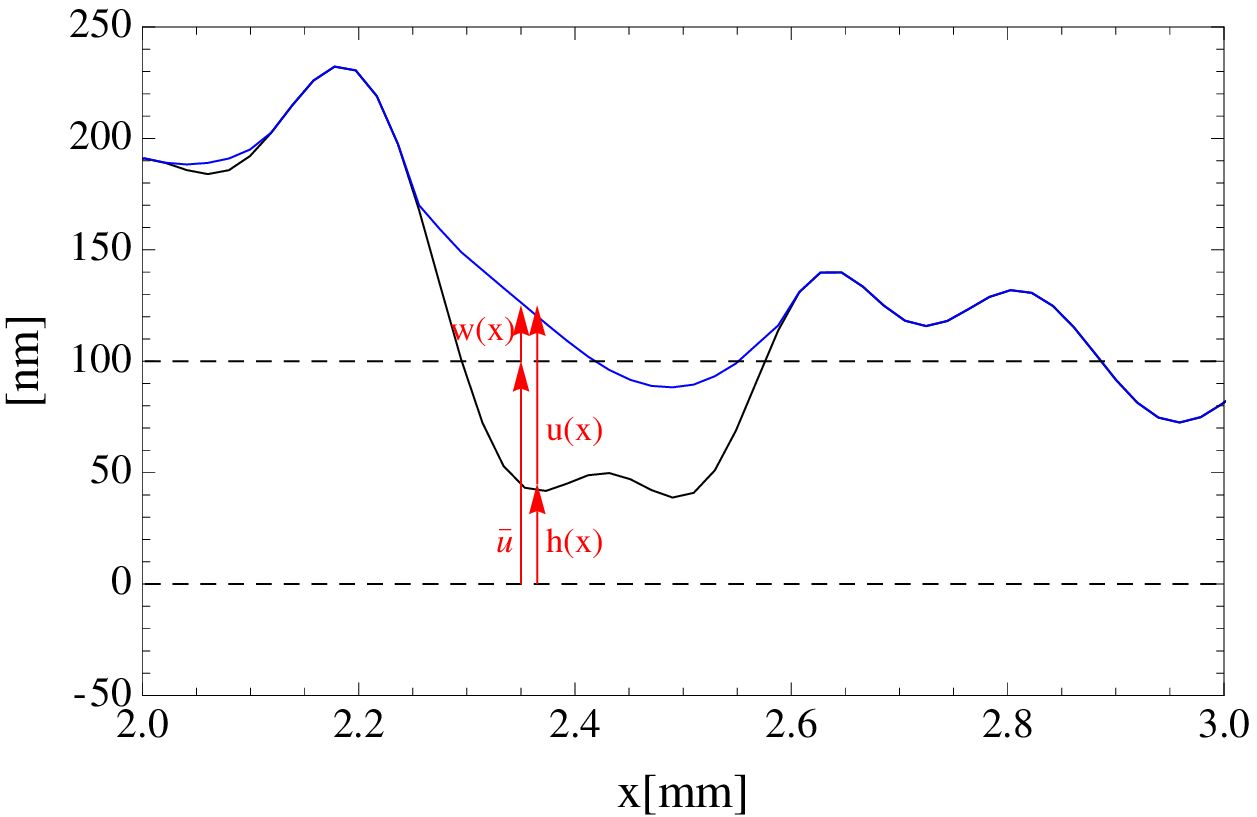} \label{geometry1.eps}
                }
\caption{a) Cross section of a generic contact interface. b) Magnified view
of the encircled area (in the left figure), with indication of the gap
equation \protect\autoref{gap} terms. Schematics.}
\label{cross.section}
\end{figure}
In Fig. \ref{cross.section} we show a schematic representation of the
contact interface, in a reference moving with the rough sliding surface. In
such a reference, the local interfacial separation $u\left( \mathbf{x}%
\right) $ can be agreed to be:%
\begin{equation}
u\left( \mathbf{x}\right) =\bar{u}+w\left( \mathbf{x}\right) -h\left( 
\mathbf{x}\right) ,  \label{gap}
\end{equation}%
where $\bar{u}$ is the average interfacial separation, $w\left( \mathbf{x}%
\right) $ the surface out-of-average-plane displacement and $h\left( \mathbf{%
x}\right) $ the surface roughness, with $\left\langle w\left( \mathbf{x}%
\right) \right\rangle =\left\langle h\left( \mathbf{x}\right) \right\rangle
=0$. We can define the following:%
\begin{equation*}
w\left( \mathbf{q}\right) =\left( 2\pi \right) ^{-2}\int d^{2}\mathbf{x\ }%
w\left( \mathbf{x}\right) e^{-i\mathbf{q}\cdot \mathbf{x}}
\end{equation*}%
and 
\begin{equation*}
\sigma \left( \mathbf{q}\right) =\left( 2\pi \right) ^{-2}\int d^{2}\mathbf{%
x\ }\sigma \left( \mathbf{x}\right) e^{-i\mathbf{q}\cdot \mathbf{x}},
\end{equation*}%
where $\sigma \left( \mathbf{x}\right) =\Delta \sigma \left( \mathbf{x}%
\right) +\sigma _{0}$ is the distribution of interfacial pressures [$\sigma
_{0}=\left\langle \sigma \left( \mathbf{x}\right) \right\rangle $ is the
average contact pressure] in the moving reference. Following the discussion
reported in Sec. \ref{navier.section}, $w\left( \mathbf{x}\right) $ can be
related to $\sigma \left( \mathbf{x}\right) $ through a simple equation in
the Fourier space%
\begin{equation}
w\left( \mathbf{q}\right) =M_{zz}\left( \mathbf{q},\omega =\mathbf{q}\cdot 
\mathbf{v}\right) \sigma \left( \mathbf{q}\right) ,  \label{A1.fourier}
\end{equation}%
where $M_{zz}\left( \mathbf{q},\omega \right) $ is the complex surface
responce of the block in the frequency domain, and $\mathbf{v}$ the sliding
velocity [in this work $\mathbf{v}=\left( v,0\right) $ without any losss of
generality]. $M_{zz}\left( \mathbf{q},\omega \right) $ depends on the
rheological and geometrical properties of the block, and its formulation
will be specifically presented in Sec. \ref{navier.section}. We observe that
in the simplest case of bulk viscoelastic contact with frequency-independent
Poisson ratio, $M_{zz}\left( \mathbf{q},\omega \right) =2/\left[ \left\vert 
\mathbf{q}\right\vert E_{\mathrm{r}}\left( \omega \right) \right] $, where $%
E_{\mathrm{r}}\left( \omega \right) =E\left( \omega \right) /\left( 1-\nu
^{2}\right) $ is the frequency-dependent (complex) reduced Young's modulus, $%
\nu $ is the Poisson ratio. In this work the assumption of constant Posson
ratio will not be adopted (unless differently explicited), thus the theory
developed hereinafter is of more general applicability, e.g. it can be
easily adapted to existing mean field contact mechanics formulations as
well. \autoref{A1.fourier} is obtained by considering that the stress in the
fixed reference $\sigma \left( \mathbf{q},\omega \right) =\sigma \left( 
\mathbf{q}\right) \delta \left( \omega -\mathbf{q}\cdot \mathbf{v}%
_{0}\right) $ is related to the displacement in the fixed reference $w\left( 
\mathbf{q},\omega \right) =w\left( \mathbf{q}\right) \delta \left( \omega -%
\mathbf{q}\cdot \mathbf{v}_{0}\right) $ through the constitutive
relationship $w\left( \mathbf{q},\omega \right) =M_{\mathrm{zz}}\left(\mathbf{
q},\omega \right) \sigma \left( \mathbf{q},\omega \right) $ (see Sec. \ref%
{navier.section}) resulting, after integration over $\omega $, in \ref%
{A1.fourier}.

Finally, the relation between separation $u(\mathbf{x})$ and interaction
pressure $\sigma (\mathbf{x})$ is calculated within the Derjaguin's
approximation \cite{Derjaguin1934}, and it can be written in term of a
generic interaction law \cite{Persson2014,scaraggi.in.prep}:

\begin{equation}
\sigma (u)=f(u).  \label{sigma_from_f}
\end{equation}%
In this work we have adopted the (integrated) repulsive term of the L-J
potential in \autoref{sigma_from_f} to simulate the adhesionless interaction.
However, the theory can be easily extended to other interface laws \cite%
{Persson2014}. \autoref{gap}, \autoref{A1.fourier} and \autoref{sigma_from_f} are
discretized on a regular square mesh of grid size $\delta $ in term of a
residuals molecular dynamics process \cite{scaraggi.in.prep.true,Persson2014}%
, resulting in the following set of equations:

\begin{align}
L_{ij}& =-u_{ij}+\left( \bar{u}+w_{ij}-h_{ij}\right)  \label{Lij} \\
\sigma _{ij}& =f\left( u_{ij}\right)  \label{sigmaij} \\
\sigma _{ij}& \rightarrow \Delta \sigma
(q_{hk})=M_{zz}^{-1}w(q_{hk})\rightarrow w\left( x_{ij}\right)
\label{delta.sigmaij}
\end{align}%
where $L_{ij}$ is the generic residual (related to the generic iterative
solution $u_{ij}$). \autoref{Lij} are solved in $u_{ij}$ though a Verlet
intergation scheme, whereas the solution accuracy is set by requiring

\begin{align*}
\left\langle L_{ij}^{2}/u_{ij}^{2}\right\rangle ^{1/2}& <\varepsilon _{%
\mathrm{L}} \\
\left\langle \left[ \left( u_{ij}^{n}-u_{ij}^{n-1}\right) /u_{ij}^{n-1}%
\right] ^{2}\right\rangle ^{1/2}& <\varepsilon _{\mathrm{u}},
\end{align*}%
where both errors are typically of order $10^{-4}$.

Among the mean physical quantities which can be extracted from the solution
fields, the one of particular relevance for this work is the micro-rolling
friction coefficient $\mu _{\mathrm{r}}=F_{\mathrm{r}}/F_{\mathrm{N}}$,
where the micro-rolling force $F_{\mathrm{r}}$ reads%
\begin{equation*}
F_{\mathrm{r}}=\left\vert \mathbf{v}\right\vert
^{-1}\int_{A_{0}}d^{2}x~\sigma (\mathbf{x})\left[ \mathbf{\nabla }h\left( 
\mathbf{x}\right) \cdot \mathbf{v}\right] ,
\end{equation*}%
and the normal load $F_{\mathrm{N}}=\int_{A_{0}}d^{2}x\ \sigma (\mathbf{x})$.

\section{$M_{zz}\left( \mathbf{q},\protect\omega \right) \ $for a bulk with
stepwise-graded viscoelastic rheology}

\label{navier.section}In this section we will show how to calculate $%
M_{zz}\left( \mathbf{q},\omega \right) $ for a stepwise-graded viscoelastic
composite. The case of continuously-graded viscoelastic composite will be
discussed in Appendix \ref{appendix.2}. 
\begin{figure}[tbh]
\centering
\includegraphics[
                        width=0.5\textwidth,
                        angle=0
                ]{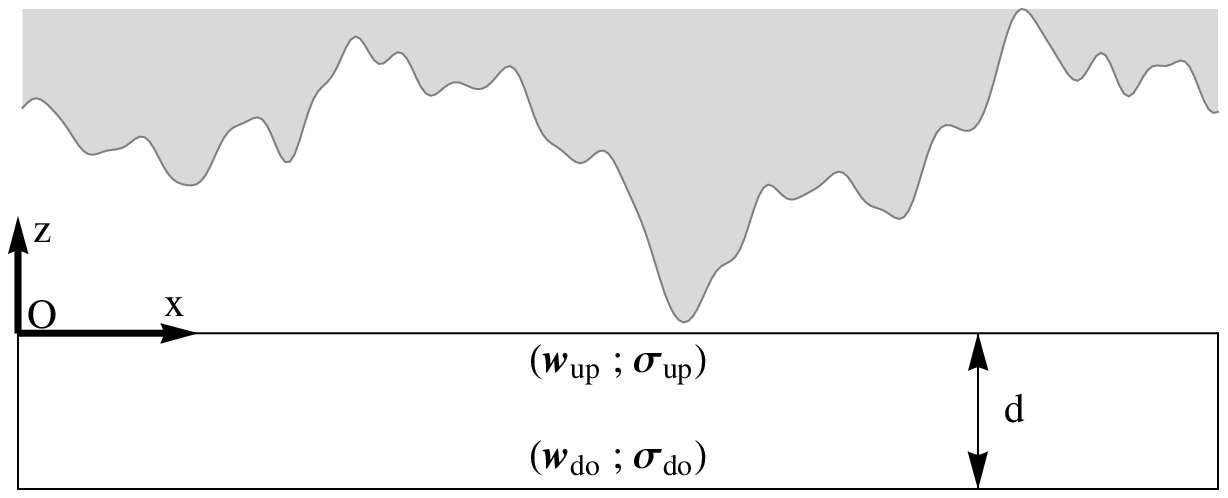}
\caption{Schematic of a infinitely-wide slab of finite thickness $d$,
characterized by a linear viscoelstic rheology. $\protect\sigma _{\mathrm{up}%
}$ ($\protect\sigma _{\mathrm{do}}$) and $w_{\mathrm{up}}$ ($w_{\mathrm{do}}$%
) are, respectively, the stress and displacement fields on the top $z=0$
(bottom $z=-d$) surface.}
\label{singlelayer.eps}
\end{figure}
In particular, we first consider the case of a linearly-viscoelastic
infinitely-wide slab of thickness $d$, see Fig. \ref{singlelayer.eps}. By
considering the following Fourier transform ($t\rightarrow \omega $ and $%
\mathbf{x}\rightarrow \mathbf{q}$)%
\begin{eqnarray*}
\mathbf{w}\left( \mathbf{q},z,\omega \right) &=&\left( 2\pi \right)
^{-3}\int dt\int d^{2}x~\mathbf{w}\left( \mathbf{x},z,t\right) e^{-i\left( 
\mathbf{q}\cdot \mathbf{x}-\omega t\right) } \\
\mu \left( \omega \right) &=&\int dt~\mu \left( t\right) e^{-i\left( -\omega
t\right) }
\end{eqnarray*}%
and, inversely,%
\begin{eqnarray*}
\mu \left( t\right) &=&\left( 2\pi \right) ^{-1}\int d \omega ~e^{i\left(
-\omega t\right) }\mu \left( \omega \right) \\
\mathbf{w}\left( \mathbf{x},z,t\right) &=&\int dt\int d^{2}x~\mathbf{w}%
\left( \mathbf{q},\omega \right) e^{i\left( \mathbf{q}\cdot \mathbf{x}%
-\omega t\right) },
\end{eqnarray*}%
the relation between the stress and displacement fields on the top ($z=0$)
and bottom surface ($z=-d$), in the limit of quasi-static interaction [i.e. $%
\omega /\left( qc\right) =v/c\ll 1$, see Appendix \ref{appendix.1}, where $c$
is the generic sound speed] reads in matrix form%
\begin{equation}
\left[ 
\begin{array}{c}
\boldsymbol{\sigma }_{\mathrm{up}}/\left[ E_{\mathrm{r}}\left( \omega
\right) q/2\right] \\ 
\mathbf{w}_{\mathrm{up}}%
\end{array}%
\right] =\cosh \left( qd\right) \left[ 
\begin{array}{cc}
\mathbf{M}_{1} & \mathbf{M}_{2} \\ 
\mathbf{M}_{3} & \mathbf{M}_{4}%
\end{array}%
\right] \left[ 
\begin{array}{c}
\boldsymbol{\sigma }_{\mathrm{do}}/\left[ E_{\mathrm{r}}\left( \omega
\right) q/2\right] \\ 
\mathbf{w}_{\mathrm{do}}%
\end{array}%
\right] ,  \label{the.slab.matrix}
\end{equation}%
where $\mathbf{M}_{j}\left[ \mathbf{q}d,\nu \left( \omega \right) \right] $
is a 3 by 3 matrix. $\boldsymbol{\sigma }_{\mathrm{up}}$ ($\boldsymbol{%
\sigma }_{\mathrm{do}}$) and $\mathbf{w}_{\mathrm{up}}$ ($\mathbf{w}_{%
\mathrm{do}}$) are, respectively, the stress and displacement fields on the
top (bottom) surface, see Fig. \ref{singlelayer.eps}. $\mathbf{M}_{j}$ are
determined in Appendix \ref{appendix.1} for the most general case of
frequency-dependent Poisson ratio $\nu \left( \omega \right) $, as well as
reported for the limiting case of constant $\nu $. \autoref{the.slab.matrix} can
be conveniently rephrased depending on the adopted boundary conditions (BCs)
on the bottom surface ($z=-d$), e.g.%
\begin{equation}
\mathbf{w}_{\mathrm{up}}\left( \mathbf{q},\omega \right) =\mathbf{M}_{3}%
\mathbf{M}_{1}^{-1}\frac{\boldsymbol{\sigma }_{\mathrm{up}}\left( \mathbf{q}%
,\omega \right) }{E_{\mathrm{r}}\left( \omega \right) q/2}+\cosh \left(
qd\right) \left[ \mathbf{M}_{4}-\mathbf{M}_{3}\mathbf{M}_{1}^{-1}\mathbf{M}%
_{2}\right] \mathbf{w}_{\mathrm{do}}\left( \mathbf{q},\omega \right)
\label{BC1}
\end{equation}%
or%
\begin{equation}
\mathbf{w}_{\mathrm{up}}\left( \mathbf{q},\omega \right) =\mathbf{M}_{4}%
\mathbf{M}_{2}^{-1}\frac{\boldsymbol{\sigma }_{\mathrm{up}}\left( \mathbf{q}%
,\omega \right) }{E_{\mathrm{r}}\left( \omega \right) q/2}+\cosh \left(
qd\right) \left[ \mathbf{M}_{3}-\mathbf{M}_{4}\mathbf{M}_{2}^{-1}\mathbf{M}%
_{1}\right] \frac{\mathbf{\sigma }_{\mathrm{do}}\left( \mathbf{q},\omega
\right) }{E_{\mathrm{r}}\left( \omega \right) q/2}.  \label{BC2}
\end{equation}%
\begin{figure}[tbh]
\centering
\includegraphics[
                        width=0.5\textwidth,
                        angle=0
                ]{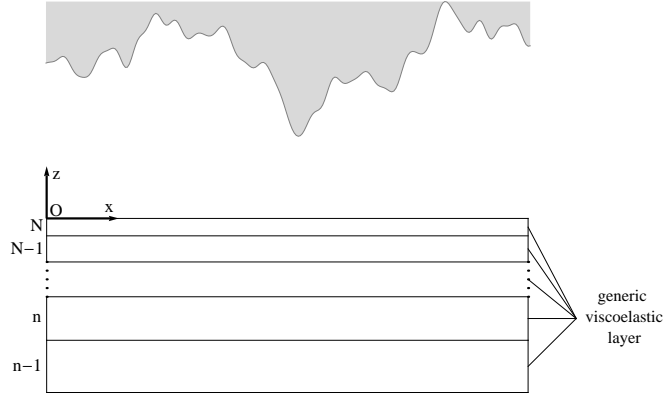}
\caption{Schematic of an infinitely-wide slab of finite thickness $%
d=\sum_{j}d_{j}$, characterized by a step-wise graded linear viscoelstic
rheology.}
\label{stepwise.schematic}
\end{figure}

Now, in Fig. \ref{stepwise.schematic} we show the schematic of the generic
composite slab with a step-wise graded rheology, with $j=1..n..N$ bonded
layers. We first assume the generic layer $\left( n-1\right) $ to be
described by the general stress-displacement relation%
\begin{equation*}
\mathbf{w}_{\mathrm{up}}=\left[ \mathbf{M}\right] ^{\left( n-1\right) }\frac{%
\boldsymbol{\sigma }_{\mathrm{up}}}{\left[ E_{\mathrm{r}}\left( \omega
\right) \right] ^{\left( n-1\right) }q/2},
\end{equation*}%
where $\mathbf{M}$ is a 3 by 3 matrix. Imposing the continuity of stress and
displacement between layer $\left( n-1\right) $ and $\left( n\right) $, and
by using \autoref{the.slab.matrix}, we get for the layer $\left( n\right) $%
\begin{equation*}
\mathbf{w}_{\mathrm{up}}=\left[ \mathbf{M}_{3}+\frac{E_{\mathrm{r}}\left(
\omega \right) }{\left[ E_{\mathrm{r}}\left( \omega \right) \right] ^{n-1}}%
\mathbf{M}_{4}\left[ \mathbf{M}\right] ^{n-1}\right] \left[ \mathbf{M}_{1}+%
\frac{E_{\mathrm{r}}\left( \omega \right) }{\left[ E_{\mathrm{r}}\left(
\omega \right) \right] ^{n-1}}\mathbf{M}_{2}\left[ \mathbf{M}\right] ^{n-1}%
\right] ^{-1}\frac{\boldsymbol{\sigma }_{\mathrm{up}}}{E_{\mathrm{r}}\left(
\omega \right) q/2},
\end{equation*}%
where the index $\left( n\right) $ has been dropped for simplicity. Thus $%
\mathbf{M}$ for the layer $\left( n\right) $ reads%
\begin{equation}
\left[ \mathbf{M}\right] ^{\left( n\right) }=\left[ \mathbf{M}_{3}+\frac{E_{%
\mathrm{r}}\left( \omega \right) }{\left[ E_{\mathrm{r}}\left( \omega
\right) \right] ^{\left( n-1\right) }}\mathbf{M}_{4}\left[ \mathbf{M}\right]
^{\left( n-1\right) }\right] \left[ \mathbf{M}_{1}+\frac{E_{\mathrm{r}%
}\left( \omega \right) }{\left[ E_{\mathrm{r}}\left( \omega \right) \right]
^{\left( n-1\right) }}\mathbf{M}_{2}\left[ \mathbf{M}\right] ^{\left(
n-1\right) }\right] ^{-1},  \label{graded.stepped}
\end{equation}%
with again%
\begin{equation*}
\mathbf{w}_{\mathrm{up}}=\left[ \mathbf{M}\right] ^{\left( n\right) }\frac{%
\boldsymbol{\sigma }_{\mathrm{up}}}{\left[ E_{\mathrm{r}}\left( \omega
\right) \right] ^{\left( n\right) }q/2}.
\end{equation*}%
\autoref{graded.stepped} shows that the surface responce of a stepwise-graded
composite can be easily determined with a recursive calculation.

Finally, for the stepwise graded composite with $N$-layers%
\begin{equation}
M_{zz}\left( \mathbf{q},\omega \right) =\frac{2}{q}\frac{\left[ \mathbf{M}%
\left( \mathbf{q},\omega \right) \right] _{3,3}^{\left( N\right) }}{\left[
E_{\mathrm{r}}\left( \omega \right) \right] ^{\left( N\right) }},
\label{final}
\end{equation}%
where $\left[ \mathbf{M}\right] ^{\left( 0\right) }$ [innermost layer,
needed to initialize \autoref{graded.stepped}] is obtained from \autoref{BC1} or \ref%
{BC2}, depending on the adopted BCs $\mathbf{w}_{\mathrm{do}}\left( \mathbf{q%
},\omega \right) =0$ (thus $\left[ \mathbf{M}\right] ^{\left( 0\right) }=%
\mathbf{M}_{3}\mathbf{M}_{1}^{-1}$) or $\mathbf{\sigma }_{\mathrm{do}}\left( 
\mathbf{q},\omega \right) $ (thus $\left[ \mathbf{M}\right] ^{\left(
0\right) }=\mathbf{M}_{4}\mathbf{M}_{2}^{-1}$), for $q\neq 0$. In the
simplest case of a bulky $\left( 0\right) $-layer (corresponding to a half
space, i.e.\ $d\rightarrow \infty $), $\left[ \mathbf{M}\right] ^{\left(
0\right) }=\mathbf{M}_{3}\mathbf{M}_{1}^{-1}=\mathbf{M}_{4}\mathbf{M}%
_{2}^{-1}$ reads%
\begin{equation}
\left[ \mathbf{M}\right] ^{\left( 0\right) }=%
\begin{bmatrix}
1+\frac{\nu }{p}\frac{q_{y}^{2}}{q^{2}} & -\frac{\nu }{p}\frac{q_{x}q_{y}}{%
q^{2}} & i\frac{2\nu -1}{2p}\frac{q_{x}}{q} \\ 
-\frac{\nu }{p}\frac{q_{x}q_{y}}{q^{2}} & 1+\frac{\nu }{p}\frac{q_{x}^{2}}{%
q^{2}} & i\frac{2\nu -1}{2p}\frac{q_{y}}{q} \\ 
i\frac{1-4\nu \left( 1-\nu _{0}\right) }{1-2\nu _{0}}\frac{q_{x}}{2pq} & i%
\frac{1-4\nu \left( 1-\nu _{0}\right) }{1-2\nu _{0}}\frac{q_{y}}{2pq} & 
\frac{p_{0}}{p}\frac{1-2\nu }{1-2\nu _{0}}%
\end{bmatrix}%
.  \label{bulk.new}
\end{equation}%
In \autoref{bulk.new} $p=1-\nu \left( \omega \right) $ and $p_{0}=1-\nu _{0}$,
where $\nu _{0}=\nu \left( \omega =0\right) $ is the Posson coefficient in
the rubbery regime. Observe that for the most general case ($p\neq p_{0}$) $%
\left[ \mathbf{M}\right] _{3,3}^{\left( 0\right) }$ in \autoref{bulk.new} is not
equal to 1, as (indirectly) expected from the theory of the
elastic-viscoelastic correspondance. This also suggests that our model can
easily overtake the restrictions imposed by the adoption of the
elastic-viscoelastic correspondance principle (frequency-independent Poisson
ratio) to the rubber rheological characteristics which can be modelled. In
the simplest case where $\nu \left( \omega \right) =\nu =\nu _{0}$, \ref%
{bulk.new} simplifies to the well known%
\begin{equation}
\left[ \mathbf{M}\right] ^{\left( 0\right) }=%
\begin{bmatrix}
1+\frac{\nu }{p}\frac{q_{y}^{2}}{q^{2}} & -\frac{\nu }{p}\frac{q_{x}q_{y}}{%
q^{2}} & i\frac{2\nu -1}{2p}\frac{q_{x}}{q} \\ 
-\frac{\nu }{p}\frac{q_{x}q_{y}}{q^{2}} & 1+\frac{\nu }{p}\frac{q_{x}^{2}}{%
q^{2}} & i\frac{2\nu -1}{2p}\frac{q_{y}}{q} \\ 
i\frac{1-2\nu }{2p}\frac{q_{x}}{q} & i\frac{1-2\nu }{2p}\frac{q_{y}}{q} & 1%
\end{bmatrix}%
.  \label{bulk.old}
\end{equation}

Finally, the linear viscoelastic complex modulus (which can be measured
recurring to standard techniques \cite{Lorenz2014565}) $E\left( \omega
\right) $ can be related to the creep spectrum through a Prony series\cite%
{Lorenz2014565,Scaraggi201415,Scaraggi2014}, obtaining%
\begin{equation}
\frac{1}{E(\omega )}\approx \frac{1}{E_{\infty }}+\sum_{j=1}^{N}\frac{H(\tau
_{j})}{1-i\omega \tau _{i}}  \label{prony}
\end{equation}%
where $N$ is the number of relaxation times, $H(\tau _{j})$ the discrete
creep function, and $\tau _{j}$ the generic relaxation time. $E_{\infty }$
is the elastic modulus in the glassy regime. 
\begin{figure}[tbh]
\centering
\includegraphics[
                        width=0.4\textwidth,
                        angle=0
                ]{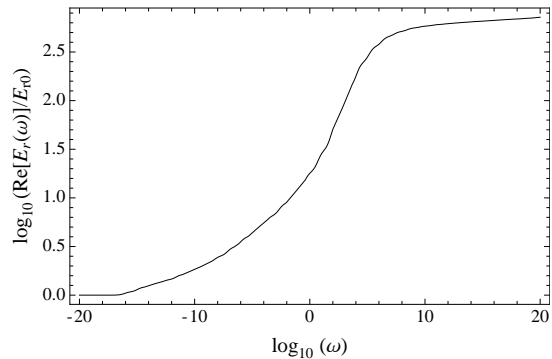}
\caption{Real part of the (dimensionless) reduced viscoelastic modulus $E_{%
\mathrm{r}}(\protect\omega )/E_{\mathrm{r}0}$ adopted in this work, as a
function of the frequency (in \textrm{s}$^{-1}$), in \textrm{log}$_{10}$-%
\textrm{log}$_{10}$. For a car tire rubber compound at a low external
temperature.}
\label{RealPart.eps}
\end{figure}
In Fig. \ref{RealPart.eps} we show the real part of the (dimensionless)
reduced viscoelastic modulus $E_{\mathrm{r}}(\omega )/E_{\mathrm{r}0}$
adopted in this work, as a function of the frequency (in \textrm{s}$^{-1}$),
in a \textrm{log}$_{10}$-\textrm{log}$_{10}$ scale. The adopted viscoelastic
modulus corresponds to a car tire tread-block compound under low operating
temperatures (see e.g. \cite{scaraggi.in.prep}).

\section{Numerical results}

\label{results}In this section we investigate for the first time the rough
contact mechanics of a rubber block (see Fig. \ref{RealPart.eps} for the rheological properties) covered by
a surface layer with modified rheological properties (with respect to the
bulk), with particular focus to hysteretic friction (i.e. micro-rolling
friction) and contact area (directly related to the adhesive contribution to
friction). We observe that a modified surface layer (MSL) of thickness order 
$\approx 1\mathrm{\mu m}$ usually occurs as a consequence of rubber wear 
\cite{schipper} in e.g. tire tread-road contacts, or in dynamic rubber
seals. The MSL thickness is clearly expected to introduce a high frequency
(physical) cut-off to the roughness spectral content which can be probed by
the bulk, making the contact mechanics unaffected by the small-wavelength
roughness regime beyond such a threshold. We further observe that without
such a physical cut-off mechanism, the hysteretic friction (normalized
contact area) increases (decreases) theoretically unbouded in an ideal
randomly rough interaction \cite{Persson20013840,scaraggi.in.prep}, thus
(classically) making the quantitative prediction of friction and contact
area to be strongly dependent on the arbitrary (or fitted) choise of such
threshold parameter.

The following results are obtained by applying the numerical model developed
in Sec. \ref{BEM} and \ref{navier.section} to the case of an elastic coating
bonded onto a viscoealstic half space, in steady sliding contact with a
rigid isotropically-rough surface. The bulk is characterized by the complex
reduced viscoelastic modulus $E_{\mathrm{r}}\left( \omega \right) $ of the
tread rubber compound reported in Sec. \ref{navier.section}, whereas the
elastic coating is assumed here to be characterized by the reduced Young's
modulus $E_{\mathrm{r0}}=E_{\mathrm{r}}(\omega =0)$, i.e. given by the
rubber relaxed elastic modulus (in qualitative agreement with the
experimental observations \cite{schipper})\footnote{%
We stress that the graded rheological formulation we have developed can be
applied to any, continuous or stepwise, formulation of the composite, whose
rheological characteristics as a function of the bulk depth can be obtained
e.g. through a sub-surface differential measurements of mechanical
properties. However, within the theoretical purposes of this contribution,
which has been intentionally limited to the fundamental understanding of
graded rheology (as e.g. induced by wear-driven MSL formation) on the rough
contact mechanics, we have numerically simulated a composite formulation
described by the smallest set of interaction parameters (e.g. elastic
coating onto rubber bulk), yet complete enough to capture the basic physics
occurring in the rough interaction between graded solids.}. 
\begin{figure}[tbh]
\centering
\includegraphics[
                        width=0.4\textwidth,
                        angle=0
                ]{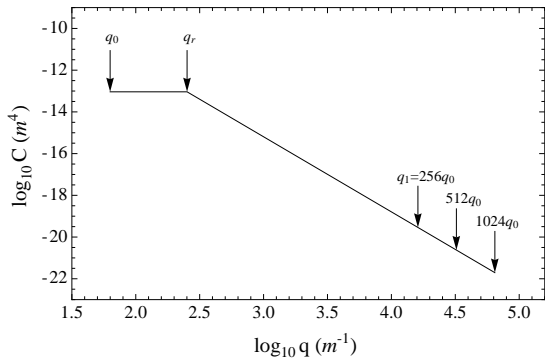}
\caption{Roughness power spectral density adopted in the present study, in $%
\log _{\mathrm{10}}$-$\log _{\mathrm{10}}$ scale. The power spectra have a
low wave vector cut-off for $q_{\mathrm{0}}=0.63\cdot 10^{2}~\mathrm{m}^{-1}$%
, and a roll-off for $q_{\mathrm{r}}=4q_{\mathrm{0}}$. For $q>q_{\mathrm{r}}$
the power spectra correspond to self-affine fractal surfaces with Hurst
exponent $H=0.8$. We consider three cases where the large wave vector
cut-off is $q_{\mathrm{1}}=256q_{\mathrm{0}}$, $512q_{\mathrm{0}}$, and $%
1024q_{\mathrm{0}}$ (corresponding to a root mean square slope $s_{\mathrm{%
rms}}=\left\langle \left\vert \protect\nabla h\right\vert ^{2}\right\rangle
^{1/2}=0.077$, $0.095$, and $0.11$, respectively. The root mean square
roughness, which is mostly determined by large wavelengths content, is
similar for all cases to $h_{\mathrm{rms}}=0.16~\mathrm{mm}$). All
calculations have been performed with $n=8$ divisions at the small
wavelength $\protect\lambda _{1}=2\protect\pi /q_{1}$.}
\label{PSD.eps}
\end{figure}

Fig.\autoref{PSD.eps} shows the generic roughness power spectral density adopted
in the present study. The power spectra have a low wave vector cut-off for $%
q_{\mathrm{0}}=0.63\cdot 10^{2}~\mathrm{m}^{-1}$, and a roll-off for $q_{%
\mathrm{r}}=4q_{\mathrm{0}}$. For $q>q_{\mathrm{r}}$ the power spectra
correspond to self-affine fractal surfaces with Hurst exponent $H=0.8$
(related to the fractal dimnsion $D_{\mathrm{F}}=3-H$)\footnote{%
We stress that whilst roughness self-affine characteristics are often found
in several man- and nature-made surfaces \cite{Persson2006b}, the
self-affine behaviour is here adopted only for convenience (as discussed
before), in order to reduce the number of parameters characterizing the
contact interface. However, there is no particular limitation in the
deterministic or statistically complexity of the rough surfaces to be
simulated.}. We consider three cases where the large wave vector cut-off is $%
q_{\mathrm{1}}=256q_{\mathrm{0}}$, $512q_{\mathrm{0}}$, and $1024q_{\mathrm{0%
}}$ (corresponding to a root mean square slope $s_{\mathrm{rms}%
}=\left\langle \left\vert \nabla h\right\vert ^{2}\right\rangle ^{1/2}=0.077$%
, $0.095$, and $0.11$, respectively). The root mean square roughess, which
is mostly determined by large wavelength content, is similar for all cases
to $h_{\mathrm{rms}}=0.16~\mathrm{mm}$. 
\begin{figure}[tbh]
\centering
\subfigure[]{\includegraphics[
                        width=0.40\textwidth,
                        angle=0
                ]{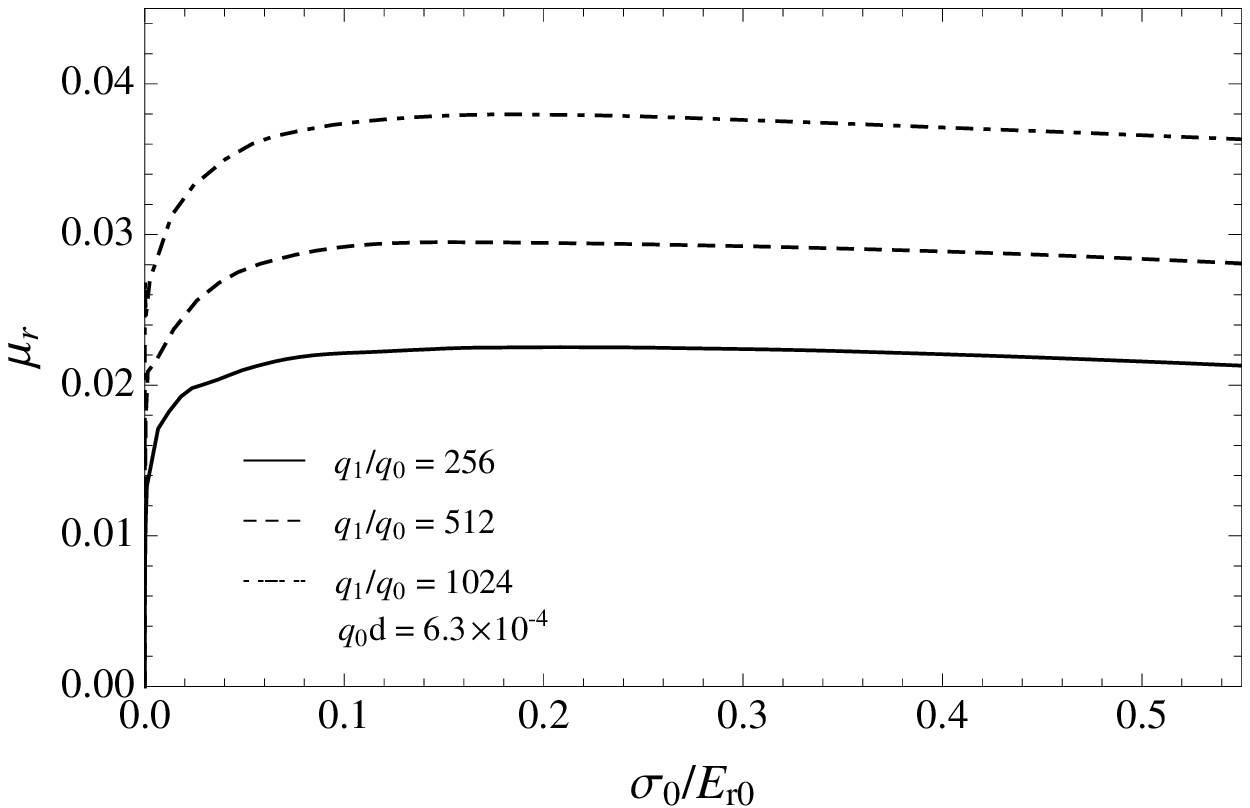} \label{frictionPavg1.eps}
                } \qquad 
\subfigure[]{\includegraphics[
                        width=0.40\textwidth,
                        angle=0
                ]{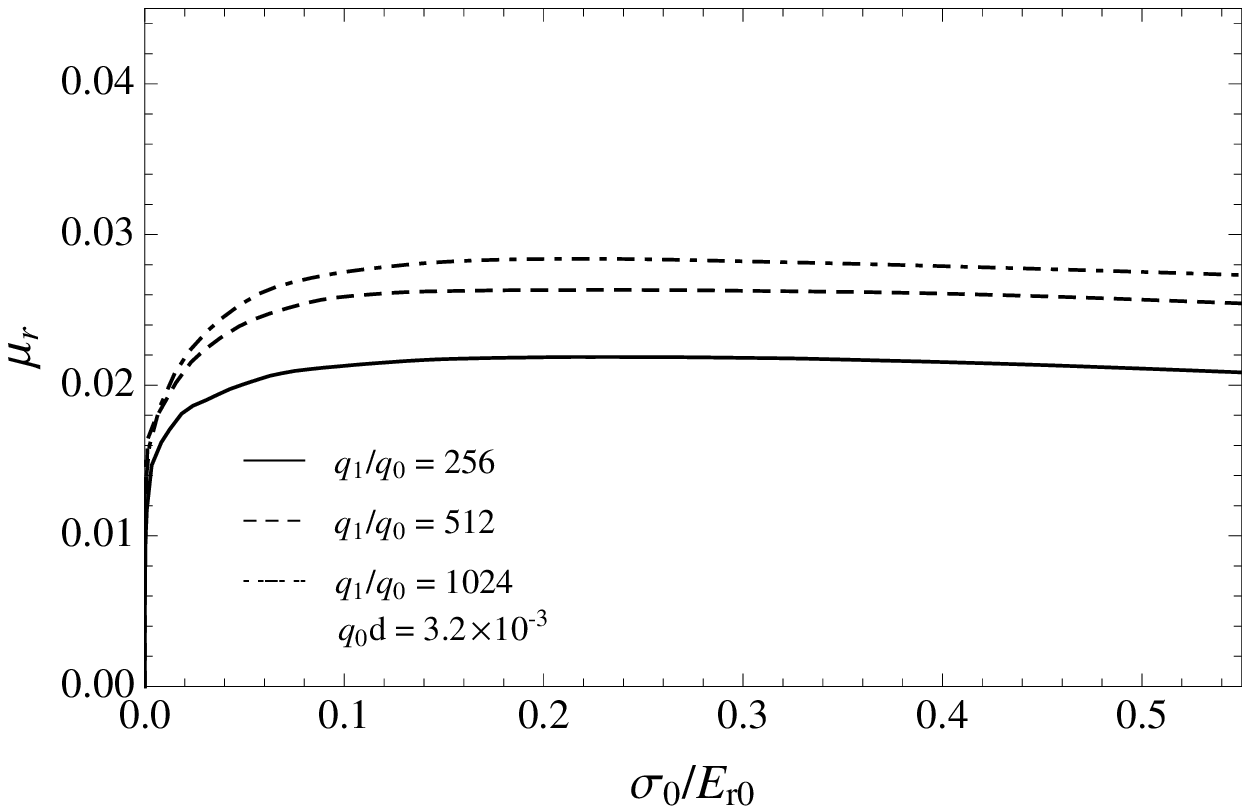} \label{frictionPavg3.eps}
                }
\\
\subfigure[]{\includegraphics[
                        width=0.40\textwidth,
                        angle=0
                ]{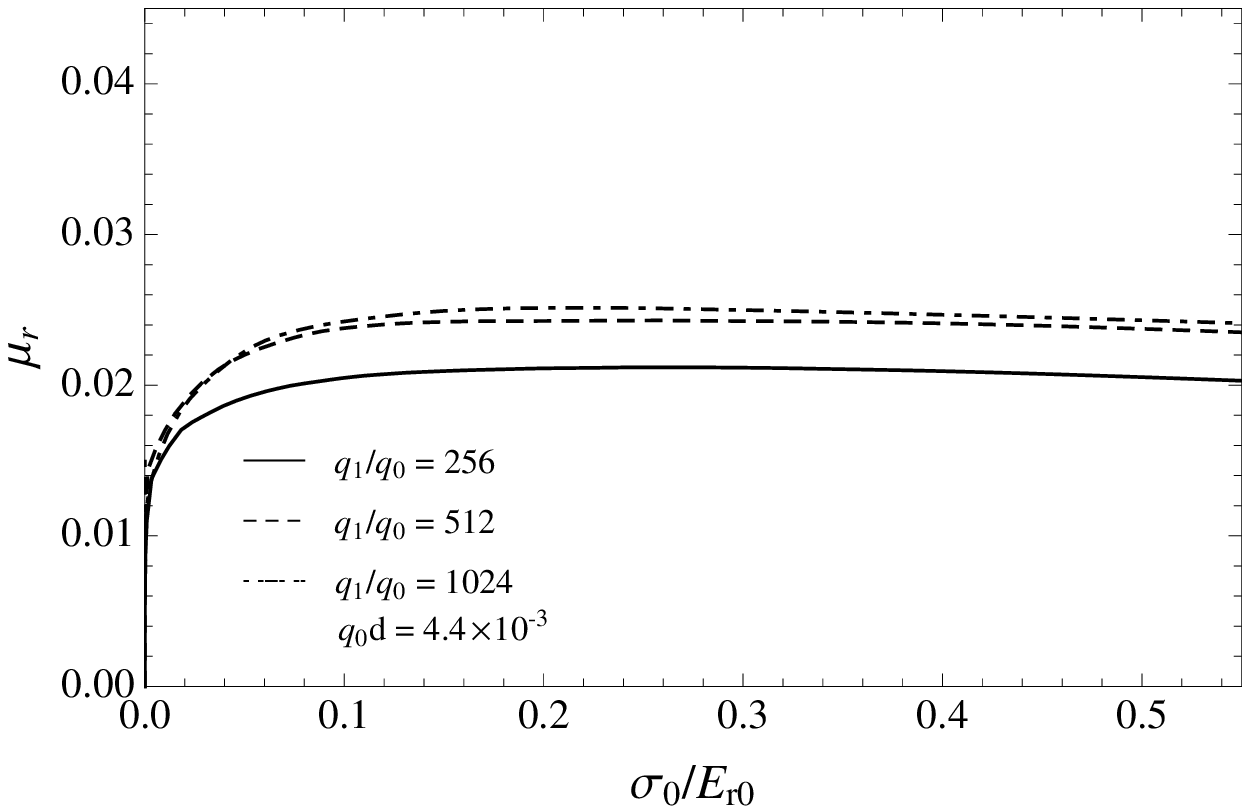} \label{frictionPavg4.eps}
                } \qquad 
\subfigure[]{\includegraphics[
                        width=0.40\textwidth,
                        angle=0
                ]{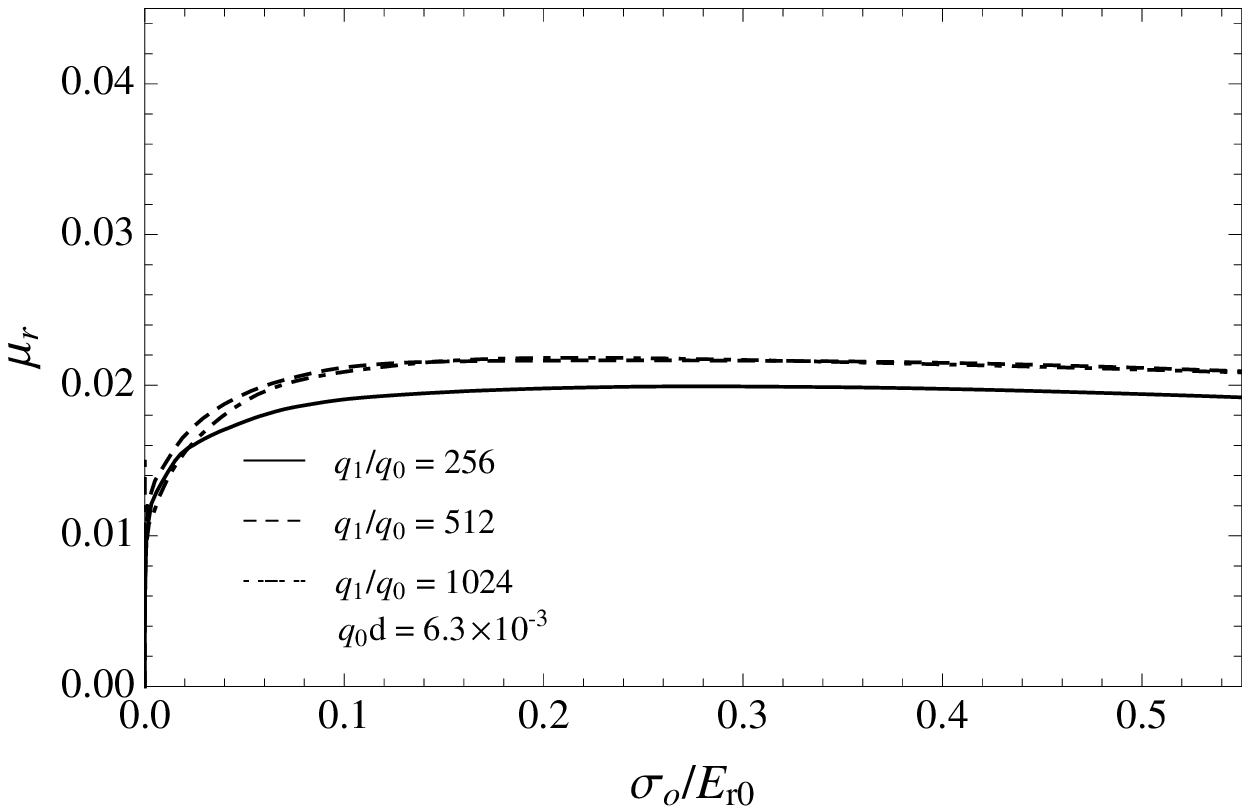} \label{frictionPavg2.eps}
                }
\caption{Micro-rolling friction $\protect\mu _{\mathrm{r}}$ as a function of
the dimensionless contact pressure $\sigma
_{0}/E_{\mathrm{r}0}$, for $q_{1}/q_{0}=2^{8}$, $2^{9}$
and $2^{10}$. The sliding velocity $v$ is set to $0.1$~$\mathrm{m/s}$. From
(a) to (d) $q_{\mathrm{0}}d\ 10^{3}=0.63$, $3.2$, $4.4$, $6.3$.}
\label{micro.rolling.friction}
\end{figure}
\begin{figure}[tbh]
\centering
\subfigure[]{\includegraphics[
                        width=0.40\textwidth,
                        angle=0
                ]{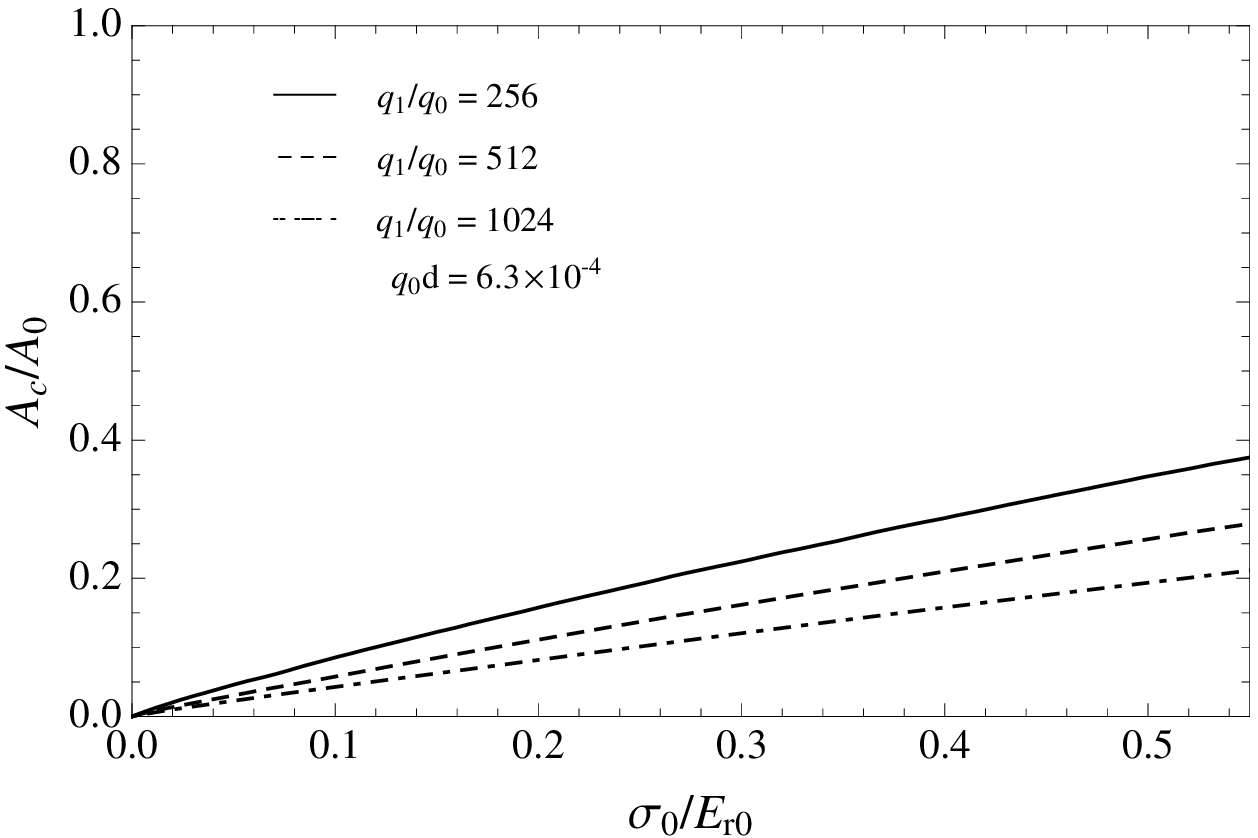} \label{AcPavg1.eps}
                } \qquad 
\subfigure[]{\includegraphics[
                        width=0.40\textwidth,
                        angle=0
                ]{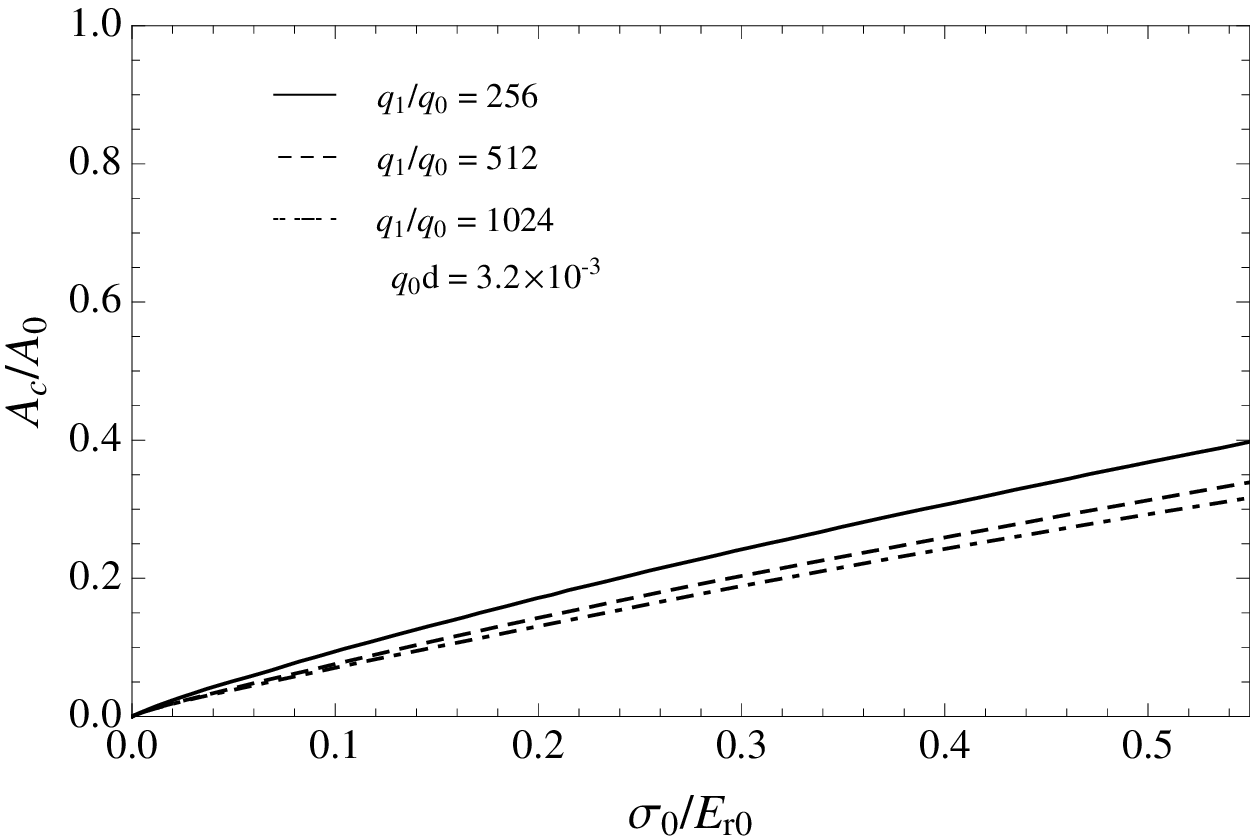} \label{AcPavg3.eps}
                }
\\
\subfigure[]{\includegraphics[
                        width=0.40\textwidth,
                        angle=0
                ]{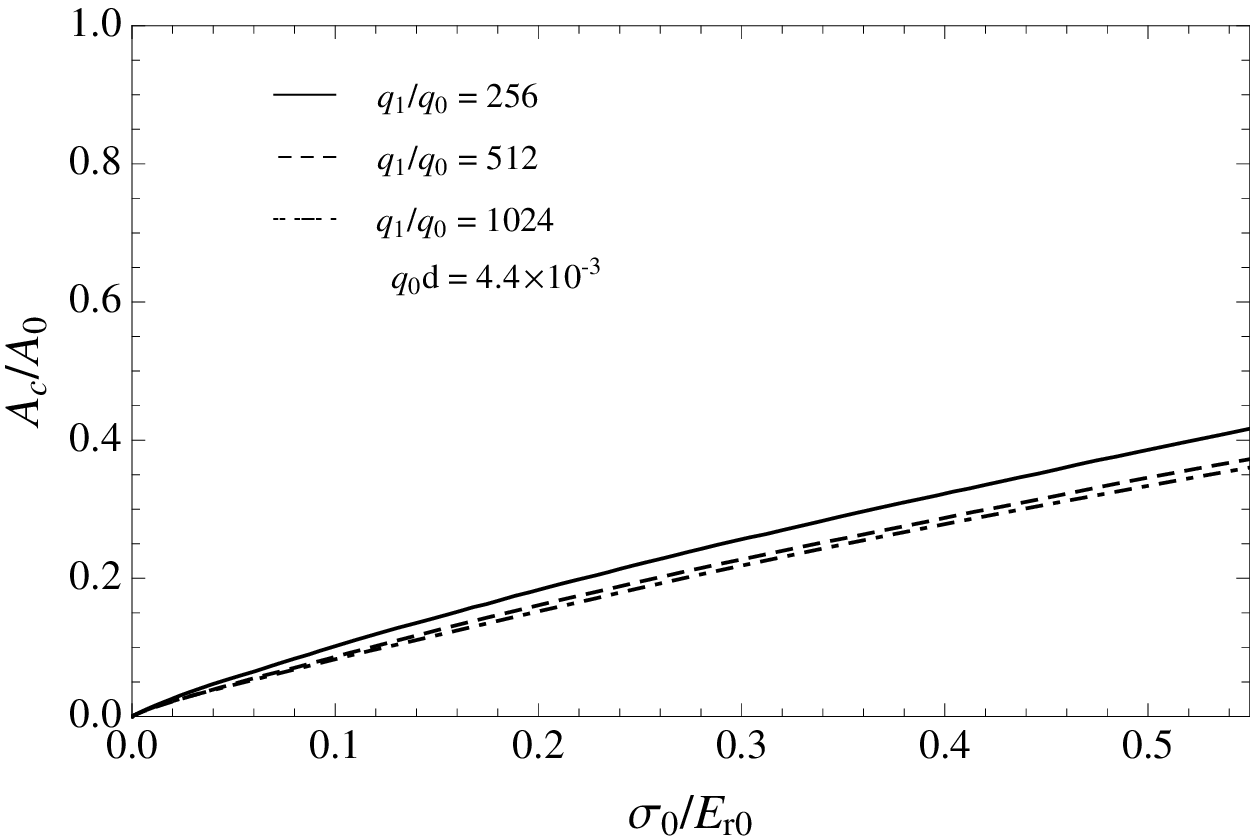} \label{AcPavg4.eps}
                } \qquad 
\subfigure[]{\includegraphics[
                        width=0.40\textwidth,
                        angle=0
                ]{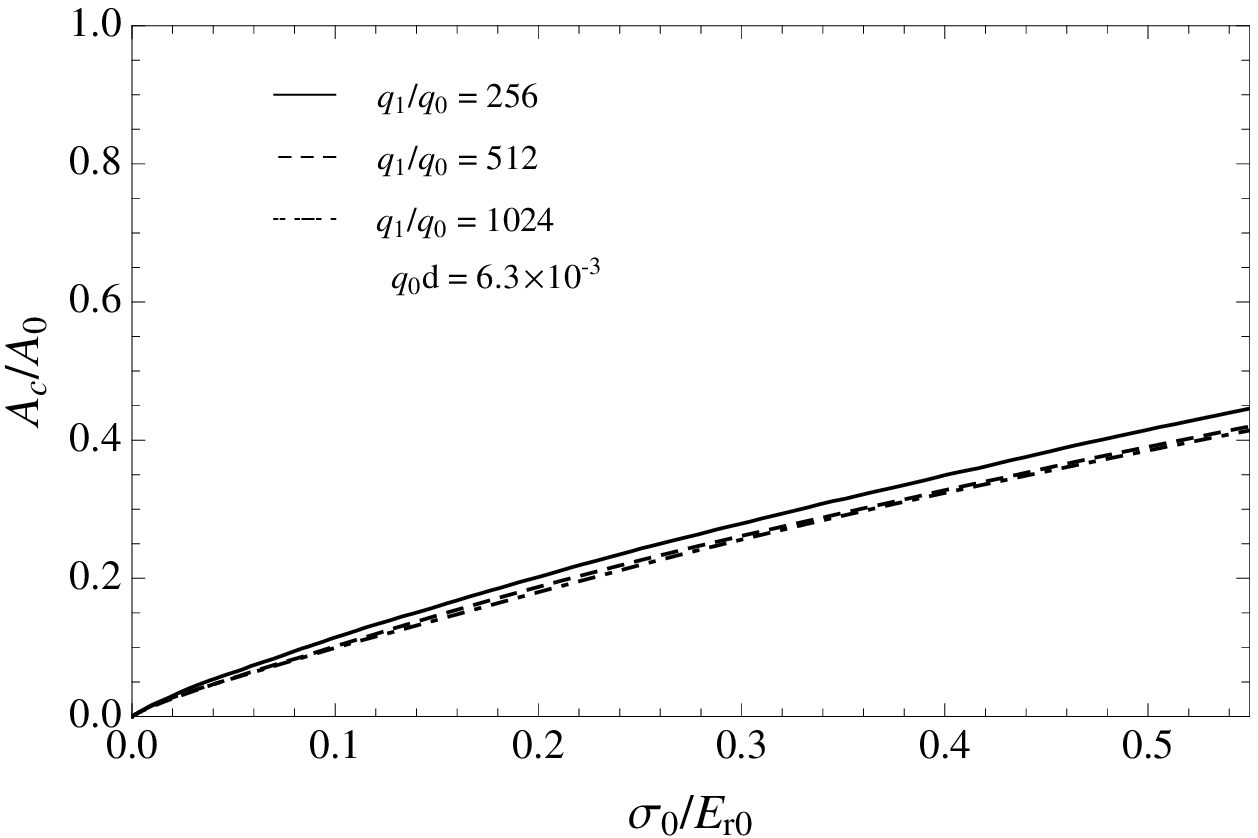} \label{AcPavg2.eps}
                }
\caption{Normalized (projected) contact area $A_{\mathrm{c}}/A_{0}$ as a
function of the dimensionless contact pressure $\sigma
_{0}/E_{\mathrm{r}0}$, for $%
q_{1}/q_{0}=2^{8}$, $2^{9}$ and $2^{10}$. The sliding velocity $v$ is set to 
$0.1$~$\mathrm{m/s}$. From (a) to (d) $ q_{\mathrm{0}}d\
10^{3}=0.63$, $3.2$, $4.4$, $6.3$.}
\label{contact.area}
\end{figure}

In Fig \ref{micro.rolling.friction} and \ref{contact.area} we show,
respectively, the micro-rolling friction $\mu _{\mathrm{r}}$ and the contact
area $A_{\mathrm{c}}/A_{0}$ as a function of the dimensionless contact pressure $\sigma
_{0}/E_{\mathrm{r}0}$, for $q_{1}/q_{0}=2^{8}$, $2^{9}$ and $2^{10}$. The sliding velocity $%
v $ is set to $0.1$~$\mathrm{m/s}$. From (a) to (d) the coating thickness is
increased from $q_{\mathrm{0}}d=0.63\cdot 10^{-4}$ to $6.3\cdot 10^{-3}$
(i.e. $d=10$ to $100~\mathrm{\mu m}$ for our system). In particular, Figs. %
\ref{frictionPavg1.eps} and \ref{AcPavg1.eps} show, respectively, $\mu _{%
\mathrm{r}}$ and $A_{\mathrm{c}}/A_{0}$ for the thinner coating ($d=10~%
\mathrm{\mu m}$). We observe that increasing the roughness high frequency
content determines an increase (decrease) of the hysteretic friction (true
contact area). This could be expected from classical mean field theories
when observing that the coating is thin enough to allow the whole range of
asperities, down to the smallest wavelengths (i.e. to $q_{1}/q_{0}=2^{10}$),
to probe the rubber bulk and, therefore, to effectively contribute
generating the stored (responsible for the contact area) and dissipated
interfacial energy. Moreover, in accordance with classical results \cite%
{Persson20013840,scaraggi.in.prep}, even a small increase in the roughness
spectral content in the high-frequency regime non-negligibly affects both
friction and contact area, as due to the strong dependence of such physical
quantities on the $s_{\mathrm{rms}}$ (which increases from $0.07$ to $0.11$
from $q_{1}/q_{0}=2^{8}$ to $2^{10}$). Figs. \ref{frictionPavg2.eps} and \ref%
{AcPavg2.eps} show, respectively, $\mu _{\mathrm{r}}$ and $A_{\mathrm{c}%
}/A_{0}$ for the thicker coating ($d=100~\mathrm{\mu m}$). In this case the
contact prediction for $q_{1}/q_{0}=2^{9}$ and $2^{10}$ overlap, i.e. an
asymptotic friction and contact area are obtained for $q_{1}/q_{0}=2^{9}$ 
\textit{in the entire set of investigated contact pressures}. Interestingly,
such asymptotes markedly differ from the corresponding curves of Figs. \ref%
{frictionPavg1.eps} and \ref{AcPavg1.eps} and are now closer to the $%
q_{1}/q_{0}=2^{8}$ curve; instead, the predictions at $q_{1}/q_{0}=2^{8}$
are almost unaffected by the coating thickness. This can be easily justified
with the following arguments. In particular, by increasing the coating size
the smallest roughness wavelengths are no more able to probe the
viscoelastic bulk, hence the hysteretic friction is unaffected by the
smallest asperities resulting into a negligible dissipation increase with
respect to the $q_{1}/q_{0}=2^{8}$ roughness case. Furthermore, given the
soft rheological characteristics of the elastic coating, the smallest
asperities are in full contact with the substrate (at increasing coating
thickness), with no relevant effect in term of contact area reduction, even
at small contact area values (i.e. in the linear regime\footnote{%
Note that in the linear contact regime, the average effective contact
pressure $\bar{\sigma}$ [$\bar{\sigma}\left( q_{1}\right) =\sigma
_{0}A_{0}/A_{\mathrm{c}}\left( q_{1}\right) $, where $A_{\mathrm{c}}\left(
q_{1}\right) $ is the true contact area when the power spectral density
contains roughness down to $q_{1}$ wavenumber], which is responsible for the
local asperity contact condition, is a magnification-only dependent value,
i.e. given $A_{\mathrm{c}}\left( q_{1}\right) \approx A_{0}k\left(
q_{1}\right) m_{2}^{-1/2}\left( q_{1}\right) \sigma _{0}/E_{\mathrm{r}%
}\left( vq_{1}\right) $, one obtains $\bar{\sigma}\left( q_{1}\right)
k\left( q_{1}\right) m_{2}^{-1/2}\left( q_{1}\right) /E_{\mathrm{r}}\left(
vq_{1}\right) \approx 1$. Thus, locally, the asperities will
be in partial or full contact depending on the actual value of $k\left( q_{1}\right)$.}). Indeed the smallest wavelength asperities probe
a locally soft solid, which is even not subjected to the sliding induced
viscoelastic stiffening, as instead was the case of Fig. \ref{AcPavg1.eps}.
Finally, for intermediate coating thicknesses we find, as expected, an
intermediate scenario for both friction and contact area, see Figs. \ref%
{frictionPavg3.eps}, \ref{frictionPavg4.eps}, \ref{AcPavg3.eps} and \ref%
{AcPavg4.eps}. 
\begin{figure}[tbh]
\centering
\includegraphics[
                        width=0.4\textwidth,
                        angle=0
                ]{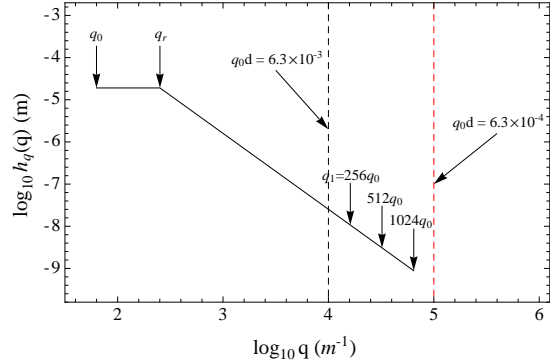}
\caption{Wavelength representative amplitude $h_{q}\left( q\right) \approx
q_{0}\protect\sqrt{C\left( q\right) }$ as a function of the wavenumber $q$,
in $\log _{\mathrm{10}}$-$\log _{\mathrm{10}}$ scale, corresponding to the
PSD\ of Fig. \protect\ref{PSD.eps}. The vertical dashed lines indicate the
roughness frequencies $q$ corresponding to $\bar{q}=qd=1$ for the thicker
(black line, with $q_{0}d=6.3\cdot 10^{-3}$) and thinner (red, with $%
q_{0}d=6.3\cdot 10^{-4}$) coatings adopted in the simulations, where $d$ is
the coating thickness.}
\label{amplitude.wave}
\end{figure}

\begin{figure}[tbh]
\centering
\subfigure[Thinner coating]{\includegraphics[
                        width=0.40\textwidth,
                        angle=0
                ]{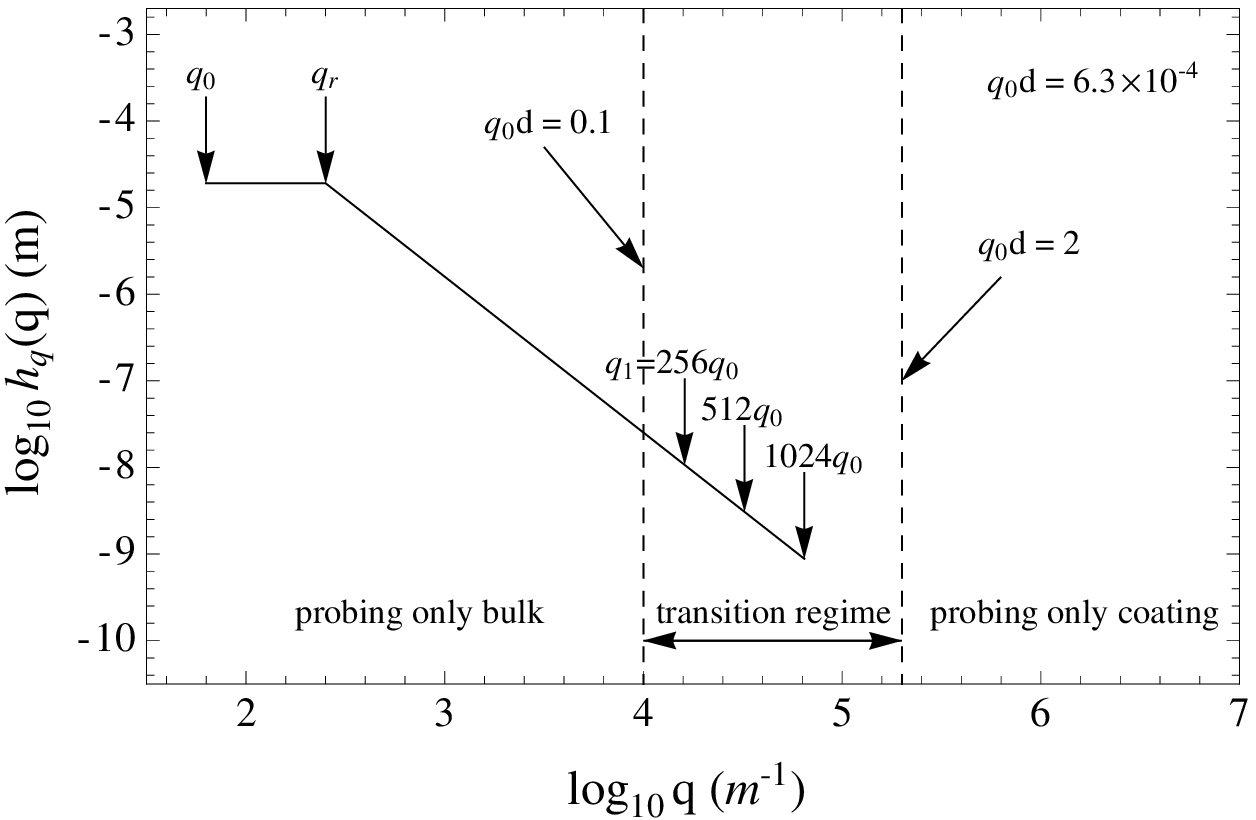} \label{amplitude.wave.t}
                } \qquad 
\subfigure[Thicker coating]{\includegraphics[
                        width=0.40\textwidth,
                        angle=0
                ]{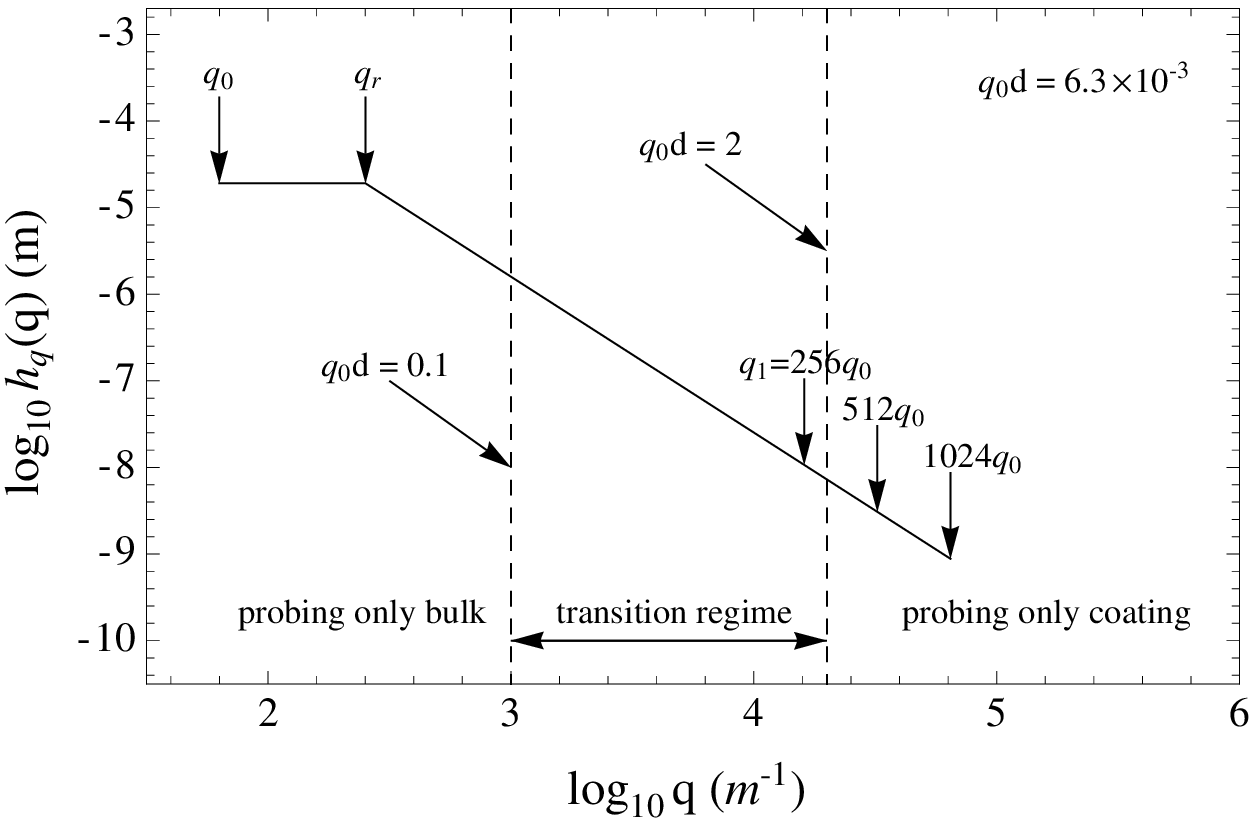} \label{amplitude.wave.T}
                }
\caption{Wavelenth representative amplitude $h_{q}\left( q\right) \approx
q_{0}\protect\sqrt{C\left( q\right) }$ as a function of the vavenumber $q$,
in $\log _{\mathrm{10}}$-$\log _{\mathrm{10}}$ scale, corresponding to the
PSD\ of Fig. \protect\ref{PSD.eps}. The vertical dashed lines indicate the
roughness frequencies $q$ corresponding to $\bar{q}=0.1$ and $2$. (a) for
the thinner ($q_{0}d=6.3\cdot 10^{-4})$, and (b) thicker ($q_{0}d=6.3\cdot
10^{-3})$ coating.}
\label{psd.test}
\end{figure}

In Fig. \ref{amplitude.wave} we show the wavelength representative amplitude $%
h_{q}\left( q\right) \approx q_{0}\sqrt{C\left( q\right) }$ as a function of
the wavenumber $q$, in $\log _{\mathrm{10}}$-$\log _{\mathrm{10}}$ scale,
corresponding to the PSD\ of Fig. \ref{PSD.eps}. The vertical dashed lines
indicate the roughness frequencies $q$ corresponding to $\bar{q}=qd=1$ for
the thicker (black line, with $q_{0}d=6.3\cdot 10^{-3}$) and thinner (red,
with $q_{0}d=6.3\cdot 10^{-4}$) coatings adopted in the simulations. From the
theory developed in Appendix \ref{appendix.1}, the layer thickness enters
the theory mainly through $\tilde{q}=\tanh \left( \bar{q}\right) $, where
again $\bar{q}=qd$. Thus we can qualitatively observe that for frequencies $%
\bar{q}\geq \bar{q}_{\mathrm{high}}=2$, $\tilde{q}\approx 1$ so that the
roughness wavelengths approximately smaller than the coating thickness are
unable to probe the sub-coating composite rheology, whereas for $\bar{q}\leq 
\bar{q}_{\mathrm{low}}=0.1$, $\tilde{q}\approx \bar{q}$ so that the
roughness asperities do mainly probe the bulk. However, whilst we expect
the exact values of $%
\bar{q}_{\mathrm{high}}$ (and $\bar{q}_{\mathrm{low}}$) to be quantitatively
affected not only by the geometrical composite characteristics, but also
from its rheological properties (see e.g. the discussion in Sec. \ref%
{discussion}), it is worth in the present context (where both coating and
bulk show a Young's modulus of similar order of magnitude) to show $\bar{q}_{%
\mathrm{high}}$ (and $\bar{q}_{\mathrm{low}}$) in Fig. \ref{amplitude.wave}
for both the thinner [Fig. \ref{amplitude.wave.t}] and thicker [Fig. \ref%
{amplitude.wave.T}] coating. 

\begin{figure}[tbh]
\centering
\subfigure[$\sigma _{0}/E_{\mathrm{r}0}=0.3$]{\includegraphics[
                        width=0.40\textwidth,
                        angle=0
                ]{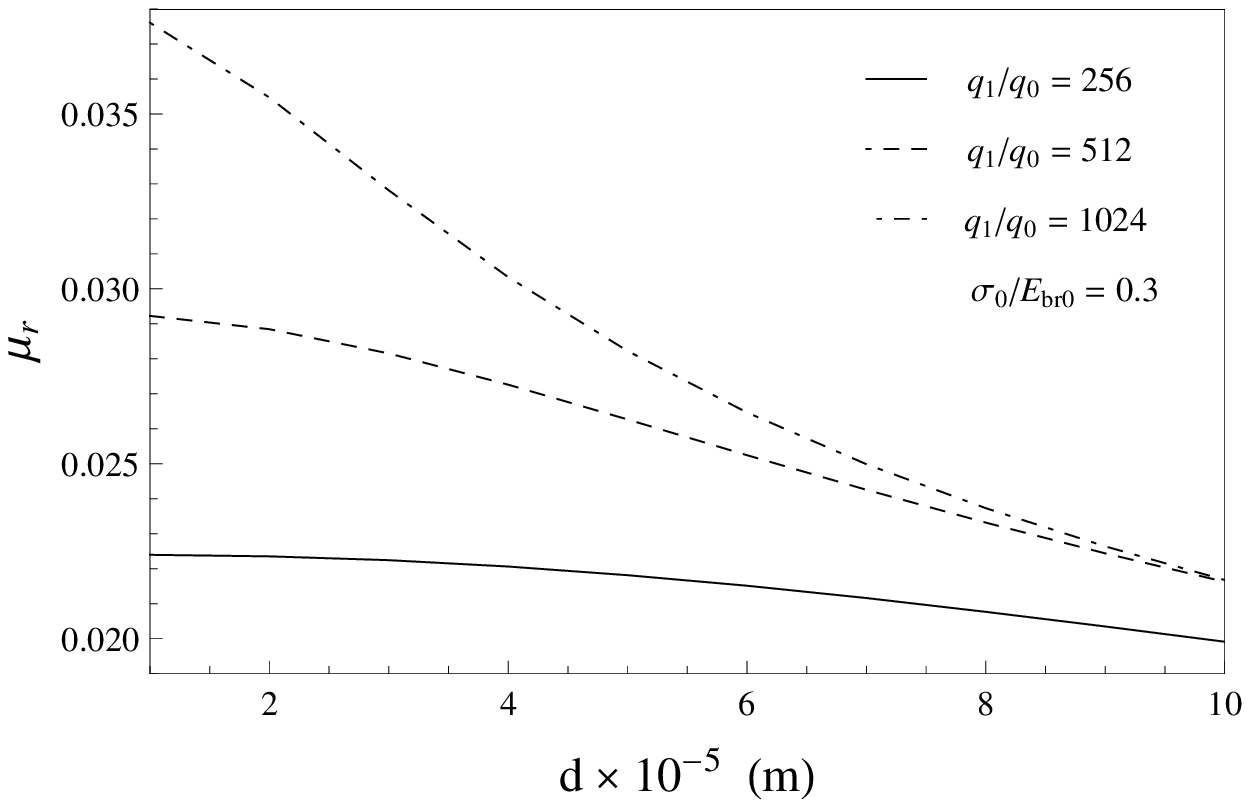} \label{friction_varthick.eps}
                } \qquad 
\subfigure[$\sigma _{0}/E_{\mathrm{r}0}=0.3$]{\includegraphics[
                        width=0.40\textwidth,
                        angle=0
                ]{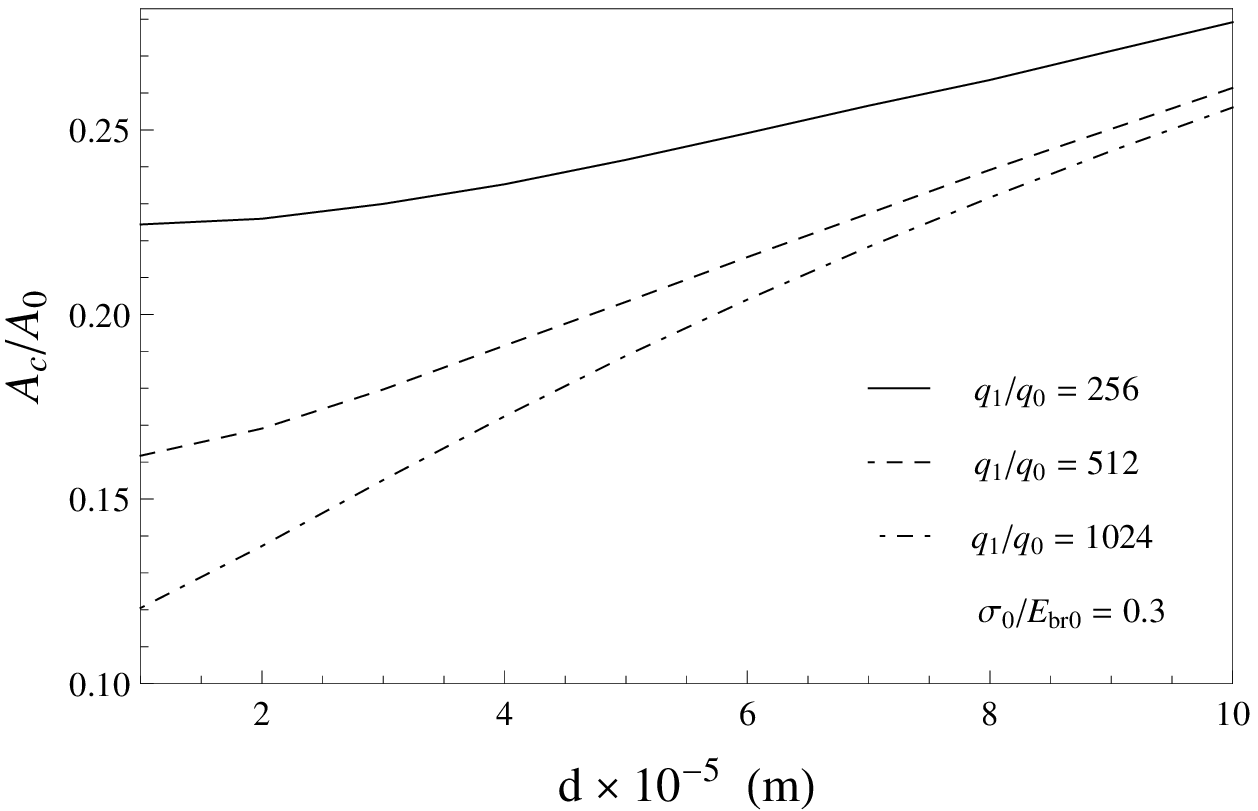} \label{area_varthick.eps}
                }
\\
\subfigure[$\sigma _{0}/E_{\mathrm{r}0}=0.1$]{\includegraphics[
                        width=0.40\textwidth,
                        angle=0
                ]{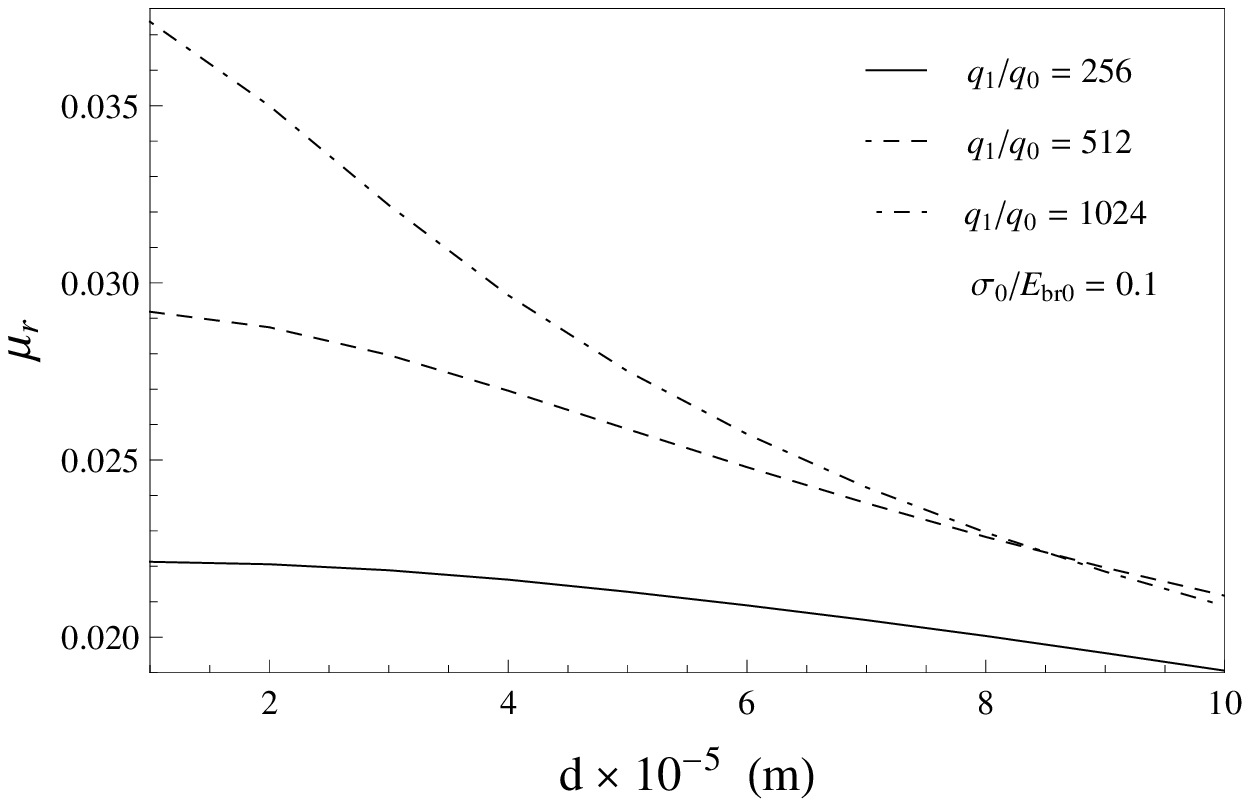} \label{friction_varthick_2.eps}
                } \qquad 
\subfigure[$\sigma _{0}/E_{\mathrm{r}0}=0.1$]{\includegraphics[
                        width=0.40\textwidth,
                        angle=0
                ]{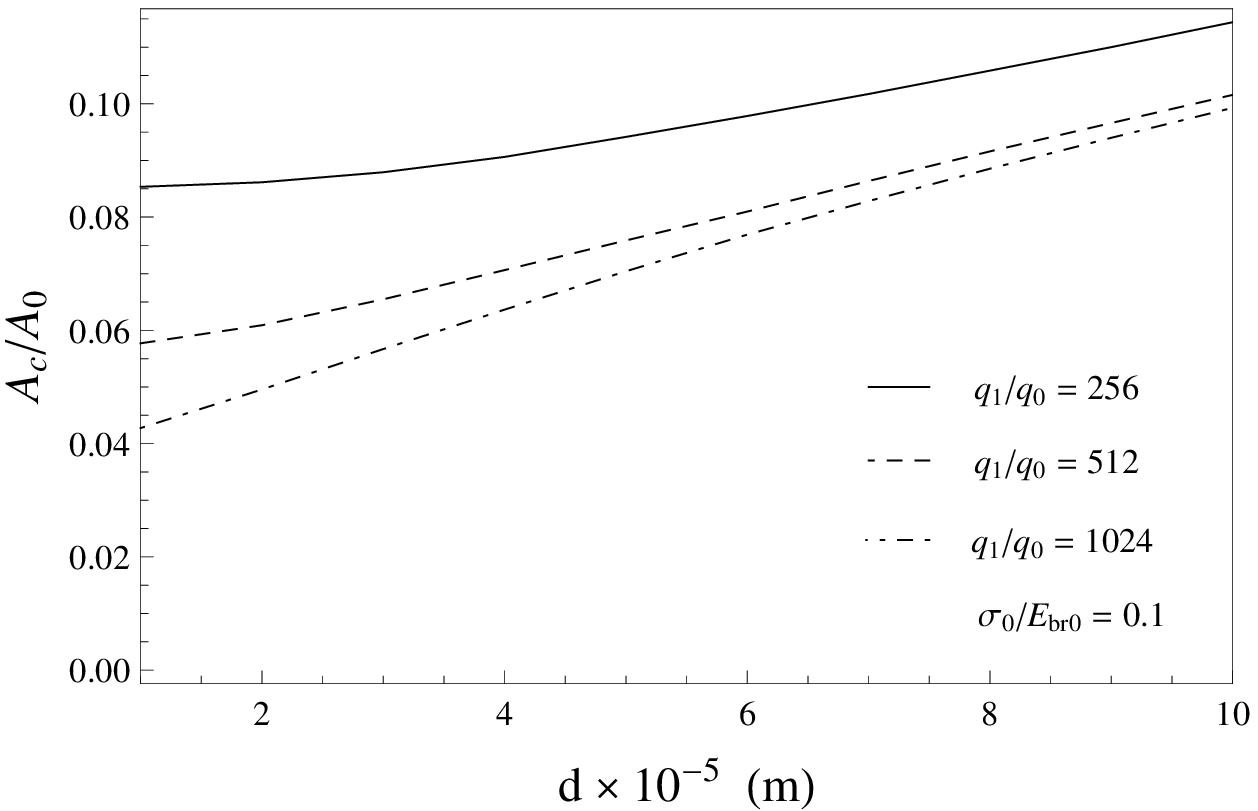} \label{area_varthick_2.eps}
                }
\caption{(a) Micro-rolling friction and (b) normalized projected contact
area as a function of the coating thickness, for $q_{1}/q_{0}=2^{8}$, $2^{9}$
and $2^{10}$. The sliding velocity $v$ is set to $0.1$~$%
\mathrm{m/s}$.}
\label{asymptote}
\end{figure}
We observe in Fig. \ref{amplitude.wave.t} (thinner coating) that,
accordingly to the previous arguments, the whole range of roughness
frequencies can probe the bulk, whereas for the thicker coating [Fig. \ref%
{amplitude.wave.T}] the smallest wavelengths are not aware of the bulk,
confirming the friction and contact area results reported in Figs. \ref%
{micro.rolling.friction} and \ref{contact.area}.

Finally, in Fig. \ref{asymptote} we show (a,c) the micro-rolling friction $\mu
_{\mathrm{r}}$ and (b,d) the contact area $A_{\mathrm{c}}/A_{0}$ as a function
of the coating thickness, for $q_{1}/q_{0}=2^{8}$, $2^{9}$ and $2^{10}$
(sliding velocity $v$ set to $0.1$~$\mathrm{m/s}$), and for
a contact pressure $\sigma _{0}/E_{\mathrm{r}0}=0.1$ and $0.3$ (qualitatively similar
behaviours characterize the interaction at different contact pressures, thus
not shown here for the sake of briefness). It is interesting to observe that
the friction (and contact area) curve for $q_{1}/q_{0}=2^{10}$ converges to
the $q_{1}/q_{0}=2^{9}$ curve at increasing values of coating thickness,
again as expected from the previous arguments. A similar conclusion applies
for the contact area, see Figs. \ref{area_varthick.eps} and \ref{area_varthick_2.eps}\footnote{The contact area
increases at larger coating thickness $d$ since, for the adopted composite,
by increasing $d$ a wider range of PSD wavelengths is allowed to probe only the coating,
which is not subjected to a sliding-induced viscoelastic stiffening. Hence, an increasing
amount of roughness wavelengths is in full-contact with the substrate.}. Thus, roughness
frequencies larger than $q_{1}/q_{0}=2^{9}$ do not affect the interfacial
contact mechanics at such coating size, supporting the general statement
that a physically meaningful characterization (and prediction) of the
friction and contact area properties of a generic interaction can only be
obtained provided that both confinement rheology and surface physics, as
well as surface roughness, are fully characterized to a same degree of
completeness.

\section{Discussion}

\label{discussion}The multiscale nature of the hysteretic friction $\mu _{%
\mathrm{r}}$ and contact area $A_{\mathrm{c}}/A_{0}$ for randomly rough
interactions is nowdays well accepted among contact mechanics researchers,
mainly thanks to the theoretical achievements of the Persson \cite%
{Persson20013840}. At a contact scale of representative size $\lambda =2\pi
/q$, where $\zeta =q/q_{0}$ is the magnification at which the contact is
observed with respect to a contact macroscale $L_{0}=2\pi /q_{0}$, the
dissipation is confined in a bulk volume $\lambda ^{3}$ and corresponds
approximately to a friction force $F_{\mathrm{T}}\approx \lambda ^{3}q~%
\mathrm{Im}[E(\omega )]h_{\mathrm{\lambda }}^{2}/\lambda ^{2}$. Here $\omega
=\mathbf{q}\cdot \mathbf{v}$ is the angular frequency of excitation with $%
\mathbf{v}$ as the sliding velocity, $E(\omega )$ is the complex Young's
modulus of rubber, and $h_{\mathrm{\lambda }}$ is the amplitude of the
roughness wavelength $\lambda $. Thus, the frictional shear stress is $\tau
_{\mathrm{T}}\approx q^{2}~\mathrm{Im}[E(\omega )]h_{\mathrm{\lambda }}^{2}$%
. Hence, for a rough surface with self affine characteristics, one has that
the contribution to friction $\Delta \tau _{\mathrm{T}}$ related to the
roughness wavelength $\lambda $ is%
\begin{equation}
\Delta \tau _{\mathrm{T}}/\Delta q\approx \mathrm{Im}[E(\omega
)]q^{3}C(q)\propto \mathrm{Im}[E(\mathbf{q}\cdot \mathbf{v})]q^{-2H+1},
\label{discussion1}
\end{equation}%
where%
\begin{equation*}
C(q)=\left( 2\pi \right) ^{-2}\int d^{2}\mathbf{x\ }\left\langle h(\mathbf{x}%
)h(\mathbf{0})\right\rangle e^{-i\mathbf{q}\cdot \mathbf{x}}
\end{equation*}%
is the power spectral density of the surface roughness, $h(\mathbf{x})$ is
the substrate height measured from the average surface plane, and $H$ is the
Hurst coefficient. \autoref{discussion1} does not show any cut-off mechanism of
friction, and moreover for fractal dimensions $D_{\mathrm{f}}>2.5$ the
contribution to dissipation generated by decreasing roughness length scales
is even unbounded for ideal (infinite) systems. Similar considerations apply
for the contact area. In particular, by observing that $A\left( \zeta
\right) \bar{\sigma}\left( \zeta \right) =\sigma _{0}A_{0}=F_{\mathrm{N}}$,
where $A\left( \zeta \right) $ is the contact area obtained when an
arbitrary high-frequency cut-off is applied to the PSD (i.e. $C\left(
q>\zeta q_{0}\right) =0$), one simply has%
\begin{equation*}
\frac{dA\left( \zeta \right) }{d\zeta }\propto -\frac{d\bar{\sigma}\left(
\zeta \right) }{d\zeta }.
\end{equation*}%
By approximating $d\bar{\sigma}\left( \zeta \right) /d\zeta $ with $\sqrt{%
d\left\langle \sigma ^{2}\right\rangle /d\zeta }$, where%
\begin{equation*}
d\left\langle \sigma ^{2}\right\rangle =\frac{E_{\mathrm{r0}}^{2}}{4}d\left[
m_{2,\mathrm{eff}}\left( \zeta \right) \right] 
\end{equation*}%
with%
\begin{eqnarray*}
m_{2,\mathrm{eff}}\left( \zeta \right)  &=&\int_{q_{0}}^{q_{0}\zeta
}dq^{2}q^{2}C\left( q\right) \left\vert E_{\mathrm{r,\theta }}\left( \mathbf{%
q}\cdot \mathbf{v}\right) \right\vert ^{2}/E_{\mathrm{r0}}^{2} \\
d\left[ m_{2,\mathrm{eff}}\left( \zeta \right) \right]  &\approx &d\zeta
q_{0}q^{3}C\left( q\right) \left\vert E_{\mathrm{r,\theta }}\left( \mathbf{q}%
\cdot \mathbf{v}\right) \right\vert ^{2}/E_{\mathrm{r0}}^{2}
\end{eqnarray*}%
and where $\left\vert E_{\mathrm{r,\theta }}\left( \mathbf{q}\cdot \mathbf{v}%
\right) \right\vert =\left( 2\pi \right) ^{-1}\int d\theta ~\left\vert
E_{\mathrm{r,\theta }}\left( qv\cos \theta \right) \right\vert $ (assuming $%
v_{y}=0$). Thus, approximately,%
\begin{equation}
\Delta A\left( \zeta \right) /\Delta q\propto -\left\vert E_{\mathrm{%
r,\theta }}\left( \mathbf{q}\cdot \mathbf{v}_{0}\right) \right\vert
q^{\left( -2H+1\right) /2}.  \label{discussion2}
\end{equation}%
We first observe that $\Delta A\left( \zeta \right) /\Delta q<0$, i.e. the
contact continuously decreases by increasing the small scale roughness
spectral content. The power law exponent is similar to the previous case
(thus, similar considerations apply here), and no cut-off mechanism of
contact area appears. Thus, accordingly to both \autoref{discussion1} and \ref%
{discussion2}, a small-scale cut-off mechanism can only be introduced
through the effective rheological response of the composite $E\left( \omega
\right) $. Whilst this has been already numerically prooved in Sec. \ref%
{results} for the case of an elastic coating bonded onto a rubber bulk,
however, we will show (a more interesting feature) in the following that the
multiscale design of the effective complex modulus $E\left( \omega \right) $
(i.e. the choise of the composite materials arrangement) can be adopted to
provide extremely tailored contact mechanics properties, such as (but not limited to) an enhanced
micro-rolling friction.
\begin{figure}[tbh]
\centering
\subfigure[]{\includegraphics[
                        width=0.5\textwidth,
                        angle=0
                ]{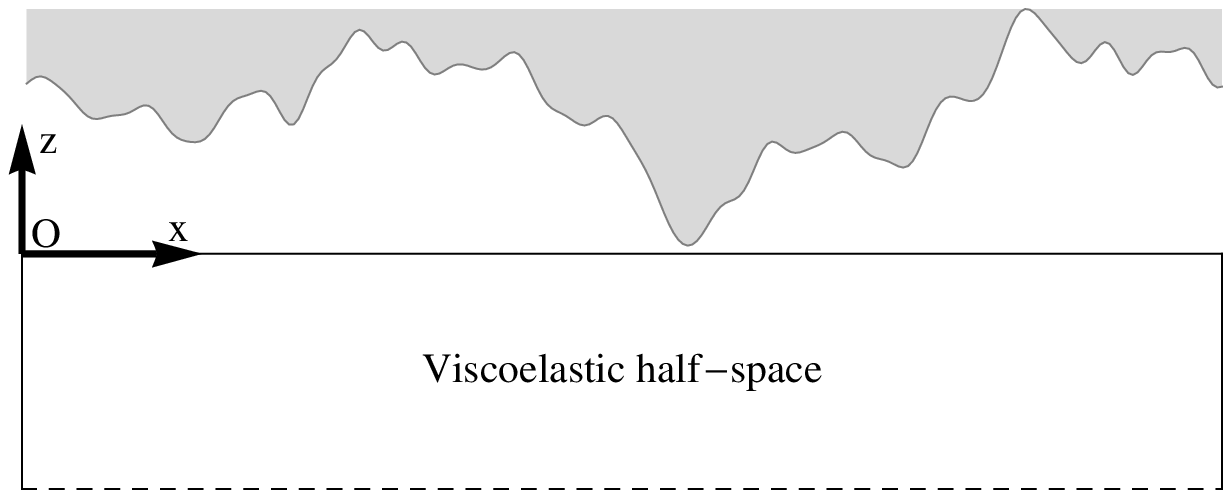} \label{Half_Space.eps}
               }
\\
\subfigure[]{\includegraphics[
                        width=0.4\textwidth,
                        angle=0
                ]{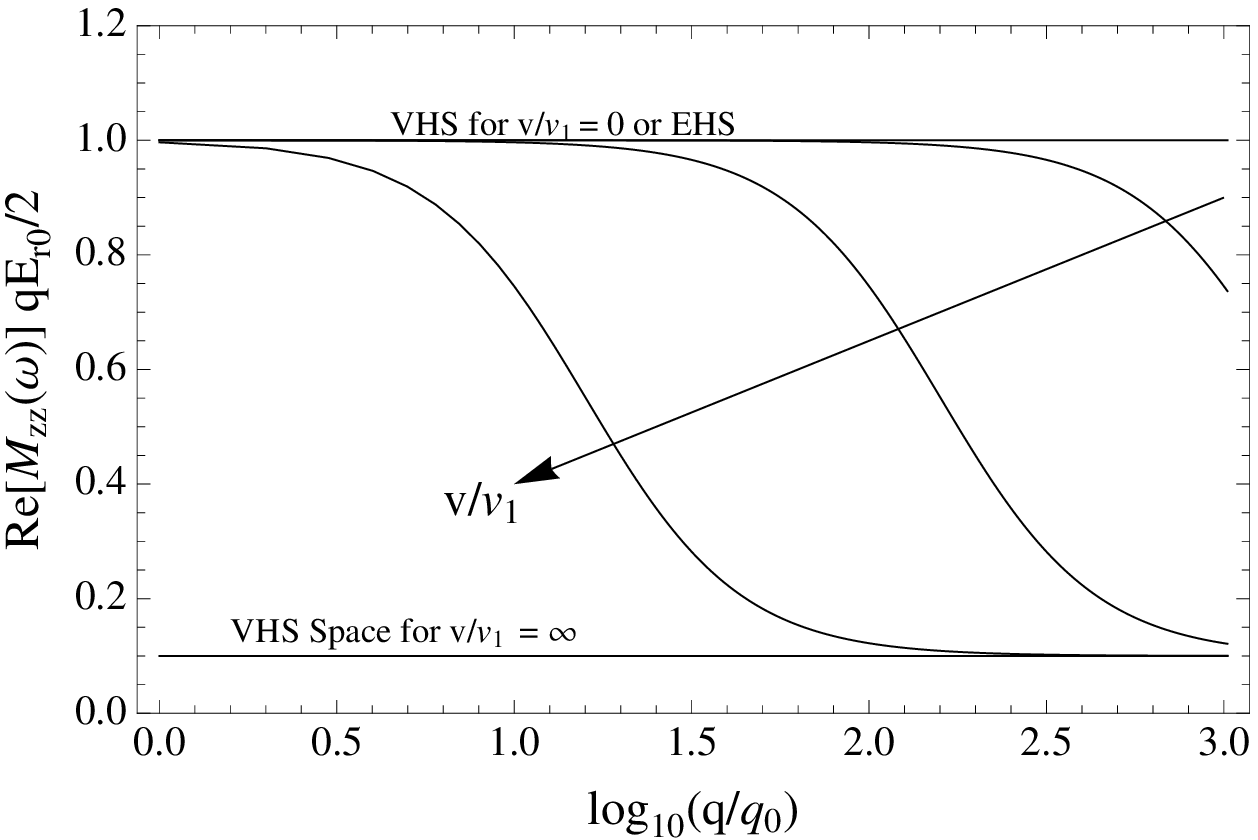}\label{grafico1.eps}
                } \qquad 
\subfigure[]{\includegraphics[
                        width=0.4\textwidth,
                        angle=0
                ]{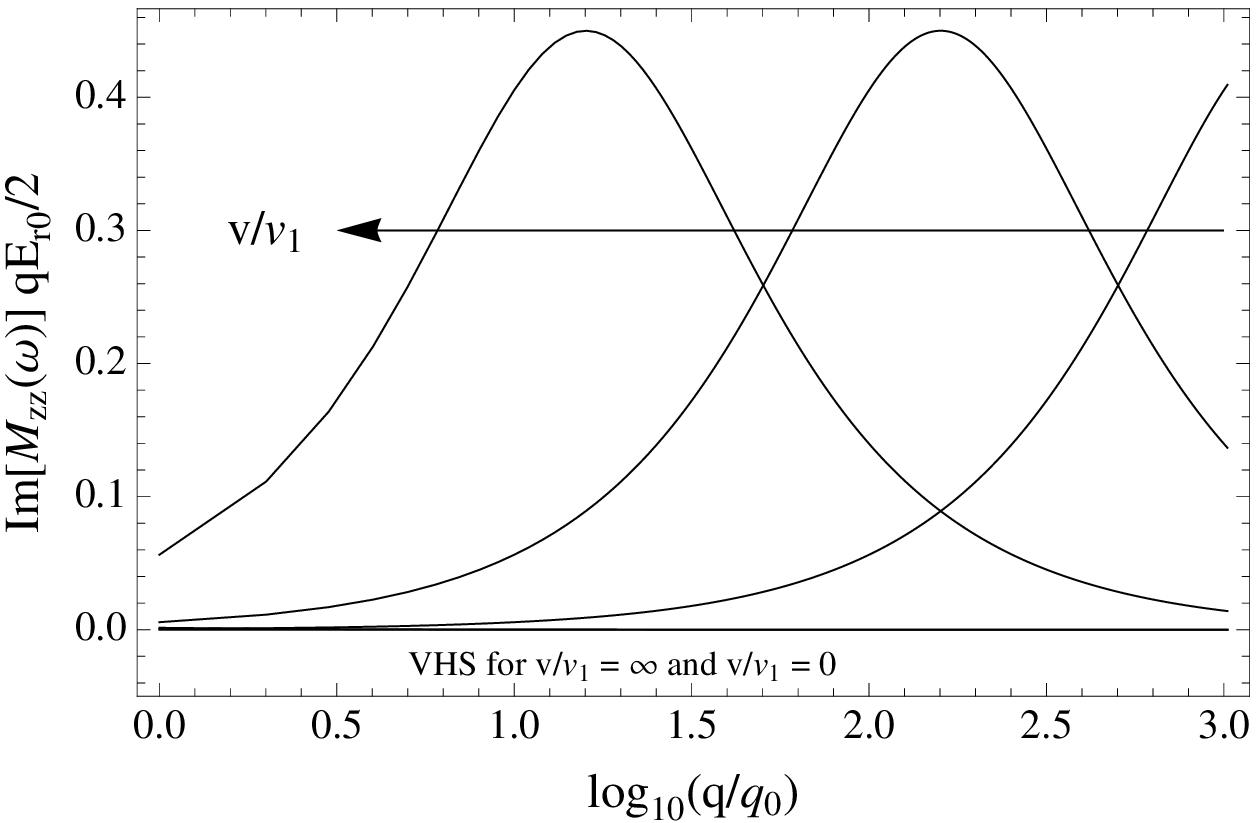}\label{grafico1_1.eps}
		}
\caption{(a) Schematic of a viscoelastic half space (VHS) in sliding contact with
a rigid rough surface. (b) Real and (c) immaginary part of the dimensionless
surface response $M_{zz}\left( \protect\omega \right) qE_{\mathrm{r0}}/2$
(with $\protect\omega =qv$ and $q_{y}=0$) as a function of the wave number $%
q/q_{0}$ ($q_{0}=2\protect\pi /L_{0}$). The bulk is characterized by a
single relaxation time $\protect\tau =L_{0}/v_{1}$ and $E_{\mathrm{r}\infty
}/E_{\mathrm{r}0}=10$ (with $\protect\nu \left( \protect\omega \right) =%
\protect\nu _{0}$). For the dimensionless sliding velocities $v/v_{1}$ in
the set $\left[ 0,10^{-4},10^{-3},10^{-2},\infty \right] $. EHS is for elastic half space.}
\label{bulk}
\end{figure}

\begin{figure}[tbh]
\centering
\subfigure[]{\includegraphics[
                        width=0.5\textwidth,
                        angle=0
                ]{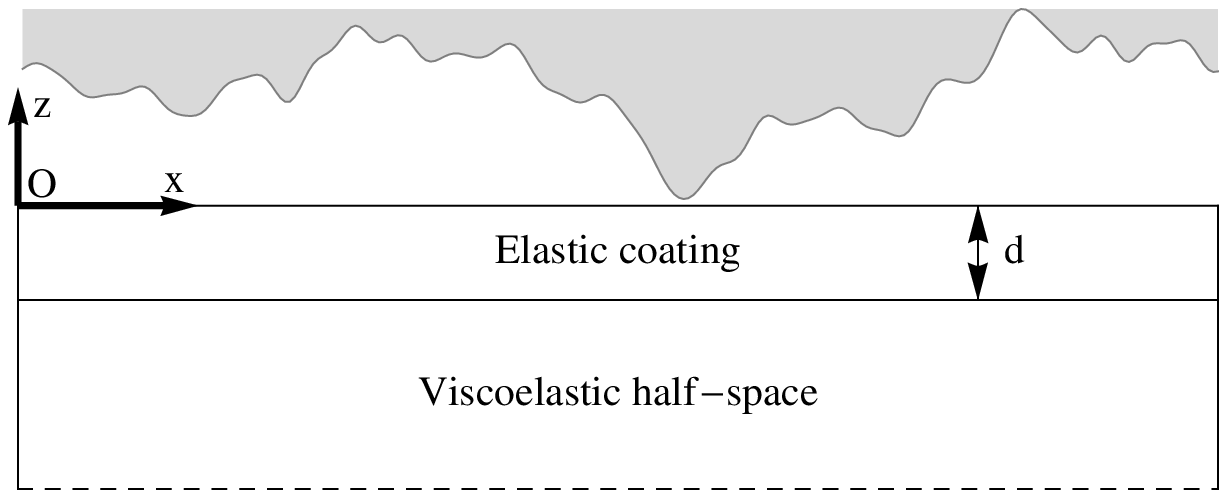} \label{coating_halfpsace.eps}
               }
\\
\subfigure[]{\includegraphics[
                        width=0.4\textwidth,
                        angle=0
                ]{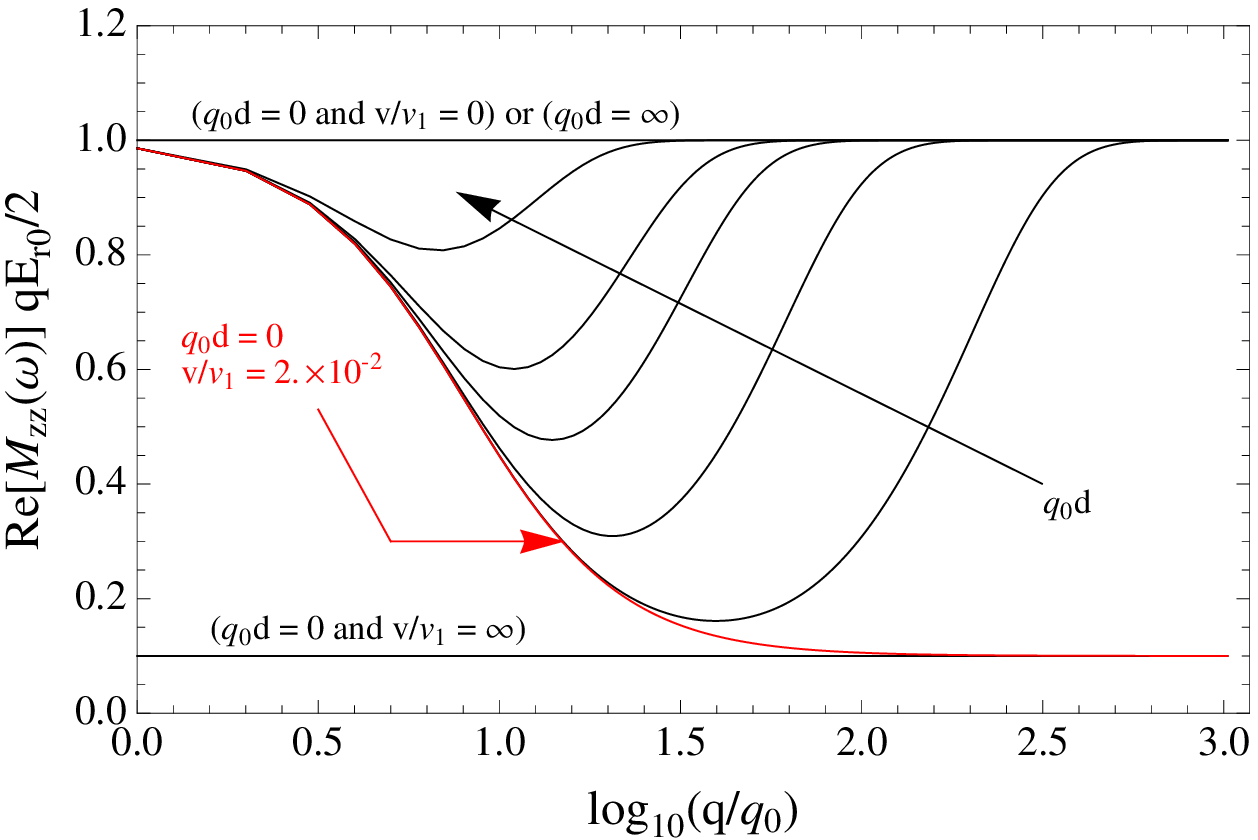} \label{grafico2.eps}
               } \qquad 
\subfigure[]{\includegraphics[
                        width=0.4\textwidth,
                        angle=0
                ]{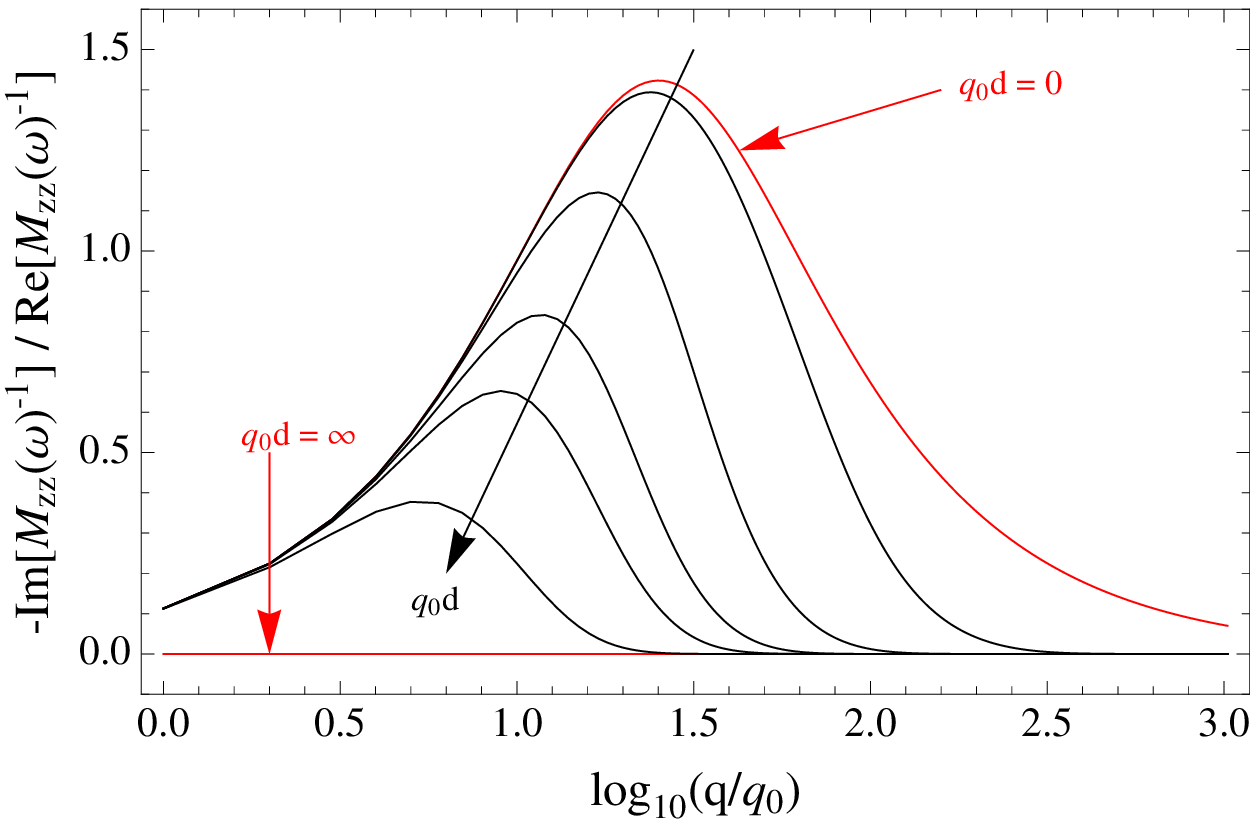}\label{grafico3.eps}
                }
\caption{(a) Schematic of a viscoelastic half space coated with an elastic
layer in sliding contact with a rigid rough surface. (b) Real part of the
dimensionless surface response $M_{zz}\left( \protect\omega \right) qE_{%
\mathrm{r0}}/2$ and (c) effective composite loss tangent $\mathrm{Im}\left[
M_{zz}\left( \protect\omega \right) ^{-1}\right] /\mathrm{Re}\left[
M_{zz}\left( \protect\omega \right) ^{-1}\right] $ (with $\protect\omega =qv$
and $q_{y}=0$) as a function of the wave number $q/q_{0}$ ($q_{0}=2\protect%
\pi /L_{0}$). The bulk is characterized by a single relaxation time $\protect%
\tau =L_{0}/v_{1}$ and $E_{\mathrm{r}\infty }/E_{\mathrm{r}0}=10$ (with
constant Poisson ratio). The coating has a reduced elastic modulus $E_{%
\mathrm{r}0}$. For the dimensionless coating thickness $q_{0}d$ in the set $%
\left[ 0,9.4,31,63,94,190,\infty \right] 10^{-3}$, with $v/v_{1}=0.02$
(unless differently specified).}
\label{coating}
\end{figure}
\begin{figure}[tbh]
\centering
\subfigure[]{\includegraphics[
                        width=0.45\textwidth,
                        angle=0
                ]{multilayer1.eps} \label{multilayer1.eps}
               } \qquad 
\subfigure[]{\includegraphics[
                        width=0.4\textwidth,
                        angle=0
                ]{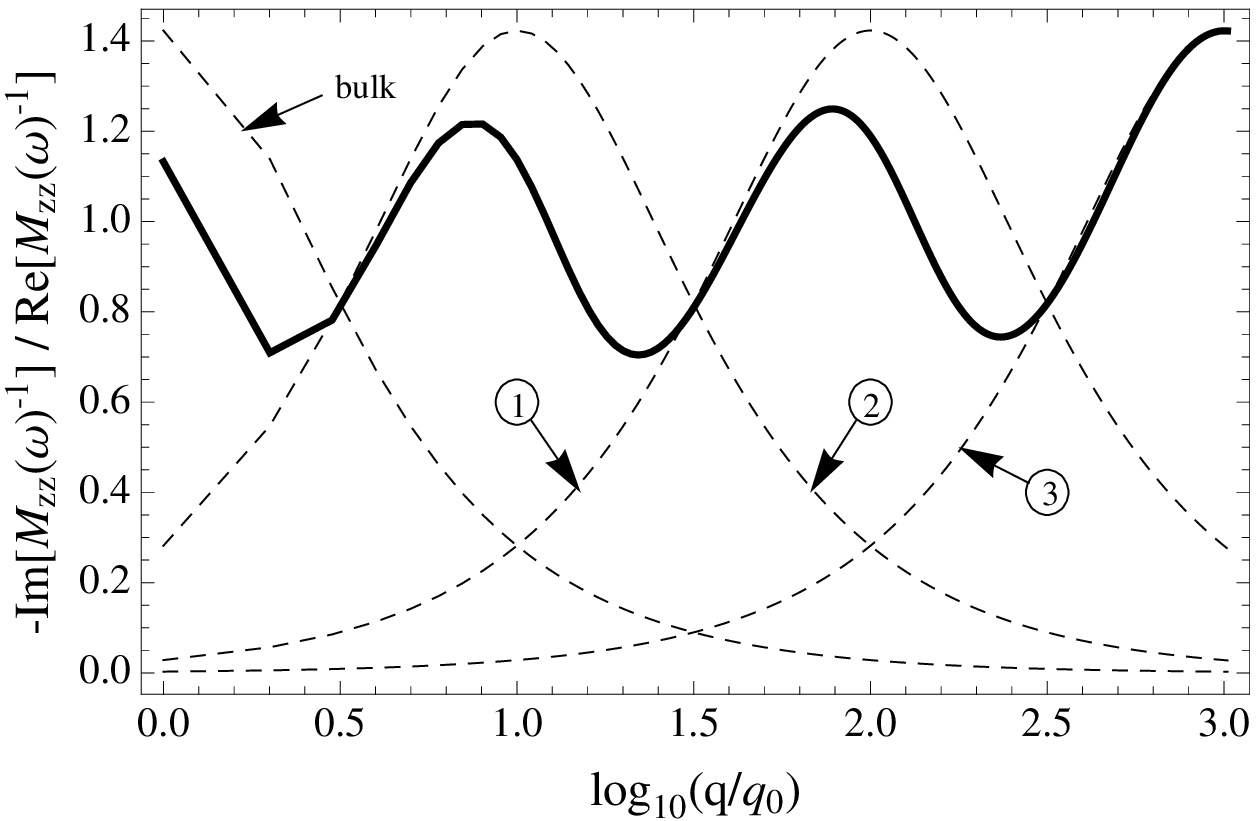} \label{grafico4.eps}
               }
\caption{(a) Schematic of a composite block constituted by a rubber bulk
coated by three rubber layers (with different rheologies) in sliding contact
with a rigid rough surface. (b) Effective composite loss tangent $\mathrm{Im}%
\left[ M_{zz}\left( \protect\omega \right) ^{-1}\right] /\mathrm{Re}\left[
M_{zz}\left( \protect\omega \right) ^{-1}\right] $ (with $\protect\omega =qv$
and $q_{y}=0$) as a function of the wave number $q/q_{0}$ ($q_{0}=2\protect%
\pi /L_{0}$). The bulk is characterized by a single relaxation time $\protect%
\tau =L_{0}/v_{1}$ and $E_{\mathrm{r}\infty }/E_{\mathrm{r}0}=10$. The
coatings are viscoelastic layers with the same rubbery ($E_{\mathrm{r}0}$)
and glassy ($E_{\mathrm{r}\infty }$) modulus of the bulk, but with different
relaxation time $\protect\tau _{j}$, where $j=1$ for the innermost layer. In
particular, $\protect\tau _{j}q_{0}v_{1}=2s_{j}\protect\pi $, with $s_{j}=%
\left[ 10^{-1},10^{-2},10^{-3}\right] $, whereas the coating size follows
the rule $q_{0}d_{j}=2s_{j}\protect\pi $. For sliding velocities $v/v_{1}=1$%
. In (b) the dashed curves represent the loss tangent of each composite
layer, as shown in descriptive balloons (corresponding to $s_j$), whereas the continuous line is the
effective loss tangent.}
\label{composite}
\end{figure}

In Fig.\ref{bulk} we show, for a viscoelastic half space in sliding contact
with a generic rigid rough surface, the cross section (at $q_{y}=0$) of the (b) real
and (c) immaginary part of the dimensionless surface response $M_{zz}\left(
\omega \right) /\left[ 2/\left( qE_{\mathrm{r0}}\right) \right] \ $(with $%
\omega =qv$) as a function of the wave number $q/q_{0}$ (with $q_{0}=2\pi
/L_{0}$). The bulk is characterized by a single relaxation time $\tau
=L_{0}/v_{1}$ and by $E_{\mathrm{r}\infty }/E_{\mathrm{r}0}=10$. Several
dimensionless sliding velocities $v/v_{1}$ are reported, beloging to the set 
$\left[ 0,10^{-4},10^{-3},10^{-2},\infty \right] $. For $v/v_{1}\rightarrow 0
$ ($\infty $), the solid is elastically probed in its rubbery or relaxed
(glassy) regime, see also Fig. \ref{grafico1.eps}. For intermediate sliding
velocities, the roughness wavelengths probe the rubber at different degree
of stiffening, and in particular a monotonic rubber stiffening occurs at
increasing roughness frequencies [Fig. \ref{grafico1.eps}]. Fig. \ref%
{grafico1_1.eps} shows the immaginary part corresponding to Fig. \ref%
{grafico1.eps}. We observe as expected that, by varying the sliding speed, a
different range of roughness wavelengths can probe the rubber at the highest
dissipation. However, no cut-off mechanism occurs in both Figs. \ref%
{grafico1.eps} and \ref{grafico1_1.eps}, as previously discussed.

In Fig. \ref{coating} the case of a viscoelastic half space coated with an
elastic layer in sliding contact with a rigid rough surface is reported. In
particular, we show the cross section (at $q_{y}=0$) of the (b) real part of
the dimensionless surface response $M_{zz}\left( \omega \right) /\left[
2/\left( qE_{\mathrm{r0}}\right) \right] $ and (c) effective composite loss
tangent $\mathrm{Im}\left[ M_{zz}\left( \omega \right) ^{-1}\right] /\mathrm{%
Re}\left[ M_{zz}\left( \omega \right) ^{-1}\right] $ (with $\omega =qv$) as
a function of the wave number $q/q_{0}$ (with $q_{0}=2\pi /L_{0}$). The bulk
is characterized by a single relaxation time $\tau =L_{0}/v_{1}$ and $E_{%
\mathrm{r}\infty }/E_{\mathrm{r}0}=10$, whereas the coating has a reduced
elastic modulus $E_{\mathrm{r}0}$. The dimensionless coating thickness $%
q_{0}d$ is let to vary in the set $\left[ 0,9.4,31,63,94,190,\infty \right]
10^{-3}$, with $v/v_{1}=0.02$ (unless differently specified). 
In Fig. \ref{grafico2.eps}, the red curve corresponds to the limiting case
of an elastic coating with a negligible thickness; thus, the composite
undergoes a monotonic stiffening at increasing roughness frequencies, as
previously reported in Fig. \ref{grafico1.eps}. For increasing coating size $%
q_{0}d$, interestingly, the effective surface response shows a minimum for
intermediate frequencies, whereas both large and small wavelengths probe the
composite in its compliant regime (corresponding to the rubber relaxed
modulus $E_{\mathrm{r}0}$). Thus, for such a composite, increasing the
small-scale roughness content is expected not to affect the true contact
area, as previously demonstrated with the arguments of Sec. \ref{results}.
In term of effective loss tangent, Fig. \ref{grafico3.eps} shows as red
curves the two limiting cases of coating with negligible thickness ($q_{0}d=0
$) and with infinite thickness ($q_{0}d\rightarrow \infty $). For $%
q_{0}d\rightarrow \infty $\ the contact regime occurs under pure elasticity,
obviously determining a null dissipation and, correspondingly, a null loss
tangent. For intermediate coating sizes, the loss tangent loses the
classical bell shape (in the log scale) and, interestingly, a dissipation
cut-off frequency $q_{\mu }$ appears, so that wavelenths smaller than $%
\approx q_{\mu }^{-1}$ do not quantitatively probe the viscoelastic bulk,
with no resulting contribution in term of hysteretic friction (even
considering that those small scale wavelenghts are in full contact with the
composite). Thus, this behaviour generates the physical scenario presented
in Sec. \ref{results}.

Finally, Fig. \ref{composite} shows the case of (a) a composite block
constituted by a rubber bulk coated by three rubber layers (with different
rheologies) in sliding contact with a rigid rough surface. In particular, we
show the cross section (at $q_{y}=0$) of the (b) effective composite loss
tangent $\mathrm{Im}\left[ M_{zz}\left( \omega \right) ^{-1}\right] /\mathrm{%
Re}\left[ M_{zz}\left( \omega \right) ^{-1}\right] $ (with $\omega =qv$ and $%
q_{y}=0$) as a function of the wave number $q/q_{0}$ ($q_{0}=2\pi /L_{0}$).
The bulk is characterized by a single relaxation time $\tau =L_{0}/v_{1}$
and $E_{\mathrm{r}\infty }/E_{\mathrm{r}0}=10$, whereas the coatings are
viscoelastic layers with the same rubbery ($E_{\mathrm{r}0}$) and glassy ($%
E_{\mathrm{r}\infty }$) modulus of the bulk, but with different relaxation
time $\tau _{j}$, where $j=1,2,3$ ($1$ for the innermost layer). In
particular, $\tau _{j}q_{0}v_{1}=2s_{j}\pi $, with $s_{j}=\left[
10^{-1},10^{-2},10^{-3}\right] $, whereas the coating size follows the rule $%
q_{0}d_{j}=2s_{j}\pi $. The sliding speed is set to $v/v_{1}=1$. In Fig. \ref%
{grafico4.eps} the dashed curves rapresent the loss tangent of each
composite layer, as shown in descriptive baloons, whereas the continuous
line is the effective loss tangent. We observe that the properly-designed
building of the composite layers allows to provide effective block
dissipation characteristics which are almost constant and independent from
the probing roughness wavelengths, e.g. in order to let the viscoelastic
friction be maximized. More generally, however, our model can be adopted to
determine the optimal composite packaging which provides tailored contact
properties, such as friction or adherence. Furthermore, the interface can be
designed to be roughness specific, i.e. providing contact mechanics
properties only over a windowed roughness spectral content, e.g. for
biological sensing, bio-adhesion, tire grip control, to cite some.

\section{Conclusions}

We have presented the first numerical contact mechanics model for (randomly or deterministic)
rough surfaces, to be applied for the prediction of the rough contact mechanics
of a general viscoelastic block, with graded rheology, in steady sliding contact with a rough rigid surface.
In particular, our model is able to handle both stepwise or continuously-graded block rheologies, with a (reduced) computational effort typical of
the residuals molecular dynamics scheme.
We have critically discussed on the role of small-scale wavelengths on rubber friction and contact area, and we showed for the first time that
the rough contact mechanics exhibits effective interface properties which converge to asymptotes upon increase of the small-scale roughness content,
under the adoption of some realistic description of the rheology of the confinement. Furthermore, we show that our model can be
effectively adopted for the design of the composite-layers packaging providing 
contact mechanics characteristics (such as friction and adhesion) tailored to be roughness specific,
e.g. for biological sensing, bio-adhesion, tire grip control, to cite some possible applications.

\label{conclusions}

%for amplitude of
%the single sinusoids that constitute the rough surface $A=3.07\times
%10^{-9}(m)$ (represented in the Fig.\ref{amplitude} of the red line $\log
%_{10}A=-8.51(m)$).

\begin{acknowledgments}
DC gratefully acknowledges the support the European Research Council (ERC starting researcher grant 'INTERFACES', No. 279439). 
\end{acknowledgments}
%\begin{widetext}
%\printnomenclature
%\end{widetext}

\newpage
\appendix

\section{General theory for the finite thickness slab with homogeneous
rheological properties}

\label{appendix.1} 
\begin{figure}[tbh]
\centering
\includegraphics[
                        width=0.4\textwidth,
                        angle=0
                ]{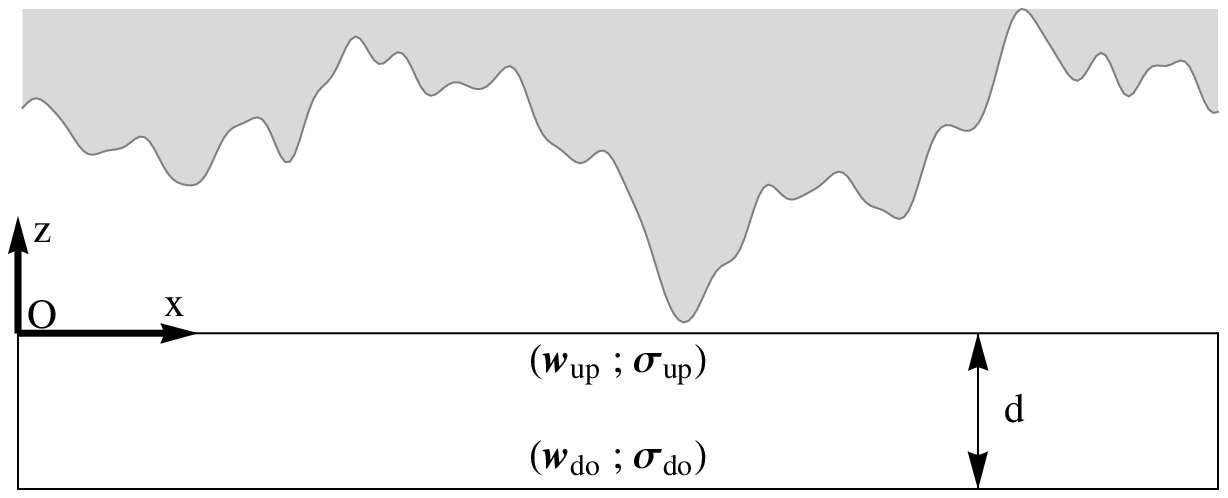}
\caption{Schematic of an infinitely-wide layer with finite thickness $d$,
with generic linearly-viscoelastic rheological properties.}
\label{slab.geometry}
\end{figure}
In this section we solve the Navier's equation for isotropic viscoelasticity
in the case of a finite-thickness infinitely wide slab. Whilst we make use
of the Persson's complex solution presented in \cite{Persson20013840}, in
this section the theory will be further developed to a more general case of
frequency-varying Poisson's ratio $\nu \left( \omega \right) $. This
newly-developed formulation will then be applied to determine the effective
surface responce $M_{zz}\left( q,\omega \right) $ of a composite with,
respectively, stepwise rheology in Sec. \ref{navier.section} and
continuously-graded rheology in Appendix \ref{appendix.2}.

In particular, we consider the case of an infinitely-wide homogeneous slab
of height $d$, with isotropic linearly-viscoelastic rheological properties.
Furthermore, we assume the contact to occurr uder isothermal conditions, and
the rough surface height $h\left( \mathbf{x}\right) $ [with $\left\langle
h\left( \mathbf{x}\right) \right\rangle =\left( L_{x}L_{y}\right) ^{-1}\int
d^{2}x~h\left( \mathbf{x}\right) =0$] to be characterized by a mean square
surface slope $m_{2}\ll 1$, where $m_{2}=\left\langle \left\vert \mathbf{%
\nabla }h\right\vert ^{2}\right\rangle $. The Navier's equation for a
viscoelastic medium reads%
\begin{equation}
\rho \frac{\partial ^{2}\mathbf{w}}{\partial t^{2}}=\mu \nabla ^{2}\mathbf{w}%
+\left( \mu +\lambda \right) \nabla \nabla \cdot \mathbf{w},  \label{navier}
\end{equation}%
where $\mu $ and $\lambda $ are the complex viscoelastic Lame' parameters, $%
\rho $ the density and $w\left( \mathbf{x},z\right) $ the displacement
field. We define the following Fourier transforms%
\begin{eqnarray*}
\mathbf{w}\left( \mathbf{q},z,\omega \right)  &=&\left( 2\pi \right)
^{-3}\int dt\int d^{2}x~\mathbf{w}\left( \mathbf{x},z,t\right) e^{-i\left( 
\mathbf{q}\cdot \mathbf{x}-\omega t\right) } \\
\mu \left( \omega \right)  &=&\int dt~\mu \left( t\right) e^{-i\left(
-\omega t\right) }
\end{eqnarray*}%
and, inversely,%
\begin{eqnarray*}
\mu \left( t\right)  &=&\left( 2\pi \right) ^{-1}\int d\omega ~e^{i\left(
-\omega t\right) }\mu \left( \omega \right)  \\
\mathbf{w}\left( \mathbf{x},z,t\right)  &=&\int dt\int d^{2}x~\mathbf{w}%
\left( \mathbf{q},\omega \right) e^{i\left( \mathbf{q}\cdot \mathbf{x}%
-\omega t\right) }.
\end{eqnarray*}%
Thus, by making use of the field decomposition suggested by Persson \cite%
{Persson20013840} (let $\mathbf{e}_{3}$ be a unit vector along the z axis, see
Fig. \ref{singlelayer.eps}) and by defining $\mathbf{\bar{\nabla}}=-i\mathbf{%
\nabla }=\left( \mathbf{q}-\mathbf{e}_{3}i\frac{\partial }{\partial z}%
\right) $, the Fourier transform ($\mathbf{x}\rightarrow \mathbf{q}$ and $%
t\rightarrow \omega $) of \autoref{navier} results in%
\begin{equation*}
\left( \omega ^{2}\rho -\mu q^{2}+\mu \frac{\partial ^{2}}{\partial z^{2}}%
\right) \mathbf{w}=\left( \mu +\lambda \right) \mathbf{\bar{\nabla}}\left[ 
\mathbf{\bar{\nabla}}\cdot \mathbf{w}\right] .
\end{equation*}%
Note that $\bar{\nabla}^{2}=q^{2}-\frac{\partial ^{2}}{\partial z^{2}}$.
Accordingly, we define $\mathbf{p}=e_{3}\times \mathbf{\bar{\nabla}}=\mathbf{%
e}_{3}\times \mathbf{q}$, and we decompose the displacement field $\mathbf{w}%
(\mathbf{q},z,\omega )$ into \cite{Persson20013840}%
\begin{equation}
\mathbf{w}=\mathbf{\bar{\nabla}}A\left( \mathbf{q},z,\omega \right) +\mathbf{%
p}B\left( \mathbf{q},z,\omega \right) +\mathbf{\bar{\nabla}}\times \mathbf{p}%
C\left( \mathbf{q},z,\omega \right) ,  \label{decomposition}
\end{equation}%
resulting in%
\begin{equation*}
\left( \omega ^{2}\rho -\mu \bar{\nabla}^{2}\right) \mathbf{w}=\left( \mu
+\lambda \right) \mathbf{\bar{\nabla}}\left[ \bar{\nabla}^{2}A\left( \mathbf{%
q},z,\omega \right) \right] ,
\end{equation*}%
where%
\begin{eqnarray*}
w_{1} &=&\left( q_{x}A-q_{y}B+iq_{x}C^{\prime }\right)  \\
w_{2} &=&\left( q_{y}A+q_{x}B+iq_{y}C^{\prime }\right)  \\
w_{3} &=&\left( q^{2}C-iA^{\prime }\right) .
\end{eqnarray*}%
This results in three independent equations in the three scalar fields%
\begin{eqnarray*}
\left[ \omega ^{2}-\frac{2\mu +\lambda }{\rho }\mathbf{\bar{\nabla}}^{2}%
\right] A\left( \mathbf{q},z,\omega \right)  &=&0 \\
\left[ \omega ^{2}-\frac{\mu }{\rho }\mathbf{\bar{\nabla}}^{2}\right]
B\left( \mathbf{q},z,\omega \right)  &=&0 \\
\left[ \omega ^{2}-\frac{\mu }{\rho }\mathbf{\bar{\nabla}}^{2}\right]
C\left( \mathbf{q},z,\omega \right)  &=&0,
\end{eqnarray*}%
and by defining $c_{T}^{2}=\mu /\rho $ and $c_{L}^{2}=\left( 2\mu +\lambda
\right) /\rho $ (where $\frac{\lambda }{\mu }=\frac{2\nu }{1-2\nu }$, $2+%
\frac{\lambda }{\mu }=\frac{2-2\nu }{1-2\nu }=c_{L}^{2}/c_{T}^{2}$)%
\begin{eqnarray*}
\left[ \omega ^{2}/c_{L}^{2}-q^{2}+\frac{\partial ^{2}}{\partial z^{2}}%
\right] A\left( \mathbf{q},z,\omega \right)  &=&0 \\
\left[ \omega ^{2}/c_{T}^{2}-q^{2}+\frac{\partial ^{2}}{\partial z^{2}}%
\right] B\left( \mathbf{q},z,\omega \right)  &=&0 \\
\left[ \omega ^{2}/c_{T}^{2}-q^{2}+\frac{\partial ^{2}}{\partial z^{2}}%
\right] C\left( \mathbf{q},z,\omega \right)  &=&0.
\end{eqnarray*}%
Given a solution $A\left( \mathbf{q},z,\omega \right) =A\left( \mathbf{q}%
,\omega \right) e^{f_{L}z}$ [and $B\left( \mathbf{q},z,\omega \right)
=B\left( \mathbf{q},\omega \right) e^{f_{T}z}$, $C\left( \mathbf{q},z,\omega
\right) =C\left( \mathbf{q},\omega \right) e^{f_{T}z}$], we get that $f=\pm q%
\sqrt{1-\alpha ^{2}+i\varepsilon }$, where $\varepsilon $ is a small
positive number (branch cut along $\pi $) and $\alpha ^{2}=\omega
^{2}/\left( q^{2}c^{2}\right) $ (with $\alpha _{L}^{2}/\alpha _{T}^{2}=\frac{%
1-2\nu \left( \omega \right) }{2-2\nu \left( \omega \right) }$) \cite%
{Persson20013840}. Hence%
\begin{eqnarray}
A\left( \mathbf{q},z,\omega \right)  &=&\left[ A_{1}\left( \mathbf{q},\omega
\right) e^{f_{L}z}+A_{2}\left( \mathbf{q},\omega \right) e^{-f_{L}z}\right] 
\label{6.scalars} \\
B\left( \mathbf{q},z,\omega \right)  &=&\left[ B_{1}\left( \mathbf{q},\omega
\right) e^{f_{T}z}+B_{2}\left( \mathbf{q},\omega \right) e^{-f_{T}z}\right] 
\notag \\
C\left( \mathbf{q},z,\omega \right)  &=&\left[ C_{1}\left( \mathbf{q},\omega
\right) e^{f_{T}z}+C_{2}\left( \mathbf{q},\omega \right) e^{-f_{T}z}\right] .
\notag
\end{eqnarray}%
Moreover, the constitutive relationship can be applied%
\begin{equation*}
\mathbf{S}\left( \mathbf{q},z,\omega \right) =\mu \left( \mathbf{\nabla w}+%
\mathbf{w\nabla }\right) +i\lambda \bar{\nabla}^{2}A~\mathbf{I},
\end{equation*}%
where $\mathbf{S}$ is the stress tensor, which along the $z$-direction
[where $\left( \mathbf{\nabla w}+\mathbf{w\nabla }\right) \mathbf{e}%
_{3}=w_{3,l}+w_{l,3}$] reads%
\begin{equation}
\sigma _{l}\left( \mathbf{q},z,\omega \right) =i\mu \left( \bar{\nabla}%
_{l}w_{3}-i\frac{\partial w_{l}}{\partial z}\right) +i\lambda \bar{\nabla}%
^{2}A\delta _{l3}.  \label{normal stress vector}
\end{equation}%
The index $l=1,~2,~3$ is used here in substitution of the reference
coordinate $x$, $y$, and $z$ respectively. From \autoref{normal stress vector}
it results%
\begin{eqnarray*}
\sigma _{1}\left( \mathbf{q},z,\omega \right)  &=&\mu \left[ 2q_{x}A^{\prime
}-q_{y}B^{\prime }+iq_{x}\left( q^{2}+f_{T}^{2}\right) C\right]  \\
\sigma _{2}\left( \mathbf{q},z,\omega \right)  &=&\mu \left[ 2q_{y}A^{\prime
}+q_{x}B^{\prime }+iq_{y}\left( q^{2}+f_{T}^{2}\right) C\right]  \\
\sigma _{3}\left( \mathbf{q},z,\omega \right)  &=&i\left[ \lambda
q^{2}-\left( 2\mu +\lambda \right) f_{L}^{2}\right] A+2\mu q^{2}C^{\prime }.
\end{eqnarray*}%
To determine the six-scalar fields of \autoref{6.scalars}, we apply the
following boundary conditions to the lower side of the slab (see Fig. \ref%
{slab.geometry})%
\begin{eqnarray*}
\sigma _{l}\left( \mathbf{q},-d,\omega \right)  &=&\sigma _{l,\mathrm{do}%
}\left( \mathbf{q},\omega \right)  \\
u_{l}\left( \mathbf{q},-d,\omega \right)  &=&u_{l,\mathrm{do}}\left( \mathbf{%
q},\omega \right) .
\end{eqnarray*}%
Hence, the stress and displacement fields on the upper surface ($z=0$) can
be easily determined, and in particular in the quasistatic regime ($\alpha
\ll 1$, i.e. $\omega /(qc)=v/c\ll 1$) $\boldsymbol{\sigma }_{\mathrm{up}}=%
\boldsymbol{\sigma }\left( \mathbf{q},z=0,\omega \right) $ and $\boldsymbol{w%
}_{\mathrm{up}}=\boldsymbol{w}\left( \mathbf{q},z=0,\omega \right) $ read%
\begin{eqnarray*}
\boldsymbol{\sigma }_{\mathrm{up}}/\left[ E_{\mathrm{r}}\left( \omega
\right) q/2\right]  &=&\cosh \left( qd\right) \left[ \mathbf{M}_{1}%
\boldsymbol{\sigma }_{\mathrm{low}}/\left[ E_{\mathrm{r}}\left( \omega
\right) q/2\right] +\mathbf{M}_{2}\mathbf{w}_{\mathrm{low}}\right]  \\
\mathbf{w}_{\mathrm{up}} &=&\cosh \left( qd\right) \left[ \mathbf{M}_{3}%
\boldsymbol{\sigma }_{\mathrm{low}}/\left[ E_{\mathrm{r}}\left( \omega
\right) q/2\right] +\mathbf{M}_{4}\mathbf{w}_{\mathrm{low}}\right] ,
\end{eqnarray*}%
i.e. in matrix form%
\begin{equation*}
\left[ 
\begin{array}{c}
\boldsymbol{\sigma }_{\mathrm{up}}/\left[ E_{\mathrm{r}}\left( \omega
\right) q/2\right]  \\ 
\mathbf{w}_{\mathrm{up}}%
\end{array}%
\right] =\cosh \left( qd\right) \left[ 
\begin{array}{cc}
\mathbf{M}_{1} & \mathbf{M}_{2} \\ 
\mathbf{M}_{3} & \mathbf{M}_{4}%
\end{array}%
\right] \left[ 
\begin{array}{c}
\boldsymbol{\sigma }_{\mathrm{low}}/\left[ E_{\mathrm{r}}\left( \omega
\right) q/2\right]  \\ 
\mathbf{w}_{\mathrm{low}}%
\end{array}%
\right] ,
\end{equation*}%
where $E_{\mathrm{r}}\left( \omega \right) =E\left( \omega \right) /\left[
1-\nu \left( \omega \right) ^{2}\right] $ is the complex reduced elastic
modulus. $\bar{q}=qd$ (and similarly for $q_{x}$ and $q_{y}$), $\tilde{q}%
=\tanh \bar{q}$, $p=1-\nu \left( \omega \right) $, $p_{0}=1-\nu _{0}$ [with $%
\nu _{0}=\nu \left( \omega \rightarrow 0\right) $] and we have defined%
\begin{align}
m &=p/p_{0},~n=\left[ 1-2\nu \left( \omega \right) \right] /\left[ 1-2\nu
_{0}\right] ,~\gamma =n/m,  \label{mnbg} \\
\beta  &=\frac{1-4\nu p_{0}}{\left[ 1-2\nu \left( \omega \right) \right] %
\left[ 1-2\nu _{0}\right] }. \notag
\end{align}%
Note that $m$, $n$, $\gamma $ and $\beta $ depends on the frequency $\omega $
through the dependence on $\nu \left( \omega \right) $. After
simplifications we obtain

\begin{equation}
M_{1}=I+\left( 2\bar{q}p_{0}\right) ^{-1}%
\begin{bmatrix}
\bar{q}_{x}^{2}\tilde{q} & \bar{q}_{x}\bar{q}_{y}\tilde{q} & -i\bar{q}%
_{x}\gamma \left[ q-\left( 1-2\nu _{0}\right) \tilde{q}\right]  \\ 
\bar{q}_{x}\bar{q}_{y}\tilde{q} & \bar{q}_{y}^{2}\tilde{q} & -i\bar{q}%
_{y}\gamma \left[ q-\left( 1-2\nu _{0}\right) \tilde{q}\right]  \\ 
-i\bar{q}_{x}\left[ \bar{q}+\beta \left( 1-2\nu _{0}\right) \tilde{q}\right] 
& -i\bar{q}_{y}\left[ \bar{q}+\beta \left( 1-2\nu _{0}\right) \tilde{q}%
\right]  & -\gamma \bar{q}^{2}\tilde{q}%
\end{bmatrix}%
,  \label{M1}
\end{equation}%
\begin{equation}
M_{2}=\bar{q}^{-2}%
\begin{bmatrix}
\left( n\bar{q}+\tilde{q}\right) \bar{q}_{x}^{2}+p\tilde{q}\bar{q}_{y}^{2} & 
\bar{q}_{x}\bar{q}_{y}\left( n\bar{q}+\nu \tilde{q}\right)  & -im\bar{q}_{x}%
\bar{q}^{2}\tilde{q} \\ 
\bar{q}_{x}\bar{q}_{y}\left( n\bar{q}+\nu \tilde{q}\right)  & p\tilde{q}\bar{%
q}_{x}^{2}+\left( n\bar{q}+\tilde{q}\right) \bar{q}_{y}^{2} & -im\bar{q}_{y}%
\bar{q}^{2}\tilde{q} \\ 
-in\bar{q}_{x}\bar{q}^{2}\tilde{q} & -in\bar{q}_{y}\bar{q}^{2}\tilde{q} & -m%
\bar{q}^{2}\left( \bar{q}-\tilde{q}/n\right) 
\end{bmatrix}%
,  \label{M2}
\end{equation}%
\begin{equation}
M_{3}=\left( 2\bar{q}mp_{0}\right) ^{-2}%
\begin{bmatrix}
4p\tilde{q}\bar{q}^{2}+m\left( \bar{q}-\tilde{q}\right) \bar{q}_{x}^{2} & m%
\bar{q}_{x}\bar{q}_{y}\left( \bar{q}-\tilde{q}\right)  & -in\bar{q}_{x}\bar{q%
}^{2}\tilde{q} \\ 
m\bar{q}_{x}\bar{q}_{y}\left( \bar{q}-\tilde{q}\right)  & 4p\tilde{q}\bar{q}%
^{2}+m\left( \bar{q}-\tilde{q}\right) \bar{q}_{y}^{2} & -in\bar{q}_{y}\bar{q}%
^{2}\tilde{q} \\ 
-im\bar{q}_{x}\bar{q}^{2}\tilde{q} & -im\bar{q}_{y}\bar{q}^{2}\tilde{q} & 
-\gamma \bar{q}^{2}\left[ m\left( \bar{q}+\tilde{q}\right) -4p\tilde{q}%
\right] 
\end{bmatrix}
\label{M3}
\end{equation}%
and%
\begin{equation}
M_{4}=I+\left( 2\bar{q}p_{0}\right) ^{-1}%
\begin{bmatrix}
\gamma \bar{q}_{x}^{2}\tilde{q} & \gamma \bar{q}_{x}\bar{q}_{y}\tilde{q} & -i%
\bar{q}_{x}\left[ q+\left( 1-2\nu _{0}\right) \tilde{q}\right]  \\ 
\gamma \bar{q}_{x}\bar{q}_{y}\tilde{q} & \gamma \bar{q}_{y}^{2}\tilde{q} & -i%
\bar{q}_{y}\left[ q+\left( 1-2\nu _{0}\right) \tilde{q}\right]  \\ 
-i\bar{q}_{x}\gamma \left[ q-\beta \left( 1-2\nu _{0}\right) \tilde{q}\right]
& -i\bar{q}_{y}\gamma \left[ q-\beta \left( 1-2\nu _{0}\right) \tilde{q}%
\right]  & -\bar{q}^{2}\tilde{q}%
\end{bmatrix}%
.  \label{M4}
\end{equation}%
Observe that $M_{j}=M_{j}\left( \omega \right) $ through the frequency
dependence of the Poisson's ratio.

In the case the frequency variation of the lateral contraction can be
nglected, i.e. $\nu \left( \omega \right) =\nu =\nu _{0}$, we have that $%
m=n=\beta =\gamma =1$ (and $p_{0}=p$) and the \autoref{M1}-\autoref{M4} simplify to%
\begin{equation*}
M_{1}=I+\left( 2\bar{q}p\right) ^{-1}%
\begin{bmatrix}
\bar{q}_{x}^{2}\tilde{q} & \bar{q}_{x}\bar{q}_{y}\tilde{q} & -i\bar{q}_{x}%
\left[ \bar{q}-\left( 2p-1\right) \tilde{q}\right]  \\ 
\bar{q}_{x}\bar{q}_{y}\tilde{q} & \bar{q}_{y}^{2}\tilde{q} & -i\bar{q}_{y}%
\left[ \bar{q}-\left( 2p-1\right) \tilde{q}\right]  \\ 
-i\bar{q}_{x}\left[ \bar{q}+\left( 2p-1\right) \tilde{q}\right]  & -i\bar{q}%
_{y}\left[ \bar{q}+\left( 2p-1\right) \tilde{q}\right]  & -\bar{q}^{2}\tilde{%
q}%
\end{bmatrix}%
\end{equation*}%
\begin{equation*}
M_{2}=\bar{q}^{-2}%
\begin{bmatrix}
\left( \bar{q}+\tilde{q}\right) \bar{q}_{x}^{2}+p\tilde{q}\bar{q}_{y}^{2} & 
\bar{q}_{x}\bar{q}_{y}\left( \bar{q}+\nu \tilde{q}\right)  & -i\bar{q}_{x}%
\bar{q}^{2}\tilde{q} \\ 
\bar{q}_{x}\bar{q}_{y}\left( \bar{q}+\nu \tilde{q}\right)  & p\tilde{q}\bar{q%
}_{x}^{2}+\left( \bar{q}+\tilde{q}\right) \bar{q}_{y}^{2} & -i\bar{q}_{y}%
\bar{q}^{2}\tilde{q} \\ 
-i\bar{q}_{x}\bar{q}^{2}\tilde{q} & -i\bar{q}_{y}\bar{q}^{2}\tilde{q} & -%
\bar{q}^{2}\left( \bar{q}-\tilde{q}\right) 
\end{bmatrix}%
\end{equation*}%
\begin{equation*}
M_{3}=\left( 2\bar{q}p\right) ^{-2}%
\begin{bmatrix}
4p\tilde{q}\bar{q}^{2}+\left( \bar{q}-\tilde{q}\right) \bar{q}_{x}^{2} & 
\bar{q}_{x}\bar{q}_{y}\left( \bar{q}-\tilde{q}\right)  & -i\bar{q}_{x}\bar{q}%
^{2}\tilde{q} \\ 
\bar{q}_{x}\bar{q}_{y}\left( \bar{q}-\tilde{q}\right)  & 4p\tilde{q}\bar{q}%
^{2}+\left( \bar{q}-\tilde{q}\right) \bar{q}_{y}^{2} & -i\bar{q}_{y}\bar{q}%
^{2}\tilde{q} \\ 
-i\bar{q}_{x}\bar{q}^{2}\tilde{q} & -i\bar{q}_{y}\bar{q}^{2}\tilde{q} & -%
\bar{q}^{2}\left[ \left( \bar{q}+\tilde{q}\right) -4p\tilde{q}\right] 
\end{bmatrix}%
,
\end{equation*}%
with $M_{1}=M_{4}^{T}$.

\subsection{Limiting cases for the single slab}

For the \textbf{slab constrained onto a rigid substrate} [given by $\mathbf{M%
}_{3}\mathbf{M}_{1}^{-1}$, see \autoref{BC1}] we have after simplifications%
\begin{equation}
\frac{M_{zz}\left( \mathbf{q},\omega \right) }{E_{\mathrm{r}}\left( \omega
\right) q/2}=m^{-1}\frac{\left( 4p_{0}-1\right) \tilde{q}-\bar{q}\left( 1-%
\tilde{q}^{2}\right) }{4p_{0}^{2}/\gamma -\bar{q}\tilde{q}\left[ \left(
1-2p_{0}\right) \left( \beta -1\right) +2p_{0}\left( \gamma -1\right)
/\gamma \right] +\bar{q}^{2}\left( 1-\tilde{q}^{2}\right) -\beta \tilde{q}%
^{2}(1-2p_{0})^{2}},  \label{on.rigid.ni}
\end{equation}%
whereas for the case of \textbf{free-standing slab} [given by $\mathbf{M}_{4}%
\mathbf{M}_{2}^{-1}$, see \autoref{BC2}] we have%
\begin{equation}
\frac{M_{zz}\left( \mathbf{q},\omega \right) }{E_{\mathrm{r}}\left( \omega
\right) q/2}=\gamma \frac{1+nq-q\tilde{q}^{2}\left[ \beta n+\left( 1-\beta
n\right) /\left( 2p_{0}\right) \right] }{\tilde{q}^{2}-n^{2}q^{2}\left( 1-%
\tilde{q}^{2}\right) }.  \label{on.air.ni}
\end{equation}

In the case the frequency variation of the lateral contraction can be
nglected, i.e. $\nu \left( \omega \right) =\nu =\nu _{0}$, $m=n=\beta
=\gamma =1$ (and $p_{0}=p$) and \autoref{on.rigid.ni} reads%
\begin{equation}
\frac{M_{zz}\left( \mathbf{q},\omega \right) }{E_{\mathrm{r}}\left( \omega
\right) q/2}=\frac{\left( 4p_{0}-1\right) \tilde{q}-\bar{q}\left( 1-\tilde{q}%
^{2}\right) }{4p_{0}^{2}-\tilde{q}^{2}(1-2p_{0})^{2}+\bar{q}^{2}\left( 1-%
\tilde{q}^{2}\right) },  \label{on.rigid.fixedni}
\end{equation}%
whereas for the case of free-standing slab \autoref{on.air.ni} simplifies to%
\begin{equation}
\frac{M_{zz}\left( \mathbf{q},\omega \right) }{E_{\mathrm{r}}\left( \omega
\right) q/2}=\frac{1+q-q\tilde{q}^{2}}{\tilde{q}^{2}-q^{2}\left( 1-\tilde{q}%
^{2}\right) },  \label{on.air.fixedni}
\end{equation}%
corresponding to the classical results\cite{Persson2012}.

\subsection{Limiting cases for the coated half-space}

For the case of a bulk [$E_{\mathrm{b}}\left( \omega \right) $, $\nu _{%
\mathrm{b}}\left( \omega \right) $, $E_{\mathrm{rb}}\left( \omega \right)
=E_{\mathrm{b}}/\left( 1-\nu _{\mathrm{b}}^{2}\right) $] coated with a layer
[$E\left( \omega \right) $, $\nu \left( \omega \right) $, $E_{\mathrm{r}%
}\left( \omega \right) =E/\left( 1-\nu ^{2}\right) $] of thickness $d$, by
using \autoref{final} we have (after some manipulation)%
\begin{equation}
\frac{M_{zz}\left( \mathbf{q},\omega \right) }{E_{\mathrm{r}}\left( \omega
\right) q/2}=n\frac{c_{1}n_{0}+c_{2}\left(
n_{1}Mp^{2}+2n_{2}mp+n_{3}\varepsilon _{\mathrm{p}}m^{2}\right) }{%
c_{1}d_{0}+c_{2}\left( d_{1}Mp^{2}+2d_{2}mp+d_{3}\varepsilon _{\mathrm{p}%
}m^{2}\right) },  \label{coating.on.bulk.ni}
\end{equation}%
where%
\begin{eqnarray*}
c_{1} &=&4Mp^{2}\epsilon _{\mathrm{e}}\left( \tilde{q}\left( 1+mq\right)
+nq\right) +\left( 2Mp-m\epsilon _{\mathrm{p}}\right) \left( 2p+mq\tilde{q}%
\right)  \\
&&+\epsilon _{\mathrm{e}}\epsilon _{\mathrm{p}}\left[ \left( m\epsilon _{%
\mathrm{p}}-4Mp\right) mq\tilde{q}-\left( 2mp\right) \left( nq+\tilde{q}%
\right) \right]  \\
c_{2} &=&-4Mp^{2}\epsilon _{\mathrm{e}}\left( \tilde{q}\left( 1+mq\right)
+nq\right) +\left( 2Mp-m\epsilon _{\mathrm{p}}\right) \gamma \left( 2p\tilde{%
q}-m(q+\tilde{q})\right)  \\
&&+\epsilon _{\mathrm{e}}\epsilon _{\mathrm{p}}\left[ \left( 2Mp\right) mq%
\tilde{q}+\left( \left( 2Mp-m\epsilon _{\mathrm{p}}\right) +2mp\right)
\left( nq+\tilde{q}\right) \right] ,
\end{eqnarray*}%
whereas%
\begin{eqnarray*}
n_{0} &=&-2Mp\epsilon _{\mathrm{e}}\left( 2p-\epsilon _{\mathrm{p}}\right) %
\left[ mq\tilde{q}-2p\right]  \\
&&+(2Mp-m\epsilon _{\mathrm{p}})n\left[ \epsilon _{\mathrm{e}}\left(
2p-\epsilon _{\mathrm{p}}\right) \gamma \left( \left( 2p-m\right) \beta 
\tilde{q}-mq\right) +\frac{n}{m}4p\tilde{q}-n(q+\tilde{q})\right]  \\
n_{1} &=&8\epsilon _{\mathrm{e}}(\beta \gamma p\tilde{q}+p-\epsilon _{%
\mathrm{p}}) \\
n_{2} &=&-2Mp\epsilon _{\mathrm{e}}\left[ q(\gamma +\tilde{q})+\beta \gamma 
\tilde{q}\right] +\epsilon _{\mathrm{e}}\epsilon _{\mathrm{p}}\left[ 1-2%
\tilde{q}\epsilon _{\mathrm{e}}\epsilon _{\mathrm{p}}(\beta \gamma p-Mq)%
\right] -Mq\tilde{q} \\
n_{3} &=&2\gamma p\epsilon _{\mathrm{e}}\left( q+\beta \tilde{q}\right) +q%
\tilde{q}\left( 1-\epsilon _{\mathrm{e}}\epsilon _{\mathrm{p}}\right) 
\end{eqnarray*}%
and%
\begin{eqnarray*}
d_{0} &=&2Mp\epsilon _{\mathrm{e}}\left( 2p-\epsilon _{\mathrm{p}}\right)
m\left( nq-\tilde{q}\right)  \\
&&+\left( 2Mp-m\epsilon _{\mathrm{p}}\right) \left[ \epsilon _{\mathrm{e}%
}(2p-\epsilon _{\mathrm{p}})nq\tilde{q}+\gamma mq\tilde{q}-2p\right]  \\
d_{1} &=&4n\tilde{q}(\beta +nq\epsilon _{\mathrm{e}}) \\
d_{2} &=&2M\epsilon _{\mathrm{e}}\left( nq-\tilde{q}\right) \left(
p-\epsilon _{\mathrm{p}}\right) -n\tilde{q}\epsilon _{\mathrm{p}}(\beta
+nq\epsilon _{\mathrm{e}})+nM\left( q-\beta \tilde{q}\right)  \\
d_{3} &=&n\left( \beta \tilde{q}-q\right) -\epsilon _{\mathrm{e}}\epsilon _{%
\mathrm{p}}\left( \tilde{q}-nq\right) ,
\end{eqnarray*}%
with $\varepsilon _{\mathrm{e}}=E\left( \omega \right) /E_{\mathrm{b}}\left(
\omega \right) $, $\varepsilon _{\mathrm{p}}=p\left( \omega \right) /p_{%
\mathrm{b}}\left( \omega \right) $, and $M=\varepsilon _{\mathrm{p}%
}/\varepsilon _{0\mathrm{p}}=m/\left( p_{\mathrm{b}}/p_{0\mathrm{b}}\right) $%
, with $m$, $n$, $\gamma $ and $\beta $ given by \autoref{mnbg}. In the limit
where the frequency variation of the lateral contraction can be neglected,
i.e. $\nu \left( \omega \right) =\nu _{0}$ and $\nu _{\mathrm{b}}\left(
\omega \right) =\nu _{0\mathrm{b}}$\ (resulting into $\beta =\gamma =m=n=M=1$%
) we obtain the classical result \cite{Persson2012}%
\begin{equation}
\frac{M_{zz}\left( \mathbf{q},\omega \right) }{E_{\mathrm{r}}\left( \omega
\right) q/2}=\frac{n_{1}\sinh (2q)+8\alpha p^{2}\cosh (2q)+n_{2}q}{8p^{2}%
\left[ \alpha \sinh (2q)+1\right] +n_{1}\left[ \cosh (2q)-1\right]
-n_{2}q^{2}},  \label{coating.on.bulk.fixedni}
\end{equation}%
where this time%
\begin{eqnarray*}
n_{1} &=&8\varepsilon _{\mathrm{e}}p^{2}+4p\varepsilon _{\mathrm{pe}%
}(\varepsilon _{\mathrm{e}}-1)-\varepsilon _{\mathrm{pe}}^{2} \\
n_{2} &=&2\varepsilon _{\mathrm{pe}}\left( 4\varepsilon _{\mathrm{e}%
}p-\varepsilon _{\mathrm{pe}}\right) 
\end{eqnarray*}%
and $\varepsilon _{\mathrm{pe}}=\varepsilon _{\mathrm{e}}\varepsilon _{%
\mathrm{p}}-1$ [$\varepsilon _{\mathrm{e}}=E\left( \omega \right) /E_{%
\mathrm{b}}\left( \omega \right) $, $\varepsilon _{\mathrm{p}}=p/p_{\mathrm{b%
}}$, as before].
\begin{figure}[tbh]
\centering
\subfigure[]{
\includegraphics[
                        width=0.407\textwidth,
                        angle=0
                ]{coating_halfpsace.eps} \label{coating_halfpsace_persson.eps}
                } \qquad 
\subfigure[]{
\includegraphics[
                        width=0.4\textwidth,
                        angle=0
                ]{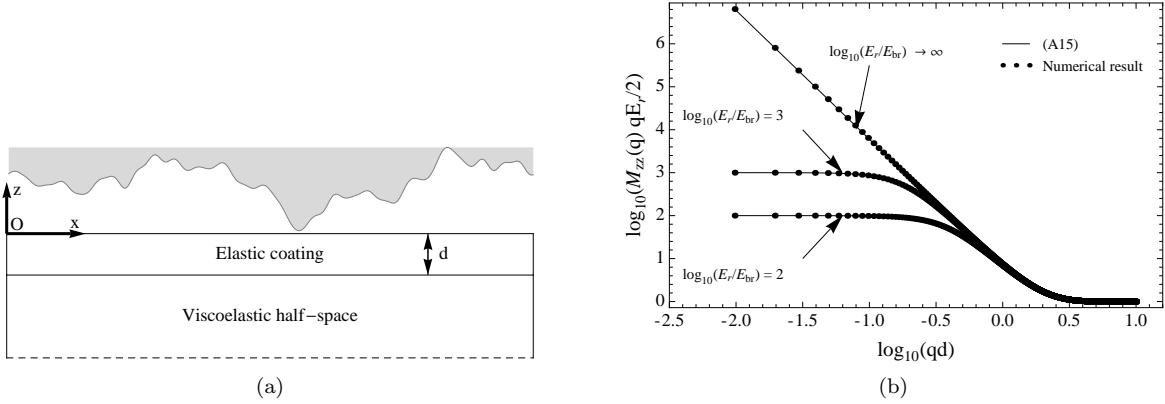} \label{comparison_persson.eps}
                }
\caption{a) Schematic of an elastic coating bonded onto an elastic half
space, b) real part of the dimensionless surface response $M_{zz}\left(
q\right) qE_{\mathrm{r0}}/2$ as a function of the wave number $qd$ ($q_{0}=2%
\protect\pi /L_{0}$). The bulk is characterized by a reduced elastic modulus 
$E_{\mathrm{rb}}$, whereas the coating by $E_{\mathrm{r}}$. For the
dimensionless coating thickness $q_{0}d=0.0195$, and for different values of 
$E_{\mathrm{r}}/E_{\mathrm{rb}}$. The continuous line is from \protect\ref%
{coating.on.bulk.fixedni}, whereas dots are from the application of \protect
\autoref{final}.}
\label{test.numerical.model}
\end{figure}
\begin{figure}[tbh]
\centering
\subfigure[$\protect\nu _{\infty \mathrm{b}}$ from $0.3$ to $0.49$, and $\protect\nu _{1}=-0.774$, resulting in $%
\protect\nu _{0 \mathrm{b}}=0.49$]{
\includegraphics[
                        width=0.407\textwidth,
                        angle=0
                ]{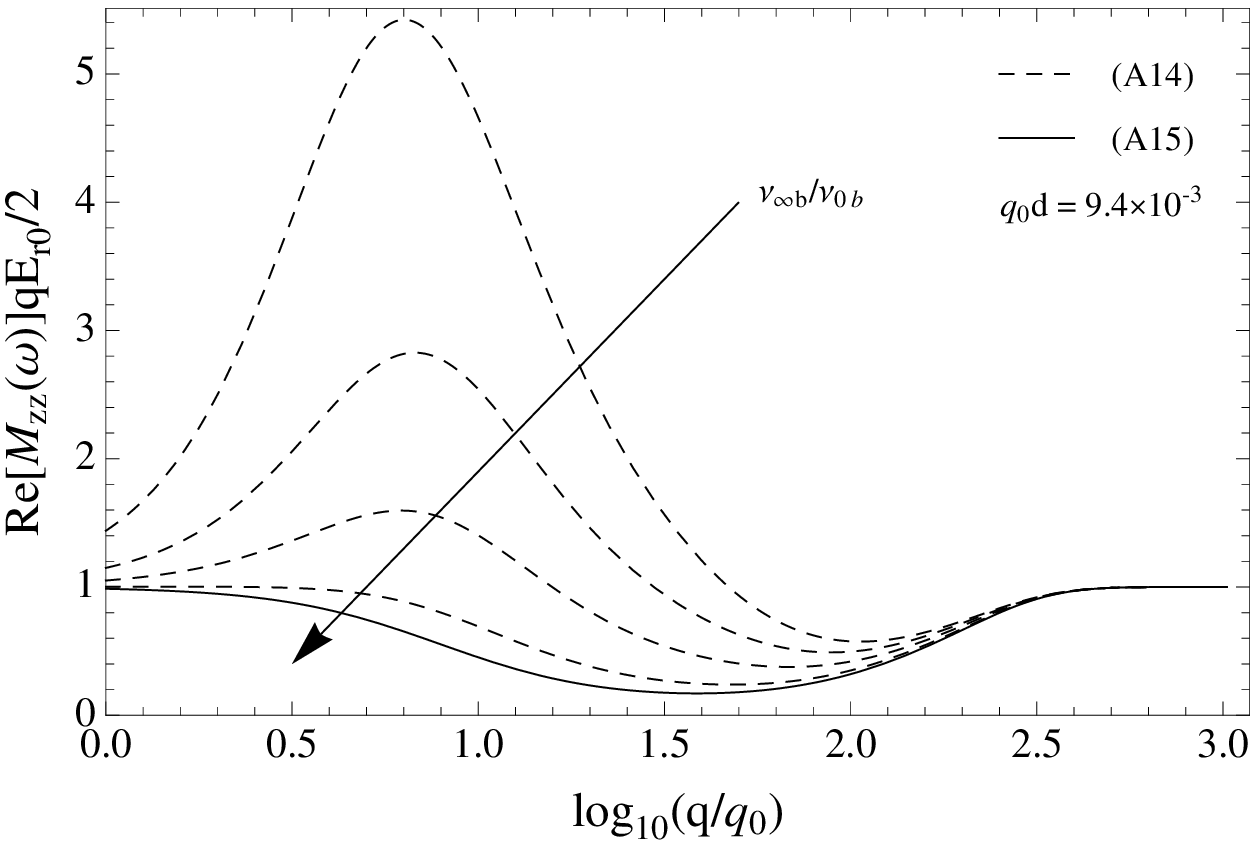} \label{graph_ni_variable_3.eps}
                } \\
\subfigure[$\protect\nu _{\infty \mathrm{b}}=0.3$ and $\protect\nu _{1}=-0.774$, resulting in $%
\protect\nu _{0 \mathrm{b}}=0.49$]{
\includegraphics[
                        width=0.4\textwidth,
                        angle=0
                ]{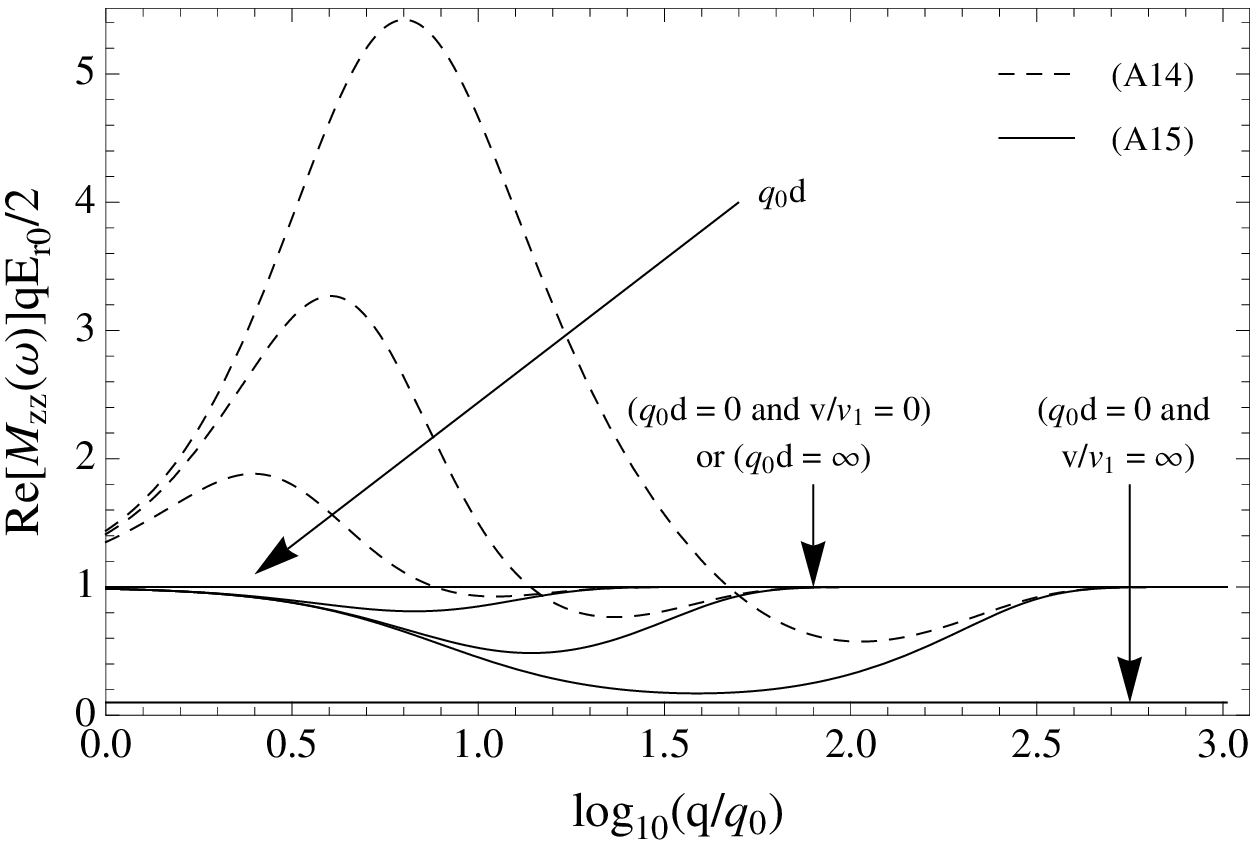} \label{graph_ni_variable_1.eps}
                } \qquad 
\subfigure[$\protect\nu _{\infty \mathrm{b}}=0.49$ and $\protect\nu _{1}=0.774$, resulting in $%
\protect\nu _{0 \mathrm{b}}=0.3$]{
\includegraphics[
                        width=0.4\textwidth,
                        angle=0
                ]{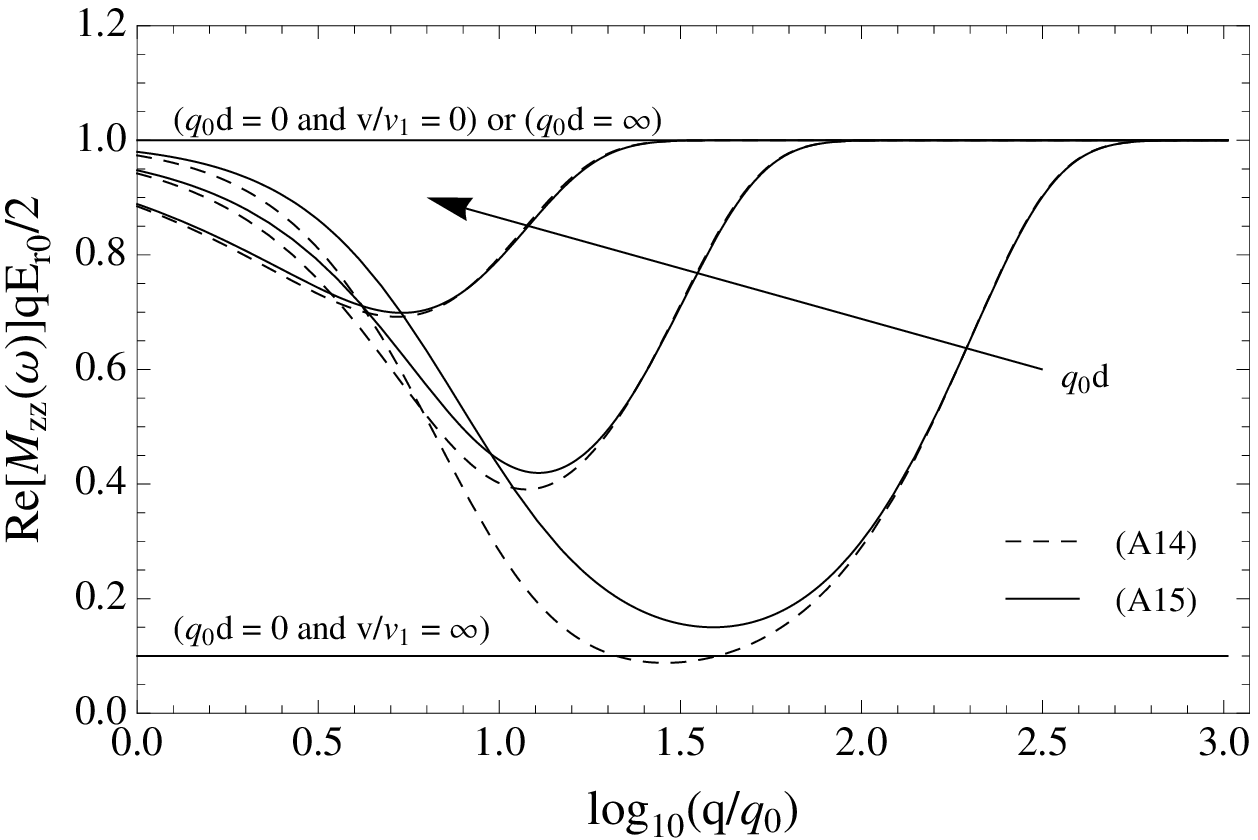} \label{graph_ni_variable_2.eps}
                }
\caption{Real part of the dimensionless surface response $M_{zz}\left( q,%
\protect\omega \right) qE_{\mathrm{r0}}/2$ (with $\protect\omega =qv$ and $%
q_{y}=0$) as a function of the wave number $q/q_{0}$ ($q_{0}=2\protect\pi %
/L_{0}$), for an elastic coating bonded onto a viscoelastic half
space [see e.g. Fig. \protect\ref{coating_halfpsace_persson.eps}]. The bulk is characterized by a single relaxation time $\protect%
\tau =L_{0}/v_{1}=0.01\mathrm{s}$ and $E_{\infty }/E_{%
0}=10$, with
$\protect\nu_{\mathrm{b}} (\protect\omega )^{-1}=\protect\nu _{\infty \mathrm{b}}^{-1}+\protect\nu %
_{1}^{-1}/\left( 1+\protect\omega ^{2}\protect\tau ^{2}\right) $. The coating has an elastic modulus $E=E_{\mathrm{0b}}$
and $\nu=0.49$. The dashed lines [$\protect\nu_{\mathrm{b}}=\protect\nu_{\mathrm{b}} (\protect\omega )$]
correspond to \autoref{coating.on.bulk.ni}, whereas the solid
lines [$\protect\nu_{\mathrm{b}}=\protect\nu_{0\mathrm{b}}$] are for \autoref{coating.on.bulk.fixedni}.
In (a) the dimensionless coating thickness $q_{0}d$ is $9.4\ 10^{-3}$, whereas in (b,c)
the $q_{0}d$ belongs to $\left[
0,9.4,63,190,\infty \right] 10^{-3}$.}
\label{davide.test.variable.nu}
\end{figure}

In Fig. \ref{test.numerical.model}, for (a) an elastic coating bonded onto
an elastic half space, we show the (b) real part of the dimensionless
surface response $M_{zz}\left( q\right) qE_{\mathrm{r0}}/2$ as a function of
the wave number $qd$ ($q_{0}=2\pi /L_{0}$). The bulk is characterized by a
reduced elastic modulus $E_{\mathrm{rb}}$, whereas the coating by $E_{%
\mathrm{r}}$. For the dimensionless coating thickness $q_{0}d=0.0195$, and
for different values of $E_{\mathrm{r}}/E_{\mathrm{rb}}$. The continuous
line is from \autoref{coating.on.bulk.fixedni}, whereas dots are from the
application of \autoref{final}, confirming the validity of the numerical tool.

Finally, in Fig. \ref{davide.test.variable.nu} for an elastic coating
bonded onto a viscoelastic half space [see e.g. Fig. \ref{coating_halfpsace_persson.eps}],
we show the real part of the dimensionless surface response $M_{zz}\left( q,%
\protect\omega \right) qE_{\mathrm{r0}}/2$ (with $\protect\omega =qv$ and $%
q_{y}=0$) as a function of the wave number $q/q_{0}$ ($q_{0}=2\protect\pi %
/L_{0}$). The bulk is characterized by a single relaxation time $\protect%
\tau =L_{0}/v_{1}=0.01\mathrm{s}$ and $E_{\infty }/E_{%
0}=10$, with
$\protect\nu_{\mathrm{b}} (\protect\omega )^{-1}=\protect\nu _{\infty \mathrm{b}}^{-1}+\protect\nu %
_{1}^{-1}/\left( 1+\protect\omega ^{2}\protect\tau ^{2}\right) $. The coating has an elastic modulus $E=E_{\mathrm{0b}}$
and $\nu=0.49$. The dashed lines [$\protect\nu_{\mathrm{b}}=\protect\nu_{\mathrm{b}} (\protect\omega )$]
correspond to \autoref{coating.on.bulk.ni}, whereas the solid
lines [given by considering $\protect\nu_{\mathrm{b}}=\protect\nu_{0\mathrm{b}}$] are for \autoref{coating.on.bulk.fixedni}.
In (a) the dimensionless coating thickness $q_{0}d$ is $9.4\ 10^{-3}$, whereas in (b,c)
the $q_{0}d$ belongs to $\left[
0,9.4,63,190,\infty \right] 10^{-3}$. The sliding velocity
is set to $v=0.02v_{1}$. Moreover, in Fig. \ref{graph_ni_variable_3.eps} and \ref{graph_ni_variable_1.eps}
we have adopted $\nu _{\infty \mathrm{b}}<\nu _{0 \mathrm{b}}$, and inversely for Fig. \ref{graph_ni_variable_2.eps}.

We note first in Fig. \ref{graph_ni_variable_1.eps} that the effective compliance of the
composite with $\nu_{\mathrm{b}}=\nu_{\mathrm{b}}(\omega)$ (dashed lines) shows
a global maximum, which is close to a range of frequencies where
$\nu_{\mathrm{b}}(\omega)$ moves from the rubbery ($0.49$) to the glassy ($0.3$) region.
Moreover such a maximum, which is even larger than 1, increases for decreasing coating thickness.
Far from the previous stationary point (in a log-scale), i.e. at small ($qv\rightarrow 0$) and large ($qv\rightarrow \infty$) roughness frequencies, the compliance
converges to the corresponding curve for the frequency-independent $\nu_{\mathrm{b}}$. This can be justified as follows.
For $qv\rightarrow 0$ ($qv\rightarrow \infty$), the asperities do only
probe the rubbery bulk (coating) of the composite, resulting in $M_{zz}\rightarrow qE_{\mathrm{r0}}/2$. For intermediate
wavelengths, the bulk undergoes a transition from an incompressible stage in the rubbery ($\nu_{\mathrm{b}}=0.49$) regime, to a compressible stage in
the glassy ($\nu_{\mathrm{b}}=0.3$) region. Since a continuity of lateral contraction must hold at the coating/bulk interface, the more $qv$ increases
the less is the lateral contraction (i.e. $\nu_{\mathrm{b}}$ decreases) coped with an increased (viscoelastic) stiffening, resulting that the composite must
show a more compliant response in order to match such an interface lateral contraction. Of course the opposite holds for Fig. \ref{graph_ni_variable_2.eps}.
Thus, we observe that neglecting the frequency-dependence of the Poisson's ratio can qualitatively and quantitatively affect the
contact mechanics predictions, given the large differences in the effective surface response in a range
of frequencies which cannot be established a priori, even for a simple composite arrangement such as the one previously adopted.

\section{General theory for the finite thickness slab with continuously
graded rheological properties}

\label{appendix.2}In this section we derive the $\mathbf{M}$ matrix for a
composite characterized by continuously-graded rheological properties. In
particular, by differentiating \autoref{graded.stepped} we obtain%
\begin{equation}
\left[ \mathbf{M}+d\mathbf{M}\right] \left[ \mathbf{M}_{1}+\left( 1+\alpha
~dz\right) \mathbf{M}_{2}\mathbf{M}\right] =\left[ \mathbf{M}_{3}+\left(
1+\alpha ~dz\right) \mathbf{M}_{4}\mathbf{M}\right] ,
\label{differentiated.graded.stepped}
\end{equation}%
with $\alpha \left( \omega ,z\right) =E_{\mathrm{r}}\left( \omega ,z\right)
^{-1}\partial E_{\mathrm{r}}\left( \omega ,z\right) /\partial z$. This
results in the following set of non-linear differential equations

\begin{eqnarray*}
\frac{pq}{m^{2}p_{0}^{2}} &=&M_{11}^{\prime }-\alpha M_{11}+M_{11}^{2}\delta
_{x}+M_{11}\left( M_{12}+M_{21}\right) \delta _{xy}+M_{12}M_{21}\delta _{y}
\\
&&+iq_{x}\delta _{1}\left( M_{31}-M_{13}\right) +q\delta _{2}M_{13}M_{31} \\
0 &=&M_{12}^{\prime }-\alpha M_{12}+M_{11}M_{12}\delta _{x}+\left(
M_{11}M_{22}+M_{12}^{2}\right) \delta _{xy}+M_{12}M_{22}\delta _{y} \\
&&+iq_{x}\delta _{1}\left( M_{32}-M_{13}\right) +q\delta _{2}M_{13}M_{32} \\
0 &=&M_{13}^{\prime }-\alpha M_{13}+M_{11}M_{13}\delta _{x}+\left(
M_{11}M_{23}+M_{12}M_{13}\right) \delta _{xy}+M_{12}M_{23}\delta _{y} \\
&&+iq_{x}\delta _{1}M_{33}+q\delta _{2}M_{13}M_{33}+\delta _{3}\left(
M_{11}q_{x}+M_{12}q_{y}\right) 
\end{eqnarray*}%
\begin{eqnarray*}
0 &=&M_{21}^{\prime }-\alpha M_{21}+M_{11}M_{21}\delta _{x}+\left(
M_{11}M_{22}+M_{21}^{2}\right) \delta _{xy}+M_{21}M_{22}\delta _{y} \\
&&+i\delta _{1}\left( M_{31}q_{y}-M_{23}q_{x}\right) +q\delta
_{2}M_{23}M_{31} \\
\frac{pq}{m^{2}p_{0}^{2}} &=&M_{22}^{\prime }-\alpha
M_{22}+M_{12}M_{21}\delta _{x}+M_{22}\left( M_{12}+M_{21}\right) \delta
_{xy}+M_{22}^{2}\delta _{y} \\
&&+iq_{y}\delta _{1}\left( M_{32}-M_{23}\right) +q\delta _{2}M_{23}M_{32} \\
0 &=&M_{23}^{\prime }-\alpha M_{23}+M_{13}M_{21}\delta _{x}+\left(
M_{13}M_{22}+M_{21}M_{23}\right) \delta _{xy}+M_{22}M_{23}\delta _{y} \\
&&+iq_{y}\delta _{1}M_{33}+q\delta _{2}M_{23}M_{33}+\left(
M_{21}q_{x}+M_{22}q_{y}\right) \delta _{3}
\end{eqnarray*}%
\begin{eqnarray*}
0 &=&M_{31}^{\prime }-\alpha M_{31}+M_{11}M_{31}\delta _{x}+\left(
M_{11}M_{32}+M_{21}M_{31}\right) \delta _{xy}+M_{21}M_{32}\delta _{y} \\
&&-iq_{x}\delta _{1}M_{33}+q\delta _{2}M_{31}M_{33}-\left(
M_{11}q_{x}+M_{21}q_{y}\right) \delta _{3} \\
0 &=&M_{32}^{\prime }-\alpha M_{32}+M_{12}M_{31}\delta _{x}+\left(
M_{12}M_{32}+M_{22}M_{31}\right) \delta _{xy}+M_{22}M_{32}\delta _{y} \\
&&-iq_{y}\delta _{1}M_{33}+q\delta _{2}M_{32}M_{33}-\left(
M_{12}q_{x}+M_{22}q_{y}\right) \delta _{3} \\
-\frac{nq(m-2p)}{2m^{3}p_{0}^{2}} &=&M_{33}^{\prime }-\alpha
M_{33}+M_{13}M_{31}\delta _{x}+\left( M_{13}M_{32}+M_{23}M_{31}\right)
\delta _{xy}+M_{23}M_{32}\delta _{y} \\
&&+q\delta _{2}M_{33}^{2}+\left[ \left( M_{31}-M_{13}\right) q_{x}+\left(
M_{32}-M_{23}\right) q_{y}\right] \delta _{3}
\end{eqnarray*}%
where we have defined%
\begin{eqnarray*}
\delta _{1} &=&\frac{\beta (2p_{0}-1)+1}{2p_{0}},~\delta _{2}=m\left( \frac{1%
}{n}-1\right) ,~\delta _{3}=\frac{i\gamma (p_{0}-1)}{p_{0}} \\
\delta _{x} &=&\frac{(n+1)q_{x}^{2}+pq_{y}^{2}}{q},~\delta _{y}=\frac{%
(n+1)q_{y}^{2}+pq_{x}^{2}}{q},~\delta _{xy}=\frac{(n+1-p)q_{x}q_{y}}{q},
\end{eqnarray*}%
and with BCs $M\left( \omega ,z=0\right) $ given by \autoref{bulk.new}.
\begin{figure}[tbh]
\centering
\subfigure[]{
\includegraphics[
                        width=0.5\textwidth,
                        angle=0
                ]{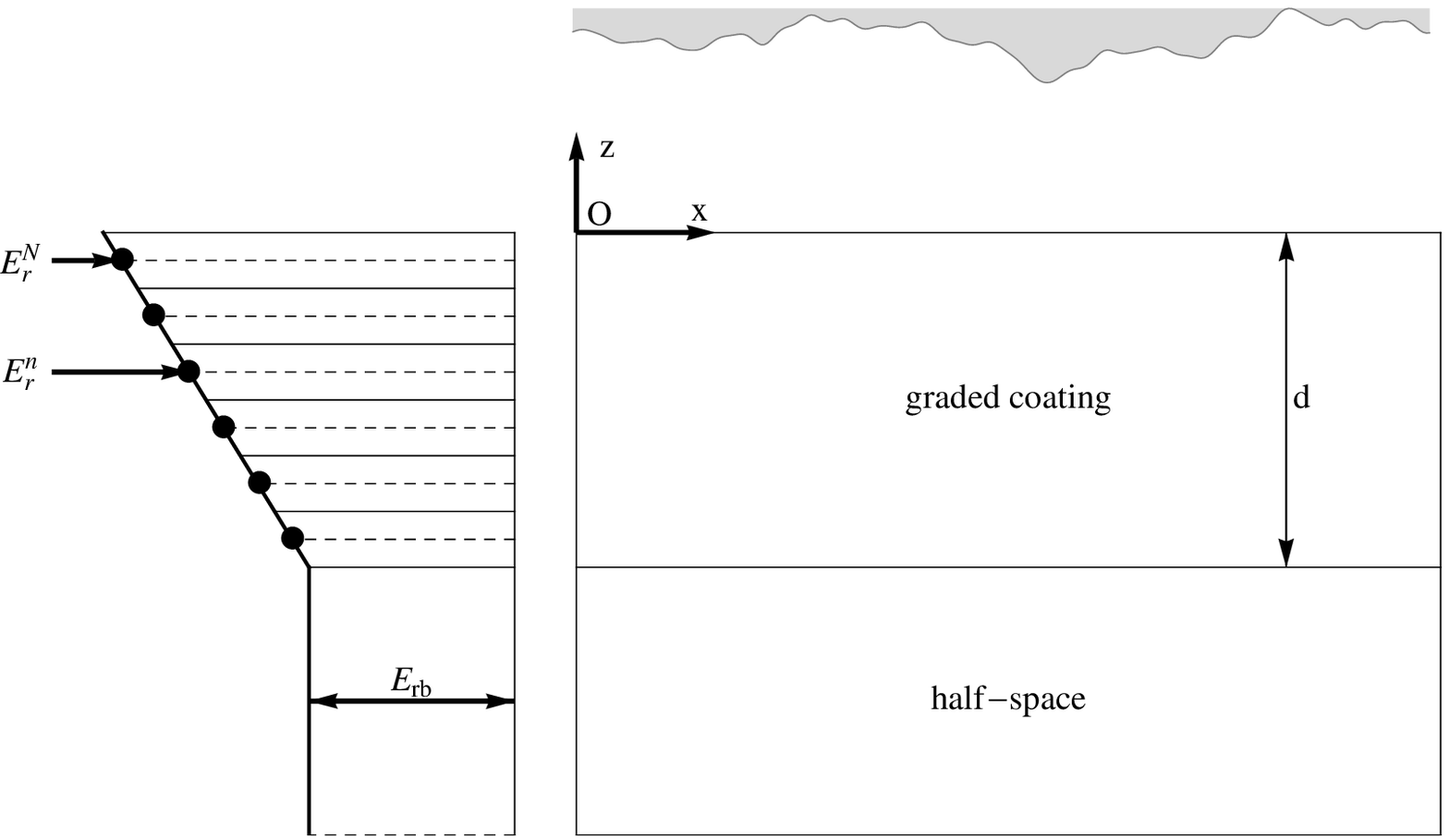} \label{other.graphs.2.eps}
                } \qquad 
\subfigure[]{
\includegraphics[
                        width=0.4\textwidth,
                        angle=0
                ]{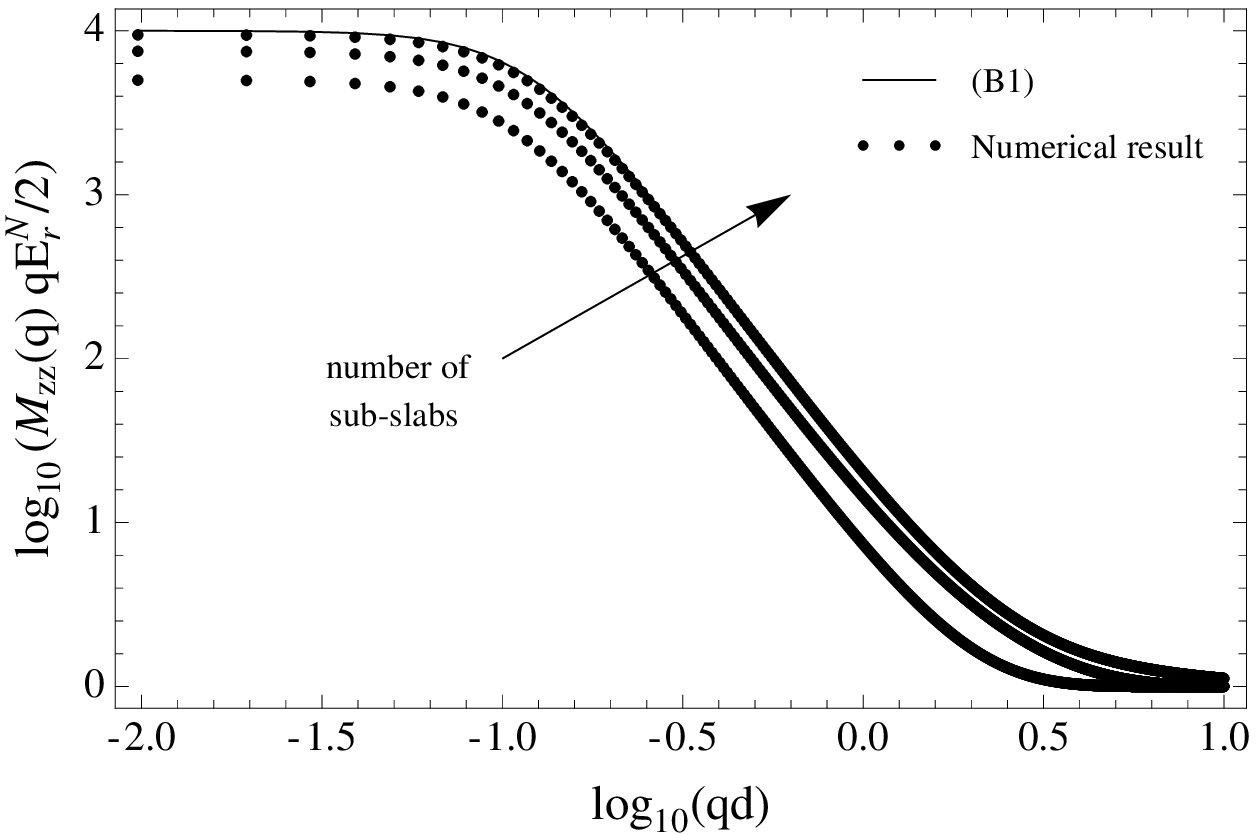} \label{comparisonscarlinear.eps}
                }
\caption{a) Schematic of a graded elastic coating bonded onto an elastic
half space, b) dimensionless surface response $M_{zz}\left(
q\right) qE_{\mathrm{r0}}/2$ (with $q_{y}=0$) as a function of the wave
number $dq$ ($q_{0}=2\protect\pi /L_{0}$). The coating is characterized by a
reduced elastic modulus $E_{\mathrm{r}}(z)=E_{\mathrm{rb}}\left[ \left(
1+10^{4}\left( z+d\right) /d\right) \right] $, whereas the bulk by $E_{%
\mathrm{rb}}$. For a dimensionless coating thickness $q_{0}d=0.0195$ and $%
L_{0}=0.1$. In (b) the continuous line is from \protect\ref%
{differentiated.graded.stepped}, whereas the dots are obtained from the
application of \protect\autoref{final} to a linear discretization of the coating 
$E_{\mathrm{r}}(z)$ into a stepwise composite of $1$, $2$ and $9$ layer
sub-divisions.}
\label{test.numerical.linear}
\end{figure}
\begin{figure}[tbh]
\centering
\subfigure[]{
\includegraphics[
                        width=0.5\textwidth,
                        angle=0
                ]{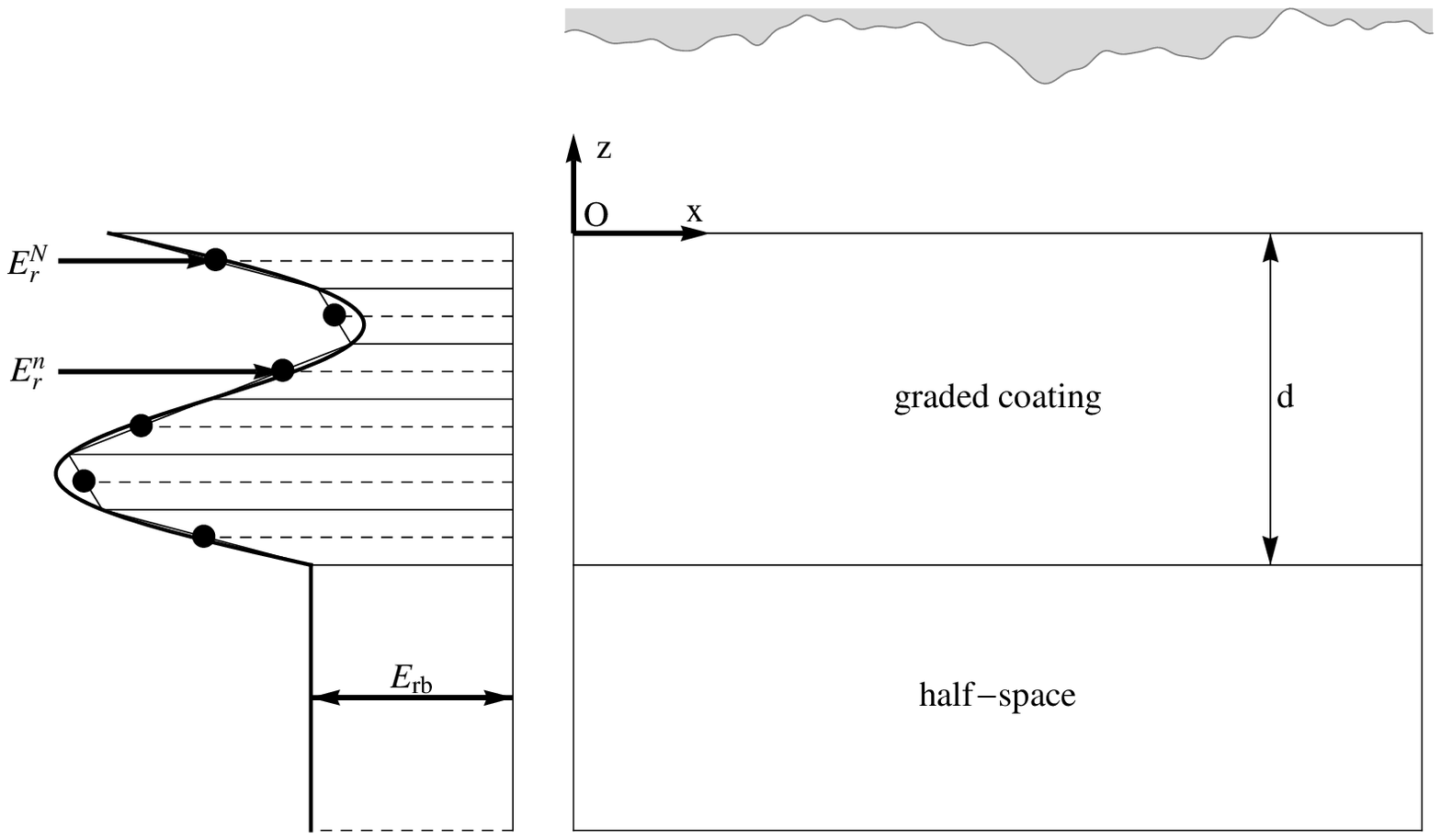} \label{other.graphs.eps}
                } \qquad 
\subfigure[]{
\includegraphics[
                        width=0.4\textwidth,
                        angle=0
                ]{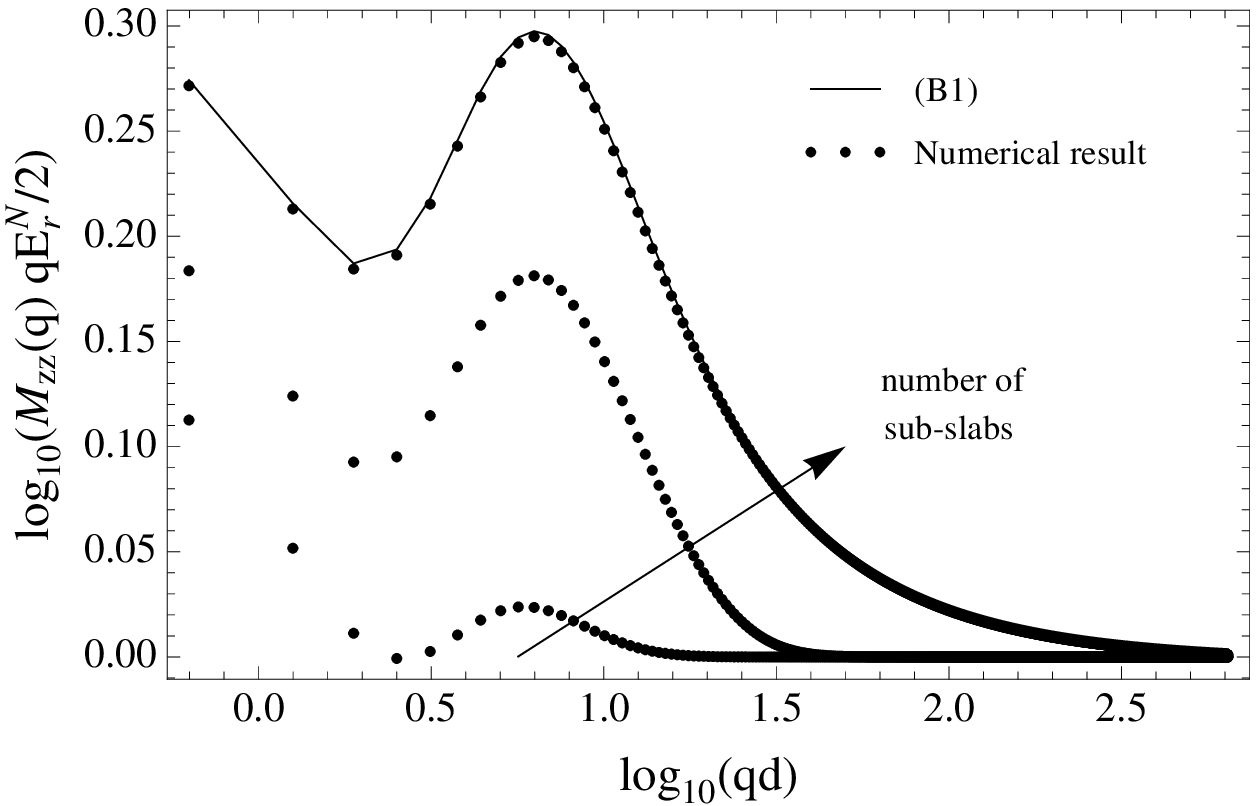} \label{comparisonscarsinusoid.eps}
                }
\caption{a) Schematic of a graded elastic coating bonded onto an elastic
half space, b) dimensionless surface response $M_{zz}\left(
q\right) qE_{\mathrm{r0}}/2$ (with $q_{y}=0$) as a function of the wave
number $dq$ ($q_{0}=2\protect\pi /L_{0}$). The coating is characterized by a
reduced elastic modulus $E_{\mathrm{r}}(z)=E_{\mathrm{rb}}\left[ 1+\left(
z+d\right) /d+\sin \left( 2\protect\pi \left( z+d\right) /d\right) \right] $%
, whereas the bulk by $E_{\mathrm{rb}}$. For a dimensionless coating
thickness $q_{0}d=0.0195$ and $L_{0}=0.1$. In (b) the continuous line is
from \protect\autoref{differentiated.graded.stepped}, whereas the dots are
obtained from the application of \protect\autoref{final} to a linear
discretization of the coating $E_{\mathrm{r}}(z)$ into a stepwise composite
of $4$, $9$ and $299$ layer sub-divisions.}
\label{test.numerical.sinus}
\end{figure}

In Fig. \ref{test.numerical.linear} and \ref{test.numerical.sinus} we show
b) the dimensionless surface response $M_{zz}\left(
q\right) qE_{\mathrm{r0}}/2$ (with $q_{y}=0$) as a function of the wave
number $dq$ ($q_{0}=2\pi /L_{0}$) for a graded elastic coating bonded onto
an elastic half space with reduced elastic modulus $E_{\mathrm{r}}(z)=E_{%
\mathrm{rb}}\left[ \left( 1+10^{4}\left( z+d\right) /d\right) \right] $ and $%
E_{\mathrm{r}}(z)=E_{\mathrm{rb}}\left[ 1+\left( z+d\right) /d+\sin \left(
2\pi \left( z+d\right) /d\right) \right] $, respectively. In the figures the continuous
line is from \autoref{differentiated.graded.stepped}, whereas the dots are
obtained from the application of \autoref{final} to a linear discretization of
the coating $E_{\mathrm{r}}(z)$ into a stepwise composite with different
layer sub-divisions (see figures caption). We observe that, whilst for the
linear graded rheology the surface response of the stepwise composite
converges to the continuously-graded in relatively few sub-layer divisions,
for the case of Fig. \ref{test.numerical.sinus} about a two orders of
magnitude refined discretization is required to reach convergence, i.e. for
complex rheological laws (as e.g. for biological applications) the adoption of \autoref{differentiated.graded.stepped}
should be computationally preferred to \autoref{final}.

\bibliographystyle{plain}
\bibliography{article_bib}

\begin{thebibliography}{10}

\bibitem{Lorenz2014565}
Lorenz B., Pyckhout-Hintzen W., and Persson B.N.J.
\newblock Master curve of viscoelastic solid: Using causality to determine the
  optimal shifting procedure, and to test the accuracy of measured data.
\newblock {\em Polymer (United Kingdom)}, 55(2):565--571, 2014.

\bibitem{Persson20013840}
Persson B.N.J.
\newblock Theory of rubber friction and contact mechanics.
\newblock {\em Journal of Chemical Physics}, 115(8):3840--3861, 2001.

\bibitem{Persson2006b}
Persson B.N.J.
\newblock Contact mechanics for randomly rough surfaces.
\newblock {\em Surface Science Reports}, 61(4):201--227, 2006.

\bibitem{Persson2012}
Persson B.N.J.
\newblock Contact mechanics for layered materials with randomly rough surfaces.
\newblock {\em Journal of Physics: Condensed Matter}, 24(9):095008, 2012.

\bibitem{Derjaguin1934}
B.~Derjaguin.
\newblock Untersuchungen über die reibung und adhäsion, iv.
\newblock {\em Kolloid-Zeitschrift}, 69(2):155--164, 1934.

\bibitem{Greenwood300}
J.~A. Greenwood and J.~B.~P. Williamson.
\newblock Contact of nominally flat surfaces.
\newblock {\em Proceedings of the Royal Society of London A: Mathematical,
  Physical and Engineering Sciences}, 295(1442):300--319, 1966.

\bibitem{Scaraggi201415}
Scaraggi M. and Persson B.N.J.
\newblock Rolling friction: Comparison of analytical theory with exact
  numerical results.
\newblock {\em Tribology Letters}, 55(1):15--21, 2014.

\bibitem{Scaraggi2014}
Scaraggi M. and Persson B.N.J.
\newblock Theory of viscoelastic lubrication.
\newblock {\em Tribology International}, 72:118--130, 2014.

\bibitem{schipper}
M.~Mokhtari and D.J. Schipper.
\newblock Existence of a tribo-modified surface layer of br/s-sbr elastomers
  reinforced with silica or carbon black.
\newblock {\em Tribology International}, 2014.
\newblock Article in Press.

\bibitem{Paggi2011696}
M.~Paggi and G.~Zavarise.
\newblock Contact mechanics of microscopically rough surfaces with graded
  elasticity.
\newblock {\em European Journal of Mechanics, A/Solids}, 30(5):696--704, 2011.

\bibitem{Persson2014}
B.N.J. Persson and M.~Scaraggi.
\newblock Theory of adhesion: Role of surface roughness.
\newblock {\em Journal of Chemical Physics}, 141(12), 2014.

\bibitem{scaraggi.in.prep.true}
M.~Scaraggi.
\newblock 2015.

\bibitem{scaraggi.in.prep}
M.~Scaraggi and B.N.J. Persson.
\newblock Friction and universal contact area law for randomly rough
  viscoelastic contacts.
\newblock {\em Journal of Physics Condensed Matter}, 27(10), 2015.

\bibitem{SCHWEITZ1986289}
JAN-ÅKE SCHWEITZ and LEIF ÅHMAN.
\newblock Chapter 9 - mild wear of rubber-based compounds.
\newblock In Klaus Friedrich, editor, {\em Friction and Wear of Polymer
  Composites}, volume~1 of {\em Composite Materials Series}, pages 289 -- 327.
  Elsevier, 1986.

\end{thebibliography}

\end{document}